\documentclass[11pt]{article}
\usepackage{amsmath, amsthm, amssymb}
\usepackage{float,epsfig}
\usepackage{color}
\usepackage{cancel}
\usepackage{graphicx}
\usepackage{rotating}
\usepackage{lscape}
\usepackage{braket}
\usepackage{subcaption}
\usepackage{tikz}
\usepackage{booktabs}
\usepackage{todonotes}
\usepackage{algpseudocode}
\usepackage{algorithm}
\usepackage{hyperref}
\usepackage{cleveref}
\usepackage{comment}
\usepackage{qcircuit}
\usepackage{xcolor}

\usepackage{lipsum}



\begin{document}

\newtheorem{theorem}{\bf Theorem}[section]
\newtheorem{proposition}[theorem]{\bf Proposition}
\newtheorem{corollary}[theorem]{\bf Corollary}
\newtheorem{lemma}[theorem]{\bf Lemma}

\theoremstyle{definition}
\newtheorem{definition}[theorem]{\bf Definition}
\newtheorem{example}[theorem]{\bf Example}
\newtheorem{exam}[theorem]{\bf Example}

\theoremstyle{remark}
\newtheorem{remark}[theorem]{\bf Remark}
\newtheorem{observation}[theorem]{\bf Observation}
\newcommand{\nrm}[1]{|\!|\!| {#1} |\!|\!|}

\newcommand{\ba}{\begin{array}}
\newcommand{\ea}{\end{array}}
\newcommand{\von}{\vskip 1ex}
\newcommand{\vone}{\vskip 2ex}
\newcommand{\vtwo}{\vskip 4ex}
\newcommand{\dm}[1]{ {\displaystyle{#1} } }

\newcommand\independent{\protect\mathpalette{\protect\independenT}{\perp}}
\def\independenT#1#2{\mathrel{\rlap{$#1#2$}\mkern2mu{#1#2}}}
\def \cnot{\mathrm{CNOT}}
\def \pf{{\bf Proof: }}

\newcommand{\be}{\begin{equation}}
\newcommand{\ee}{\end{equation}}
\newcommand{\beano}{\begin{eqnarray*}}
\newcommand{\eeano}{\end{eqnarray*}}
\newcommand{\inp}[2]{\langle {#1} ,\,{#2} \rangle}
\def\bmatrix#1{\left[ \begin{matrix} #1 \end{matrix} \right]}
\def\dmatrix#1{\left| \begin{matrix} #1 \end{matrix} \right|}
\def \noin{\noindent}
\newcommand{\evenindex}{\Pi_e}


\def \N{{\mathbb N}}
\def \Z{{\mathbb Z}}
\def \R{{\mathbb R}}
\def \C{{\mathbb C}}
\def \K{{\mathbb K}}
\def \J{{\mathcal J}}
\def \Q{{\mathbb Q}}
\def \Uf{{\mathsf U}}
\def \Sf{{\mathsf S}}
\def \sf{{\mathfrak s}}
\def \uf{{\mathfrak u}}
\def \calL{\mathcal{L}}

\def \calB{{B}}
\def \calK{\mathcal{K}}
\def \calC{\mathcal{C}}
\def \calP{\mathcal{P}}
\def \calD{{D}}
\def \calV{{V}}
\def \calG{{G}}
\def \calN{\mathcal{N}}
\def \calT{\mathcal{T}}
\def \calH{\mathcal{H}}
\def \calM{\mathcal{M}}
\def \calI{\mathcal{I}}
\def \calE{{E}}
\def \calU{{U}}
\def \norm{\nrm{\cdot}\equiv \nrm{\cdot}}

\def \cnot{\mathrm{CNOT}}

\def \diag{\mbox{diag}}
\def \tr{\mathrm{Tr}}
\def \lam{\lambda}
\def \sig{\sigma}
\def \Sig{\Sigma}
\def \Lam{\Lambda}
\def \ep{\epsilon}
\def \sgn{\mathrm{sgn}}
\def \det{\mathrm{det}}

\algrenewcommand\algorithmicrequire{\textbf{Input:}}
\algrenewcommand\algorithmicensure{\textbf{Output:}}
\newcommand{\cupdot}{\mathbin{\mathaccent\cdot\cup}}

\newcommand{\tm}[1]{\textcolor{magenta}{ #1}}
\newcommand{\tre}[1]{\textcolor{red}{ #1}}
\newcommand{\tb}[1]{\textcolor{blue}{ #1}}


\title{A quantum neural network framework for scalable quantum circuit approximation of unitary matrices \thanks{This is an extended version of a conference paper titled ``Scalable approximation of $n$-qubit unitaries", IEEE International conference on Quantum Computing and Engineering (QCE 23).}}

\author{ Rohit Sarma Sarkar\thanks{Corresponding author, Department of Mathematics, IIT Kharagpur, India, E-mail: rohit15sarkar@yahoo.com
 }\thanks{The author currently works at the International Centre for Theoretical Sciences (ICTS-TIFR), Bengaluru, India} \,\, and \,\,   Bibhas Adhikari\thanks{Department of Mathematics, Indian Institute of Technology Kharagpur, bibhas@maths.iitkgp.ac.in, bibhas.adhikari@gmail.com} \thanks{The author currently works at Fujitsu Research of America, Inc., Santa Clara, CA, USA}
  }

\date{}

\maketitle

{\small \noin{\bf Abstract.}} In this paper, we develop a Lie group theoretic approach for parametric representation of unitary matrices. This leads to develop a quantum neural network framework for quantum circuit approximation of multi-qubit unitary gates. Layers of the neural networks are defined by product of exponential of certain elements of the Standard Recursive Block Basis, which we introduce as an alternative to Pauli string basis for matrix algebra of complex matrices of order $2^n$. The recursive construction of the neural networks implies that the quantum circuit approximation is scalable i.e. quantum circuit for an $(n+1)$-qubit unitary can be constructed from the circuit of $n$-qubit system by adding a few $\cnot$ gates and single-qubit gates. \\

 

{\small \noin{\bf Key words.}} Quantum neural network, parametrized quantum circuit, quantum compilation, multi-controlled rotation gates, Lie algebra, special unitary matrices 
\section{Introduction}
Decomposing dense unitary matrices into product of sparse unitaries is a subject of interest for mathematicians, physicists and computer scientists. Specifically in quantum computing, the problem is reiterated in the form of constructing any $n$-qubit quantum gate or a circuit using only one and two qubit gates i.e. writing a $2^n\times 2^n$ unitary matrix as a product of permutations and Kronecker products of rotation gates belonging to $\mathsf{S}\mathsf{U}(2),$ the \textit{special linear group} of $2\times 2$ complex matrices. This problem of finding good approximation of unitaries is often referred to as \textit{quantum compilation problem}\cite{harrow2001,paler2017}.

The existence of such a construction is validated by the Solovay-Kitaev algorithm, which shows that any $n$-qubit quantum circuit can be approximated using a sequence of just one qubit rotation gates and CNOT gates. Hence, these gates are \textit{computationally universal} and can represent unitaries for multi-qubit systems \cite{harrow2001,barenco1995}. Mathematically, a gate set $\mathcal{G}$ is said to be computationally universal \cite{dawson2005} in $\mathsf{S}\mathsf{U}(d)$ if the group generated by $\mathcal{G}$ is dense in $\mathsf{SU}(d)$. In other words, given any quantum gate $U\in\mathsf{SU}(d)$ and any accuracy  $\epsilon>0, \exists $ a product $S\equiv g_1\hdots g_m$ of gates from $\mathcal{G}$ which is an $\epsilon$-approximation to $U$ i.e. $\|U-S\|<\epsilon$ where $\|.\|$ is the standard operator norm \cite{golub2013}.

In Solovay-Kitaev algorithm however, the approximation of unitary matrices and length of the sequence are directly correlated, which shows that longer sequences yield better approximations. Hence, one needs to do an exhaustive search over sequences of a particular length in order to find the minimal distance from the given unitary matrix, known as Solovay-Kitaev approximation algorithms \cite{kitaev2002}.  Since the search covers only a sparse region of the entire space of possible approximation sequences, several methods are proposed for optimization of the Solovay-Kitaev algorithm that finds application in \textit{fault-tolerant quantum computation} \cite{pham2013,zhiyenbayev2018}. 
It is also to be noted that the problem of quantum compilation is not limited to qubit systems and thus, can be generalized for any qudit systems as well. In such cases the problem boils down to approximating a $d\times d$ unitary matrix $U\in \mathsf{U}(d)$ via a sequence of ``instruction gates"\cite{dawson2005} from an instruction gate set $\mathcal{G}$ that satisfies the following three conditions: (a) All gates $g\in \mathcal{G}$ are in $\mathsf{S}\mathsf{U}(d)$. (b) The gates in $g\in \mathcal{G}$ are closed under inversion in $\mathcal{G}$. (c) $\mathcal{G}$ is a computationally universal set in $\mathsf{S}\mathsf{U}(d)$. Such a task is accomplished by generalizing the Solovay-Kitaev algorithm \cite{dawson2005}. However, the algorithm shares the similar drawbacks like its qubit counterpart.

There have been advancements for efficiently approximating $n$-qubit unitaries using various methods such as recursive CS decomposition and Quantum Shannon-decomposition \cite{mottonen2004,krol2022}. However, the algorithms developed, though aimed at  minimizing number of CNOT and one-qubit gates, rely on numerical algorithms to find SVD and eigen-decomposition of a matrix, which are itself challenging computational problems for large matrices. Recently, an optimization based viewpoint for the compilation problem has generated a lot of interest \cite{madden2022best,madden2022,nakajima2005}. In this approach, a unitary matrix is found that can be realized in hardware with constraints that is the closest to a target unitary with respect to a metric. Various cost functions are defined in these optimization-based approaches to achieve a good implementation of the target unitary. For example, optimizing the structure (i.e., where to place a $\mathrm{CNOT}$ gate), optimizing the rotation angles of the rotation gates, optimizing the number of $\mathrm{CNOT}$ count etc. after writing a parametric representation using matrix decomposition of the target unitary \cite{shende2005,shende2004}.

Other methods like QFAST \cite{younis2020,younis2021qfast} makes use of geometry of the unitary manifold by approximating a target unitary with help of the tangent space around the identity matrix. It is evident that the Pauli strings form a basis for the complex vector space $M_{2^n}(\mathbb{C})$ of all $2^n\times 2^n$ complex matrices. Further the Pauli strings are Hermitian and traceless, making them basis elements that are $\iota$-times the Pauli strings for the Lie algebra of the unitary manifold, where $\iota=\sqrt{-1}$. Hence, in this method one can approximate a $2^n\times 2^n$ unitary matrix using exponentials of scaling of Pauli strings. Other methods like using decomposition of isometries into single qubit rotation gates and CNOT gates helps in reducing the total number of CNOT gates while decomposing a generic unitary matrix \cite{Iten2016,malvetti2021}. An isometry is an inner-product-preserving transformation that maps between two Hilbert spaces with different dimensions \cite{Iten2016}. In a physical sense, isometries can be thought of as the introduction of
ancilla qubits in a fixed state which is generally $\ket{0}$, followed by a generic unitary on the system and ancilla qubits \cite{Iten2016}. There is no rigidity while constructing the general unitary in this method due to the fact that the action only has to be specified when the ancilla systems start in state $\ket{0}$ which in turn, helps to reduce the number of CNOT gates in the circuit \cite{Iten2016}.

A variational approach to quantum compilation problem has also been developed in the recent past. For instance, a quantum-assisted quantum compiling (QAQC) method is introduced in \cite{khatri2019} to approximate a (possibly unknown) target unitary to a trainable quantum gate sequence, which is able to optimally compile larger-scale gate sequences in contrast to classical approaches that are limited to smaller gate sequence. A recursive variational quantum compiling algorithm (RVQC) is proposed in \cite{bilek2022}. Here the target circuit is divided into several parts and each part is recursively compressed into parameterized ansatz.

From the discussions above, it is evident that the problem of approximating generic unitary matrices by ``well-known" sparse unitary matrices is of great significance in quantum computing. This has led to a surge of research in this area over the years, making it fascinating to address this problem from the perspective of the quantum circuit model of computation.

In this paper, we present an optimization-based approach to approximate a given unitary matrix comparing it with a generic parameterized unitary matrix. This leads to the development of a quantum neural network framework for implementing $n$-qubit unitaries using quantum circuits of $\cnot$ and one-qubit gates. To obtain a generic parameterized representation for unitaries,  a new Hermitian unitary basis for matrix algebra of $d\times d$ complex matrices is introduced, with the aim of expressing any unitary through product of exponentials of $\iota$-times the proposed basis elements. The new bases have Hermitian and unitary elements, with diagonal or $2$-sparse matrices, alike the Pauli string basis. The proposed bases have an advantage over the Pauli string basis as the method of constructing such basis elements is recursive. Further the matrices are permutation similar to block diagonal matrices and  making it easier to compute the exponentials of the basis elements.

First, we introduce a recursive approach for construction of a basis comprises of Hermitian unitary $1$-sparse matrices  for the matrix algebra of $d\times d$ complex matrices, $d> 2$. For $d=2,$ the basis is the Pauli basis, and hence the proposed construction may be regarded as a generalization of the Pauli basis of $2\times 2$ complex matrices. Then altering some of the basis elements, replacing them by Pauli strings formed by Kronecker product of the identity matrix of order $2$  and Pauli $Z$ matrix of order $2$, we propose a Hermitian unitary trace-less basis for algebra of $2^n\times 2^n$ complex matrices. We call this basis as \textit{Standard Recursive Block Basis} (SRBB), inspired by the recursive construction of the basis elements which have certain block structure. Then we provide a direct computable expression for the exponentials of these basis elements, which is further employed for exact synthesis of any $2$-level unitary matrix (a matrix obtained from the identity matrix by replacing a $2\times 2$ principal submatrix with a unitary block) of order $2^n$ and block-unitary matrices that correspond to multi-controlled rotation gates. It is needless to mention that any unitary matrix can be written as a product of $2$-level matrices \cite{NielsenChuang2011}.

Then utilizing the obtained basis for the Lie algebra of skew-Hermitian matrices and considering the unitary matrices as its corresponding Lie group, we develop algorithms for approximation of any unitary matrix as product of exponentials of the basis elements, which form one-parameter subgroup of unitary matrices. This formulation of the approximation can be interpreted as a quantum neural network, in which the unitary matrices represent the quantum evolution of an $n$-qubit system that can be compiled using quantum circuits of parameterized elementary gates for practical implementation in \textit{Noisy Intermediate Scale Quantum} (NISQ) computers.  Consequently, we formulate the optimization problem of estimating the values of these parameters for approximation of any target unitary matrix, very much like variational quantum algorithms (VQAs). The objective function of the optimization problem is defined as the Frobenius distance of the parameterized unitary approximation and the target unitary. 

 It may be emphasized here that due to the exponential dimension of the concerned matrices, which increases with $n,$ (which corresponds to the $n$-qubit system), it is a classically hard optimization problem for exponentially large number of parameters, in the generic case, when all the basis elements are employed to approximate the target unitary. Obviously, several basis elements need not be considered for the approximation when the target unitary is sparse and has certain sparsity pattern. Besides, the execution time may be reduced for small number of parameters, when standard classical optimization algorithms are used. The classical optimization algorithms, such as Nelder-Mead and Powell's method are usually applied as classical optimization algorithms for VQAs. In this paper, we employ Nelder-Mead method to perform  the simulation for various target unitaries which appear in quantum computation. We also report an improvement of the execution time in compiling standard quantum gates using this approach in Table \ref{Table:error2q} as compared to the same using all the basis elements in our previous simulation reported in \cite{SarmaSarkar2023}. 
 
 It may further be noted that ordering of the basis elements play a crucial role for the approximation which we address while constructing quantum circuits for unitaries that are product of exponentials of certain basis elements. Indeed, we identify and determine the basis elements such that products of their exponentials have suitable existing quantum circuit representation such as multi-controlled rotation gates. We also develop quantum circuits for exponentials of basis elements that are diagonal matrices, and of the permutations which are product of certain type of transpositions that arise during the approximation. Thus we develop a framework of a multi-layered quantum neural network defined by quantum circuit of parameterized rotation gates and $\cnot$ gates for approximating a target unitary matrix, applicable for implementation in NISQ computers. Indeed, we decide on the choice of the ordering of the basis elements such that it reduces the number of $\cnot$ gates in the quantum circuit implementation of the approximation algorithm. 

 Moreover, we show that the proposed recursive approach for the basis has the advantage that the proposed quantum circuit representation of the approximation for $n$-qubit systems is scalable. Thus, given the circuit for $n$-qubits, the circuit for $(n+1)$-qubits can be implemented using the current circuit with the addition of new $\cnot$ gates and one-qubit rotation gates. We prove that the proposed quantum circuit of one layer of approximation has the use of at most $2 \cdot 4^n + (n-5)2^n$ $\cnot$ gates, and at most $\frac{3}{2} \cdot 4^n - \frac{5}{2} \cdot 2^n + +1$ one-qubit rotation gates corresponding to $Y$ and $Z$ axes.
  
 We examine various scenarios to evaluate the effectiveness of our approximation algorithms in approximating standard and random unitary matrices for $2$-qubit, $3$-qubit, and $4$-qubit systems, and unitary matrices of order $d=3,5$. Our results indicate that the proposed algorithms perform better when the target unitaries are sparse when only one layer of approximation is used, and the error of the approximation reduces with the increase of number of layers for the approximation.  It is evident that the performance of the algorithm is influenced by the initial parameter values and the optimization technique utilized to obtain the optimal parameter values. Thus we randomize for the choice of the initialization of the optimization algorithm. Lastly, we present an algorithm that enables the implementation of the proposed quantum circuits from $n$-qubit to $(n+1)$-qubit systems.

 The remainder of the paper is structured as follows. In Section \ref{Sec:2}, recursive methods for construction of a basis consists of Hermitian unitary matrices for complex matrices of size $d\times d$ is given, whcih is further modified to obtain a suitable basis for algebra of $2^n\times 2^n$ complex matrices. In Section \ref{Sec:3}, we propose a Lie group theoretic approach for approximation of unitary matrices through proposed basis elements. Exact representation of $2$-level matrices and unitary matrices corresponding to multi-controlled rotation gates through product exponentials of certain proposed basis elements are also given. Section \ref{Sec:4} presents methods for approximating unitary matrices of order $2^n$ i.e. $n$-qubit unitaries through the use of SRBB elements. A quantum neural network framework for developing a generic parametric representation of $n$-qubit unitaries is provided via an optimization-based approximation algorithm. Numerical simulation results for examples of Haar random unitaries are given. In Section \ref{sec:circuit}, a scalable quantum circuit for the proposed approximation algorithm is established, providing a quantum circuit representation and implementation of $n$-qubit unitaries. Finally, we conclude the paper with some remarks on future research directions.
 \section{Recursive construction of Hermitian unitary basis}\label{Sec:2}

 In this section, we provide a recursive method for generation of a basis consisting of Hermitian, unitary matrices for the matrix algebra of $d\times d,$ $d\geq 3$ matrices. Then this basis is employed to define a parametric representation of unitary matrices of order $d\times d$. We denote the identity matrix of order $k$ as $I_k,$ $k\geq 0$ where $I_1=[1]$, and $I_0$ is just void which means to ignore the index from the construction. We denote the Pauli basis by $\sigma$ whose elements are given by $$ \sigma_1=\bmatrix{0&1\\ 1&0}, \sigma_2=\bmatrix{0&-i\\ i&0}, \sigma_3=\bmatrix{1&0\\ 0&-1}, \sigma_4=\bmatrix{1&0\\0&1},$$ called Pauli matrices.  Then we define a new basis of Hermitian unitary trace-less matrices for the algebra of $2^n\times 2^n$ complex matrices by changing some of the elements of the former basis.

 \subsection{Construction of Hermitian unitary basis for $d\times d$ complex matrices}\label{constructionHerm}

The following theorem describes a recursive approach for construction of Hermitian unitary basis of $\C^{d\times d}$, with $d^2-1$ of them having trace zero when $d$ is even. The proof of the theorem is given in the Appendix.

\begin{theorem}\label{thm:basis2} 
Let $\boldsymbol{\mathcal{B}^{(d)}}=\{B^{(d)}_j : 1\leq j\leq d^2\},$ $d> 2$ denote the desired ordered basis for the matrix algebra of $d\times d$ complex matrices. Then setting $\boldsymbol{\mathcal{B}^{(2)}}$ as the Pauli basis, the elements of $\boldsymbol{\mathcal{B}^{(d)}}$ can be constructed from the elements of $\boldsymbol{\mathcal{B}^{(d-1)}}$ using the following recursive procedure \begin{eqnarray*}
\hspace{-0.75cm}B^{(d)}_j=
     \begin{cases}
     \left[ 
    \begin{array}{c|c} 
     B^{(d-1)}_{j}  & 0 \\ 
      \hline 
      0 & (-1)^{d-1}
    \end{array} 
    \right]; \,\, \mbox{if} \,\, j \in \{1,\hdots,(d-1)^2-1\}, \\
      
      P_{(d-k,d-1)}\left[ 
    \begin{array}{c|c} 
      D & 0 \\ 
      \hline 
      0 & \sigma_1
    \end{array} 
    \right]  P_{(d-k,d-1)}; \,\, \mbox{if} \,\, j=(d-1)^2+(k-1),  k\in\{1,\hdots,d-1\}\\ 
       
      P_{(d-k,d-1)}\left[ 
    \begin{array}{c|c} 
      D & 0 \\ 
      \hline 
      0 & \sigma_2 
    \end{array} 
    \right]P_{(d-k,d-1)}; \,\, \mbox{if} \,\, j=(d-1)^2+(d-1)+(k-1),  k\in\{1,\hdots,d-1\}\\ 
     
     \left[ 
    \begin{array}{c|c} 
      I_{\lfloor d/2\rfloor +1} & 0 \\ 
      \hline 
      0 & -I_{\lfloor d/2\rfloor}
    \end{array} 
    \right]; \,\, \mbox{if} \,\, j = d^2-1 \,\, \mbox{and} \,\, d \,\, \mbox{is odd}\\
    
     \left[ 
    \begin{array}{c|c} 
      \Sigma & 0 \\ 
      \hline 
      0 & \sigma_3  \\
    \end{array} 
    \right]; \,\, \mbox{if} \,\, j = d^2-1 \,\, \mbox{and} \,\, d \,\, \mbox{is even} \\
    
    I_d \,\, \mbox{if} \,\, j = d^2 
     \end{cases},\end{eqnarray*}
     where $P_{k,(d-1)}$ is the permutation matrix of order $d\times d$ corresponding to the $2$-cycle $(k,d-1),$ $D=\mbox{diag}\{d_l : 1\leq l\leq d-2\},$ $d_l=(-1)^{l-1},$ and $\Sigma=\bmatrix{I_{\lfloor d/2\rfloor-1} & 0 \\ 0 & -I_{\lfloor d/2\rfloor-1}}$ Besides, $$\tr(B_j^{(d)})=\begin{cases}
      1 \,\, \mbox{if} \,\, d \,\, \mbox{is odd} \\
      0 \,\, \mbox{if} \,\, d \,\, \mbox{is even,}
     \end{cases},$$ $1\leq j\leq d^2-1,$ $\left(B_j^{(d)}\right)^2=I_d,$ and $\{B_j^{(d)} : 1\leq j\leq d^2-1\}$ forms a basis for $su(d)$ when $d$ is even. The basis elements that are diagonal matrices are given by $B^{(n)}_j$ where $j=m^2-1, 2\leq m\leq d$ and $B^{(d)}_{d^2}=I_d.$
 \end{theorem}
\pf First observe that the matrices $B^{(d)}_j, 1\leq j\leq d^2$ are Hermitian and unitary due to the construction. Also, $\tr(B_j^{(d)})=0$ when $d$ is even and $\tr(B_j^{(d)})=1$ when $d$ is odd. Now, we show that these matrices form a linearly independent subset of $\C^{d\times d}.$ Suppose $d$ is even. Then setting
 
 \begin{eqnarray}\hspace{-0.65cm}0 &=& \sum_{m=1}^{(d-1)^2-1} c_{1m}\bmatrix{B_m^{(d-1)}&0\\0& -1} + \sum_{m=1}^{(d-1)} c_{2m}P_{(m(d-1))}\bmatrix{D&0\\0&\sigma_1}P_{(m(d-1))} \nonumber \\ && + \sum_{m=1}^{(d-1)} c_{3m}P_{(m(d-1))}\bmatrix{D&0\\0&\sigma_2}P_{(m(d-1))} + c_{44}\bmatrix{\Sigma &0\\0&-\sigma_3} + c_{55} I_{d} \nonumber \\ 
 &=& \sum_{m=1}^{(d-1)-1^2} \underbrace{\bmatrix{c_{1m}B_m^{(d-1)}&0\\0&-c_{1m}}}_A + \sum_{m=1}^{(d-1)} \underbrace{P_{(m(d-1))} \bmatrix{(c_{2m}+c_{3m})D &0\\0&c_{2m}\sigma_1+c_{3m}\sigma_2} P_{(m(d-1))}}_B \nonumber \\ && + \underbrace{\bmatrix{c_{44}\Sigma +c_{55}I_{d-2}&0\\0& -c_{44}\sigma_3+c_{55}I_2}}_C. \label{eqn:pf1} \end{eqnarray}
 
It can be seen from equation (\ref{eqn:pf1}) that the first $d-1$ entries of the last column of $B$ are given by $c_{2m}-ic_{3m},$ $1\leq m\leq d-1,$ whereas these corresponding entries in $A$ and $C$ are zero. Also first $n-1$ entries (left to right) of the last row of $B$ are given by $c_{2m}+ic_{3m},$ $1\leq m\leq d-1,$ whereas these corresponding entries in $A$ and $C$ are zero. Then it immediately follows that $c_{2m}=c_{3m}=0,$ $1\leq m\leq d-1.$ Then the equation (\ref{eqn:pf1}) becomes \begin{eqnarray} 0=\sum_{m=1}^{(d-1)^2-1} \bmatrix{c_{1m}B_m^{(d-1)}&0\\0&-c_{1m}} + {\bmatrix{c_{44}\Sigma +c_{55}I_{d-2}&0\\0& -c_{44}\sigma_3+c_{55}I_2}}. \end{eqnarray} Further, since $\{B_m^{(d-1)} : 1\leq m\leq (d-1)^2-1\}\cup I_{d-1}$ is linearly independent, then using the same method described above, the matrix $\sum_{m=1}^{(d-1)^2-1}c_{1m}B_j^{(d-1)}$ has all non-diagonal entries $0$. Thus the only terms remain are diagonal matrices i.e. the equation reduces to  \small\begin{eqnarray} \label{eqn:LI3} \hspace{-1.85cm}0 &=&\sum_{m=2}^{(d-1)} \bmatrix{c_{1(m^2-1)}B_{(m^2-1)}^{(d-1)}&0\\0&-c_{1(m^2-1)}} + {\bmatrix{c_{44}\Sigma +c_{55}I_{d-2}&0\\0& -c_{44}\sigma_3+c_{55}I_2}}  \\ 
 &=&  \bmatrix{\sum_{m=2}^{(d-1)}c_{1(m^2-1)}B_{(m^2-1)}^{(d-1)}+c_{44}\bmatrix{\Sigma & 0\\ 0& -1}+c_{55}I_{d-1}&0\\0&(-\sum_{m=2}^{(d-1)}c_{1(m^2-1)})+c_{44}+c_{55}}\nonumber,
  \end{eqnarray}\normalsize where $B^{(d-1)}_{m^2-1}$ and $I_{d-1}=B^{(d-1)}_{(d-1)^2},$ $2\leq m\leq d-1$ are proposed basis elements of $\C^{(d-1)\times (d-1)}.$ 
 
 For a diagonal matrix $M$ of order $d$ with diagonal entries $m_{jj}, 1\leq j\leq d,$ set $\diag(M)=[m_{11} \, m_{22} \, \hdots \, m_{dd}]^T$ as the column vector. Then observe that equation (\ref{eqn:LI3}) can be described as a linear system $Ax=0,$ where $x=\bmatrix{c_{13}& \hdots & c_{1((d-1)^2-1)} & c_{44} & c_{55}}^T$ and \begin{eqnarray*}
     \hspace{-0.85cm}A=\bmatrix{\diag\left(B^{(d)}_{3}\right) & \diag\left(B^{(d)}_{8}\right) & \hdots & \diag\left(B^{(d)}_{(d-1)^2-1}\right) & \diag\left(\bmatrix{\Sigma&0&0\\ 0&-1&0\\ 0&0&1}\right)& \diag\left(I_{d}\right)}.
 \end{eqnarray*} 
 
 Next, we show that $A$ is non-singular i.e. the columns of $A$ form a linearly independent set. Suppose $$\sum_{m=2}^{d-1} \alpha_m\bmatrix{\diag\left(B^{(d-1)}_{m^2-1}\right) \\ -1} +\beta \bmatrix{\diag\left(\bmatrix{\Sigma&0&0\\ 0&-1&0 \\ 0&0&1}\right)} + \gamma \bmatrix{\diag\left(I_{d}\right)}=0.$$ Then multiplying the all-one vector ${\bf 1}_d^T$ from left at the above equation, we obtain $d\gamma =0$ since sum of entries of all other vectors are zero. This further implies $\gamma=0.$ Thus we have $$\sum_{m=2}^{d-1} \alpha_m\bmatrix{\diag\left(B^{(d-1)}_{m^2-1}\right) \\ -1}  + \beta \bmatrix{\diag\left(\bmatrix{\Sigma&0&0\\ 0&-1&0 \\ 0&0&1}\right)}=0.$$
 Now note that the first entry of all the vectors in the above vectors are $1.$ Then considering the first and last entries of the above vectors, we obtain $$
 \beta+\sum_{m=2}^{d-1} \alpha_m = 0 \,\, \mbox{and} \,\,
 \beta -\sum_{m=2}^{d-1} \alpha_m = 0,$$ whose only solution is $\beta=\alpha_m=0$ for all $m.$ Hence the desired result follows when $m$ is even. The proof for odd $m$ follows similarly.  \hfill{$\square$}

\begin{remark}
 
   \begin{itemize}
       \item[(a)] Note that any of the  basis elements described by the above theorem that is a non diagonal matrix, is one of the following forms $$P\bmatrix{D_1 & 0 & 0 \\ 0 & \sigma & 0 \\ 0&0& D_2}P, \,\, \,\, P\bmatrix{\sigma & 0\\ 0 & D}P, \,\,\,\, P\bmatrix{D &0 \\0 & \sigma}P$$ where $D, D_1, D_2$ are diagonal matrices with entries from $\{1,-1\}$, $\sigma\in\{\sigma_1,\sigma_2\}$ and $P$ is a $2$-cycle. Thus the basis elements are unitary, Hermitian, ans $1$-sparse matrices (alike Pauli string basis elements).
       
       \item[(b)] Then exponentials of these matrices are of the form $$P\bmatrix{\exp(D_1) & 0 & \\ 0 & \exp(\sigma) & 0 \\ 0&0& \exp(D_2)}P, \,\,\,\, P\bmatrix{\exp(\sigma) & 0\\ 0 & \exp(D)}P, \,\,\,\, P\bmatrix{\exp(D) &0 \\0 & \exp(\sigma)}P.$$
       
       \item[(c)] The indices $j$ for which the permutation matrix $P=I_d,$ and the basis elements are of the form 
       $$\bmatrix{D_1 & 0 & 0 \\ 0 & \sigma & 0 \\ 0&0& D_2}, \, \bmatrix{\sigma&0\\0&D} \, \mbox{or} \, \bmatrix{D&0\\0&\sigma}$$ with $\sigma=\sigma_1$ when $j\in \mathcal{J}_{\sigma_1}=\{(j-1)^2 | 2\leq l\leq d\},$ and $\sigma=\sigma_2$ when $j\in \mathcal{J}_{\sigma_2}=\{l^2-1 | 2\leq l\leq d\}.$ 
       
       \item[(d)] The basis elements with indices $j\in\mathcal{J}=\{l^2-1 : 2\leq l\leq d\}\cup\{d^2\}$ are diagonal matrices, which are orthogonal to each other. Obviously, $|\mathcal{J}|=d-1.$

   \end{itemize}  
\end{remark}
 \subsubsection{Hermitian unitary basis for $2^n\times 2^n$ complex matrices }
 Now, we present another Hermitian unitary basis of $\C^{2^n\times 2^n}$ which we will play a crucial role in the remainder of the paper. The idea is that we now replace the diagonal basis elements of $\boldsymbol{\mathcal{B}^{(2^n)}}$ described in Theorem \ref{thm:basis2} by another set of diagonal matrices keeping invariance of the linearly independent property of the basis. First note that the set of matrices \begin{equation}\label{eqn:diag}\mathcal{D}_{IZ}=\{A_1\otimes \hdots \otimes A_n : A_j\in\{I_2,\sigma_3\}, 1\leq j\leq n\}\end{equation} is a set of $2^n$ linearly independent diagonal matrices with trace zero except when $A_j=I_2$ for all $j$ i.e. $A_1\otimes \hdots\otimes A_n=I_{2^n}.$ We call this as the {\bf S}tandard {\bf R}ecursive {\bf B}lock {\bf B}asis (SRBB).

\begin{corollary}\label{cor:basis} ({\bf SRBB})
Let $\boldsymbol{\mathcal{B}^{(2^n)}}=\{B^{({2^n})}_j : 1\leq j\leq {2^{2n}}\}$ denote the basis described in Theorem \ref{thm:basis2}, and $\mathcal{D}_{IZ}$ is given by equation (\ref{eqn:diag}). Then the set $\boldsymbol{\mathcal{U}}^{(2^n)}=\{U^{(2^n)}_j : 1\leq j\leq 2^{2n}\}$, where $$U^{(2^n)}_j = \begin{cases}
 D \in\mathcal{D}_{IZ}\,\, \mbox{if} \,\, j\in \mathcal{J}=\{l^2-1 : 2\leq l\leq 2^n\}\cup\{2^{2n}\} \\
 B_j^{(2^n)}, \,\, \mbox{otherwise} 
\end{cases}$$ forms a Hermitian unitary basis for $\C^{2^n\times 2^n}$. Besides, $\tr(U_j^{(2^n)})=0$ when $U^{(2^n)}_j\neq I_{2^n}.$     
\end{corollary}

Observe that the non-diagonal basis matrices as defined in Corollary \ref{cor:basis} are of two types as described below.

\begin{equation}\label{eqn:nqbasis}\left[U\right]_{kl}=\begin{cases}
 (-1)^{l-1} \,\, \mbox{if} \,\, k=l\notin\{p,q\} \\
 1 \,\, \mbox{if} \,\, k=p, l=q \\
 1 \,\, \mbox{if} \,\, k=q, l=p \\
 0 \,\, \mbox{otherwise}
\end{cases} \,\, \mbox{and} \,\, \left[U\right]_{kl}=\begin{cases}
 (-1)^{l-1} \,\, \mbox{if} \,\, k=l\notin\{p,q\} \\
 -i \,\, \mbox{if} \,\, k=p, l=q \\
 i \,\, \mbox{if} \,\, k=q, l=p \\
 0 \,\, \mbox{otherwise}.
\end{cases}\end{equation} $1\leq k,l\leq 2^n.$

Now, we prove certain results which will be used in sequel. First we introduce a function which provides an ordering of the diagonal basis elements of $\boldsymbol{\mathcal{U}}^{(2^n)}.$ From now onward, we denote $A_1\otimes A_2\otimes \hdots\otimes A_m=\otimes_{i=1}^m A_i$ for some matrices or vectors $A_i.$ If $A_i=A$ for all $i$ then we denote $\otimes_{i=1}^m A_i=\otimes^m A.$ 

 \begin{definition}\label{definition2}
 Define $\chi:\{I,Z\}\rightarrow \{0,1\}$ such that $\chi{(I)}=0,\chi(Z)=1.$ For any positive integer $m,$ define $\chi_m:\{\otimes_{i=1}^m A_i \,|\, A_i\in \{I,Z\},1\leq i\leq m\}\rightarrow \{0,1,\hdots,2^m-1\}$ such that $$\chi_m\left(\otimes_{i=1}^m A_i\right)=\sum_{i=1}^m 2^{i-1}\chi(A_i).$$
 \end{definition}

The above definition is inspired from the fact that for any matrix $A=[A_0 \, A_1]\in\C^{2\times 2}$, the columns of $\otimes^nA$ are ordered according to the lexicographic ordering of binary strings, where the bits $0$ and $1$ represent the first column $A_0$ and the second column $A_1$ of $A$. The $k$-th column of $\otimes^n A$ corresponds to the the binary string representation of $k$, say $k_1k_2\hdots k_n, k_j\in\{0,1\}$, and hence it is given by $A_{k_1}\otimes A_{k_2}\otimes \hdots \otimes A_{k_n},$  $0\leq k\leq 2^n-1.$ In particular, for the Hardamard matrix $H_2,$ we can write $$H_2=\frac{1}{\sqrt{2}}\bmatrix{\diag(I_2) & \diag(\sigma_3)}.$$ Hence the $k$-th column of $2^{n/2}H_{2^n}:=2^{n/2} \left( \otimes^n H_2\right) $ is given by $\diag(A_{k_1})\otimes \diag(A_{k_2}) \otimes \hdots \otimes \diag(A_{k_n})=\diag(A_{k_1}\otimes A_{k_2}\otimes \hdots \otimes A_{k_n}),$ where $k=k_1k_2\hdots k_n$ is the binary representation of $k$ and $A_{k_l}\in\{I_2, \sigma_3\},$ $1\leq l\leq n.$ Thus there is a one-to-one correspondence between the columns of $2^{n/2}H_{2^n}$   and the diagonal basis elements of $\boldsymbol{\mathcal{U}}^{(2^n)},$ through the $\diag$ operation.\\

Now, we provide parametric representations of unitary matrices of order $d.$

\section{Lie group theoretic approach for parametric representation of unitary matrices}\label{Sec:3}
It is well-known that the set of all unitary matrices of order $d,$ denoted by $\Uf(d)$ forms a Lie group and the corresponding Lie algebra is the real vector space of all skew-Hermitian matrices of order $d$ which we denote as $\uf(d).$ A classification of unitary matrices is that: any unitary matrix can be expressed as exponentials of a skew-Hermitian matrix i.e. the map $\exp: \uf(d)\rightarrow \Uf(d)$ is surjective [Theorem 3.2, \cite{gallier2020}]. Now, we develop a parametric representation of  unitary matrices of order $d.$ 

We recall from [paper 2, \cite{varadarajan2013}] that if $\{X_1,\hdots, X_k\}$ is a basis of the Lie algebra of a Lie group $G$ then for some $\theta>0,$ the map $$\psi : (\theta_1,\theta_2, \hdots, \theta_k) \mapsto \exp(\theta_1X_1)\exp(\theta_2X_2)\hdots\exp(\theta_kX_k)$$ from $\R^k$ into $G$ is an analytic diffeomorphism of the cube $I_\theta^k=\{(\theta_1,\hdots,\theta_k) : |\theta_j|<\theta, 1\leq j\leq k\}$ of $\R^k$ onto an open subset $U$ of $G$ containing the identity element $I$ of $G.$ If $x_1,\hdots,x_k$ are the analytic functions on $U$ such that the map $y\mapsto (x_1(y),\hdots, x_k(y))$ inverts $\psi,$ then for $1\leq j\leq k,$ $$x_j(\exp \theta_1X_1, \exp \theta_2X_2, \hdots,\exp \theta_kX_k)=\theta_j, (\theta_1,\hdots,\theta_k)\in I^k_\theta.$$ Then $x_1,\hdots,x_k$ are called the canonical coordinates of the second kind around $I$ with respect to the basis $\{X_1,\hdots, X_k\}.$ 

Setting $G=\Uf(d)$, the (real) dimension of $\uf(d)$ is $d^2$ and if $\{B_j^{(d)} : 1\leq j\leq d^2\}$ denotes a basis of $\uf(d)$ then we have the following theorem. 

\begin{theorem}\label{thm:span}
There exists a $\theta>0$ such that $\left\{\prod_{j=1}^{d^2} \exp\left(i\theta_jB_j^{(d)}\right) : (\theta_1,\hdots,\theta_{d^2}) \in I^{d^2}_\theta \right\}$ generates $\Uf(d).$ 

\end{theorem}
\pf With the standard subspace topology of the matrix algebra of complex matrices, $\Uf(d)$ is a connected topological space. Then there exists $\theta>0$ such that the map $\psi:  (\theta_1,\hdots,\theta_{d^2})\mapsto \exp(\theta_1B_1^{(d)}),\hdots, \exp(\theta_{d^2}B^{(d)}_{d^2})$ is a diffeomorphism from $I^{d^2}_\theta$ onto an open neighborhood $U$ of $\Uf$ containing the identity matrix. Since $\Uf$ is connected, then the desired result follows immediately. \cite{kirillov2008}. \hfill{$\square$}

Now, we have the following proposition.
 
 \begin{proposition}\label{prop:exp}
 Let $\boldsymbol{\mathcal{B}^{(d)}}=\{B^{(d)}_j : 1\leq j\leq d^2\}$ denote a basis of Hermitian unitary matrices for $\C^{d\times d}$ as described in Theorem \ref{thm:basis2} or Corollary \ref{cor:basis}. Then $$\exp(\pm i\theta_jB^{(d)}_j)=\cos \theta_j I_d \pm i\sin \theta_jB^{(d)}_j,$$ for any $\theta_j\in \R,$ $1\leq j\leq d^2.$
 \end{proposition}
 
 \pf The proof follows from the fact that $\exp(\pm it\sigma_j)=\cos t\pm i\sin t\sigma_j,$ $j=0,1,2,3$, $t\in\R,$ and $P_{(k,d-1)}$ is a symmetric unitary matrix, as described in Theorem \ref{thm:basis2} and Corollary \ref{cor:basis}. \hfill{$\square$}
 
 Thus it follows from Proposition \ref{prop:exp} that exponentials of basis elements given in Corollary \ref{cor:basis} is either a $2$-level matrix or a diagonal matrix since the basis elements $U^{2^n}_j,$ $1\leq j\leq 2^n$ are either a $2$-level or a diagonal matrix. As mentioned above, it is a well-known result that any unitary matrix can always be written as a product of $2$-level matrices \cite{NielsenChuang2011}.  On the other hand, due to Theorem \ref{thm:span} and Proposition \ref{prop:exp}, it is clear that as a byproduct of the construction of the proposed basis, it provides such a decomposition. 
 
\subsection{Exact parametric representation of certain unitary matrices}
In the Next, section we provide parametric representation of certain unitaries for $n$-qubit systems by employing the basis $\boldsymbol{\mathcal{U}}^{(d)}, d=2^n$ proposed in Corollary \ref{cor:basis}.

\begin{theorem}\label{2levSun} 
    Any $2$-level unitary matrix $U\in \Sf\Uf(2^n)$ can be represented as $$\left(\prod_{j\in \J} \exp\left(i t_jU_{j}^{(2^n)}\right)\right) \exp\left(i t_l U_{l}^{(2^n)}\right) \left(\prod_{j\in \J} \exp\left(i t'_jU_{j}^{(2^n)}\right)\right),$$ where $l= (d-1)^2+d-1,\hdots,(d-1)^2+2(d-1)-1$ for some $d\in\{2,\hdots, 2^{n}\},$ $U_{l}^{(2^n)}, U_{j}^{(2^n)}\in \boldsymbol{\mathcal{U}}^{(2^n)},$ and  $t_j, t'_j\in\mathbb{R}, j \in \J$. 
\end{theorem}

\pf Any $2$-level matrix $U=[U_{\alpha\beta}] \in \Sf\Uf(2^n)$ of order $2^n$ is of the form $$u_{\alpha\beta}=\begin{cases}
 1 \,\, \mbox{if}\,\, \alpha=\beta, \alpha,\beta\notin\{p,q\} \\
 ae^{\iota \theta_a} \,\, \mbox{if}\,\, \alpha=p=\beta \\
 ae^{-\iota \theta_a} \,\, \mbox{if}\,\, \alpha=q=\beta\\
 -be^{\iota \theta_b} \,\, \mbox{if} \,\, \alpha=p, \beta=q \\
 be^{-\iota \theta_b} \,\, \mbox{if} \,\, \alpha=q, \beta=p \\
 0 \,\, \mbox{otherwise}
\end{cases} \,\, \mbox{i.e.} \,\, U=\bmatrix{I_{p-1} & & & & &\\ &  ae^{\iota \theta_a} & &  -be^{\iota \theta_b} & & \\ & & I_{q-p-1} & & & \\ & be^{-\iota \theta_b} & & ae^{-\iota \theta_a} & & \\ & & & & & I_{2^n-q}}$$ for some $1\leq p<q\leq 2^n,$ $a^2+b^2=1, a,b,\theta_a,\theta_b\in\R.$ Now from the Hermitian unitary basis from Corollary \ref{cor:basis} we have $B_q^{(2^n)}=P_{(p,2^{n-1})}\bmatrix{D & \\ & \sigma_2}P_{(p,2^{n}-1)},$ $D=\mbox{diag}\{(-1)^{j-1} : 1\leq j\leq 2^{n}-2\}=\mbox{diag}\{D_{q-1}, D_{q-p-1}, D_{2^n-q}\}$ for which \begin{eqnarray} \exp{(\iota t_qB_q^{(2^n)})} &=& \cos t_qI_{2^n} + i\sin t_qB_q^{(2^n)} \nonumber \\ &=&\bmatrix{\exp(\iota t_qD_{p-1}) & & & & &\\ &  \cos t_q & &  \sin t_q & & \\ & & \exp(\iota t_qD_{q-p-1}) & & & \\ & -\sin t_q & & \cos t_q & & \\ & & & & & \exp(\iota t_qD_{2^n-q})}, \nonumber\end{eqnarray} when $p$ is odd and $q$ is even, or $p$ is even and $q$ is odd.  

Now the matrix $U$ can be obtained from $\exp{(\iota t_qB_q^{(2^n)})}$ by the following transformation. Set $$D_L=\bmatrix{D_1^{(\alpha_a)} & &&& \\ &e^{\iota \alpha_a}&&& \\ && D_2^{(\alpha_a)} && \\ &&& e^{-\iota \alpha_a}& \\ &&&& D_3^{(\alpha_a)} }, \, \,\, D_L=\bmatrix{D_1^{(\alpha_b)} & &&& \\ &e^{-\iota \alpha_b}&&& \\ && D_2^{(\alpha_b)} && \\ &&& e^{\iota \alpha_b}& \\ &&&& D_3^{(\alpha_b)} }$$ as the diagonal unitary matrices of order $2^n$, where $D_1^{(\alpha_a)},D_1^{(\alpha_b)}$ are order $p-1,$ $D_2^{(\alpha_a)}, D_2^{(\alpha_b)}$ are of order $q-p-1,$ $D_3^{(\alpha_a)}, D_3^{(\alpha_b)}$ are of order $2^n-q,$ and $\alpha_a,\alpha_b\in\R$ such that $\alpha_a+\alpha_b=\theta_b,$ $\alpha_a-\alpha_b=\theta_a.$ Further if the diagonal blocks can be chosen such that $D_1^{(\alpha_a)}\exp(\iota t_qD_{p-1})D_1^{(\alpha_b)}=I_{p-1},$ $D_2^{(\alpha_a)} \exp(\iota t_qD_{q-p-1})D_2^{(\alpha_b)}=I_{q-p-1},$ and $D_3^{(\alpha_a)} \exp(\iota t_qD_{2^n-q}) D_3^{(\alpha_b)}=I_{2^n-q}$ then $$D_L\exp{(\iota t_qB_q^{(2^n)})}D_R=U$$ with $a=\cos t_q,$ $b=-\sin t_q.$

Now, since $\mathcal{D}_{IZ}=\{U_j^{(2^n)} : j\in\mathcal{J}\}$ from equation (\ref{eqn:diag}) form a basis for the (real) linear space of diagonal traceless matrices of order $2^n,$ there must exist $c_j$ and $c'_{j}$ such that \small\begin{eqnarray} \sum_{j\in\mathcal{J}} c_jU_j^{(2^n)} &=& \bmatrix{-\frac{1}{2}t_qD_{q-1} &&&& \\ & \alpha_a &&& \\ && -\frac{1}{2}t_qD_{q-p-1} && \\ &&& -\alpha_a & \\ &&&& -\frac{1}{2}t_qD_{2^n-q}} \nonumber \\ \sum_{j\in\mathcal{J}} c'_{j} U_j^{(2^n)} &=& \bmatrix{-\frac{1}{2}t_qD_{q-1} &&&& \\ & -\alpha_b &&& \\ && -\frac{1}{2}t_qD_{q-p-1} && \\ &&& \alpha_b & \\ &&&& -\frac{1}{2}t_qD_{2^n-q}}. \nonumber \end{eqnarray}\normalsize Moreover, \begin{eqnarray} && \exp\left(\sum_{j\in\mathcal{J}} ic_jU_j^{(2^n)}\right)=\prod_{j\in\mathcal{J}} \exp\left(ic_jU_j^{(2^n)}\right) =D_L \,\, \mbox{and} \nonumber \\   && \exp\left(\sum_{j\in\mathcal{J}} ic'_{j}U_j^{(2^n)}\right)=\prod_{j\in\mathcal{J}} \exp\left(ic'_{j}U_j^{(2^n)}\right) =D_R.\nonumber \end{eqnarray} When both $p$ and $q$ are odd or even, the desired result follows similarly. $\hfill{\square}$

\begin{remark}\begin{itemize}
    \item[(a)] It is well known that any matrix $U\in \Sf\Uf(2)$ has a ZYZ decomposition $U=\exp(\iota \alpha\sigma_3)\exp(\iota \beta\sigma_2)\exp(\iota \gamma\sigma_3).$ The Theorem \ref{2levSun} provides a ZYZ like decomposition for matrices in $\Sf\Uf(2^n).$ Indeed, the matrix $\prod_{j\in\mathcal{J}} \exp\left(it_{j}U_j^{(2^n)}\right)$ and $\prod_{j\in\mathcal{J}} \exp\left(it'_{j}U_j^{(2^n)}\right)$ act as the diagonal matrix which represents for the `Z' defined by $\sigma_3,$ and the matrix $\exp\left(i t_l U_{l}^{(2^n)}\right)$ and as `Y' which is defined by $\sigma_2.$
    \item[(b)] Note that the proof is valid for $2$-level special unitary matrix of any order $d.$ We write it for $d=2^n$ just to correspond to a quantum circuit representation of such matrices for $n$-qubit systems.
\end{itemize}

\end{remark}
Then we have the following corollary. 

\begin{corollary}\label{levUn}
     Any $2$-level unitary matrix $U\in \Uf(2^n)$ can be represented using $$e^{\iota \alpha}\left(\prod_{j\in \J} \exp\left(i t_jU_{j}^{(2^n)}\right)\right) \exp\left(i t_l U_{l}^{(2^n)}\right) \left(\prod_{j\in \J} \exp\left(i t'_jU_{j}^{(2^n)}\right)\right),$$ where $l= (d-1)^2+d-1,\hdots,(d-1)^2+2(d-1)-1$ for some $d\in\{2,\hdots, 2^{n}\},$ $U_{l}^{(2^n)}, U_{j}^{(2^n)}\in \boldsymbol{\mathcal{U}}^{(2^n)},$ and  $\alpha, t_j, t'_j\in\mathbb{R}, j \in \J$. 
\end{corollary}

\pf  Suppose  $p$ is odd and $q$ is even, or $p$ is even and $q$ is odd. Then any $2$-level matrix $U=[U_{\alpha\beta}]\in \Uf(d)$ of order $2^n$ is of the form 

\begin{eqnarray*}
    U=\bmatrix{I_{p-1} & & & & &\\ &   e^{\iota (\alpha-\frac{\beta}{2}-\frac{\delta}{2})}\cos\frac{\theta}{2} & &  -e^{\iota (\alpha-\frac{\beta}{2}+\frac{\delta}{2})}\sin\frac{\theta}{2} & & \\ & & I_{q-p-1} & & & \\ & e^{\iota (\alpha+\frac{\beta}{2}-\frac{\delta}{2})}\sin\frac{\theta}{2} & & e^{\iota (\alpha+\frac{\beta}{2}+\frac{\delta}{2})}\cos\frac{\theta}{2} & & \\ & & & & & I_{2^n-q}}
\end{eqnarray*} for some $1\leq p<q\leq 2^n,$ $\alpha, \beta, \delta, \theta \in\R.$ Then $U=e^{\iota \alpha}U'$, where \begin{eqnarray*}
    U' = \bmatrix{e^{-\iota \alpha}I_{p-1} & & & & &\\ &   e^{\iota (-\frac{\beta}{2}-\frac{\delta}{2})}\cos\frac{\theta}{2} & &  -e^{\iota (-\frac{\beta}{2}+\frac{\delta}{2})}\sin\frac{\theta}{2} & & \\ & & e^{-\iota \alpha}I_{q-p-1} & & & \\ & e^{\iota (\frac{\beta}{2}-\frac{\delta}{2})}\sin\frac{\theta}{2} & & e^{\iota (\frac{\beta}{2}+\frac{\delta}{2})}\cos\frac{\theta}{2} & & \\ & & & & & e^{-\iota \alpha}I_{2^n-q}}.
\end{eqnarray*} Now, there exists a basis element $U_{l}^{(2^n)}$ as described in Corollary \ref{cor:basis} such that \small\begin{eqnarray*}
   \hspace{-0.65cm} \exp\left( -i\frac{\theta}{2}U_l^{(2^n)}\right) = \bmatrix{\exp\left(-i\frac{\theta}{2}D_{p-1}\right) & & & & &\\ &   \cos\frac{\theta}{2} & &  -\sin\frac{\theta}{2} & & \\ & & \exp\left(-i\frac{\theta}{2}D_{q-p-1}\right) & & & \\ & \sin\frac{\theta}{2} & & \cos\frac{\theta}{2} & & \\ & & & & & \exp\left(-i\frac{\theta}{2}D_{2^n-q}\right)}
\end{eqnarray*}\normalsize, where $l= (d-1)^2+d-1,\hdots,(d-1)^2+2(d-1)-1$ for some $d\in\{2,\hdots, 2^{n}\}.$ Define the diagonal matrices 
\small\begin{eqnarray} 
\sum_{j\in\mathcal{J}} c_jU_j^{(2^n)} = \bmatrix{-\frac{\alpha}{2}I_{p-1} + \frac{\theta}{4}D_{p-1} &&&&& \\ &\frac{-\beta}{2} && && \\ && -\frac{\alpha}{2}I_{q-p-1} + \frac{\theta}{4}D_{q-p-1} &&& \\ &&& \frac{\beta}{2} && \\ &&&&& -\frac{\alpha}{2}I_{2^n-q} + \frac{\theta}{4}D_{2^n -q}  } \nonumber\end{eqnarray} \begin{eqnarray} \sum_{j\in\mathcal{J}} c'_jU_j^{(2^n)} = \bmatrix{-\frac{\alpha}{2}I_{p-1} + \frac{\theta}{4}D_{p-1} &&&&& \\ &\frac{-\delta}{2} && && \\ && -\frac{\alpha}{2}I_{q-p-1} + \frac{\theta}{4}D_{q-p-1} &&& \\ &&& \frac{\delta}{2} && \\ &&&&& -\frac{\alpha}{2}I_{2^n-q} + \frac{\theta}{4}D_{2^n -q}  } \nonumber 
\end{eqnarray}\normalsize where $U^{(2^n)}_j\in D_{IZ}, j\in\J.$ Then it can be easily checked that $$U' = \exp\left(\sum_{j\in\mathcal{J}} c_jU_j^{(2^n)}\right) \exp\left( -i\frac{\theta}{2}U_l^{(2^n)}\right) \exp\left(\sum_{j\in\mathcal{J}} c'_jU_j^{(2^n)}\right).$$ When both $p$ and $q$ are odd or even, the desired result follows similarly. \hfill{$\square$} 

Now, we consider $2$-sparse unitary matrices that are block diagonal matrices,  each block is a special unitary matrix. Let $R_a(\theta)$ denote a rotation gate around an axis $a$ with an angle $\theta\in \R.$ In particular, when the rotation matrices around the axes $X, Y, Z$ are defined as $$R_z(\theta)=\bmatrix{e^{\iota  \theta}&0\\0&e^{-\iota  \theta}},R_y(\theta)=\bmatrix{\cos{\theta} &\sin{\theta}\\-\sin{\theta}&\cos{\theta}},R_x(\theta)=\bmatrix{\cos{\theta} &\iota \sin{\theta}\\ \iota \sin{\theta}&\cos{\theta}}.$$

\begin{definition} \label{def:mcgate}\cite{krol2022}
For $n$-qubit systems, a multi-controlled rotation gate around an axis $a$ is defined as 
\begin{eqnarray}
    {\Qcircuit @C=1em @R=.7em {
 &\lstick{1}&\qw& \gate{\circ} &\gate{\circ} &\qw & \hdots\vdots &\gate{\circ} &\gate{\circ}&\qw\\
 &\lstick{2}&\qw&\gate{\circ}\qwx[-1] &\gate{\circ}\qwx[-1] &\qw & \hdots\vdots  &\gate{\circ}\qwx[-1] &\gate{\circ}\qwx[-1] &\qw\\
&\lstick{\vdots}&\qw&\gate{\circ}\qwx[-1] &\gate{\circ}\qwx[-1] &\qw & \hdots\vdots  &\gate{\circ}\qwx[-1] &\gate{\circ}\qwx[-1] &\qw\\
&\lstick{n-1}&\qw& \gate{\circ}\qwx[-1] &\gate{\circ}\qwx[-1] &\qw &\hdots \vdots  &\gate{\circ}\qwx[-1] &\gate{\circ}\qwx[-1] &\qw\\
&\lstick{n}&\qw& \gate{R_a(\theta_1)} \qwx[-1]&\gate{R_a(\theta_2)} \qwx[-1] &\qw& \hdots &\gate{R_a(\theta_{2^{n-2}})} \qwx[-1]&\gate{R_a(\theta_{2^{n-1}})}\qwx[-1] &\qw\\}}
\end{eqnarray} 

where \Qcircuit @C=1em @R=.7em {&\gate{\circ}&\qw\\} $\in\{$ \Qcircuit @C=1em @R=.7em {&\ctrl{0}&\qw},\hspace{.5cm}\Qcircuit @C=1em @R=.7em {&\ctrlo{0}&\qw}$\},$ and $\theta_j, 1\leq j\leq 2^{n-1}\in \R.$ Then the unitary matrix corresponding to the above circuit is given by $$F_n(R_a(\theta_1,\theta_2,\hdots,\theta_{2^{n-1}})):=F_n(R_a) =\left[ \begin{array}{c|c|c|c} 
      R_a(\theta_1) &  0 &0 & 0\\ 
      \hline 
       0 &  0&\ddots &0\\
        \hline
       0 & 0& 0& R_a(\theta_{2^{n-1}})
    \end{array} 
    \right] $$
\end{definition} 

In short, we use the circuit in Definition \ref{def:mcgate} as

\begin{eqnarray}
    {\Qcircuit @C=1em @R=.7em {
    &\lstick{1}&\qw& \gate{} &\qw\\
    &\lstick{2}&\qw&\gate{ }\qwx[-1]&\qw\\
    &\lstick{\vdots}&\qw&\gate{ }\qwx[-1]&\qw\\
    &\lstick{n-1}&\qw&\gate{ }\qwx[-1]&\qw\\
    &\lstick{n}&\qw&\gate{F_n(R_a)}\qwx[-1]&\qw\\}}
\end{eqnarray}
    
For example, setting $n=4,$ the circuit corresponding to $F_4(R_a)$ is given by 

\begin{eqnarray}
    {\Qcircuit @C=1em @R=.7em {
 &\lstick{1}&\qw& \ctrl{1} &\ctrl{1} & \ctrl{1}& \ctrlo{1} &\ctrlo{1} &\ctrl{1} &\ctrlo{1}&\ctrlo{1}&\qw\\
 &\lstick{2}&\qw& \ctrl{1} &\ctrl{1} &\ctrlo{1} & \ctrl{1}  &\ctrlo{1} &\ctrlo{1} &\ctrl{1}&\ctrlo{1}&\qw\\
&\lstick{3}&\qw& \ctrl{1} &\ctrlo{1} &\ctrl{1} &\ctrl{1}  &\ctrl{1} &\ctrlo{1} &\ctrlo{1}&\ctrlo{1}&\qw\\
&\lstick{4}&\qw& \gate{R_a(\theta_1)} &\gate{R_a(\theta_2)}  &\gate{R_a(\theta_3)}&\gate{R_a(\theta_4)}  &\gate{R_a(\theta_5)} &\gate{R_a(\theta_6)} &\gate{R_a(\theta_7)}&\gate{R_a(\theta_8)}&\qw\\}}\hspace{0.35cm}
\end{eqnarray}

Further, it can be shown that the multi-controlled rotation gates can be decomposed and implemented through $\cnot$ and single qubit gates \cite{krol2022}. Indeed, the multi-controlled rotation gate on an $n$ qubit system given by \begin{eqnarray}\label{multi1}
{\Qcircuit @C=1em @R=.7em {
    &\lstick{1}&\qw& \gate{} &\qw\\
    &\lstick{2}&\qw&\gate{ }\qwx[-1]&\qw\\
    &\lstick{\vdots}&\qw&\gate{ }\qwx[-1]&\qw\\
    &\lstick{n-1}&\qw&\gate{ }\qwx[-1]&\qw\\
    &\lstick{n}&\qw&\gate{F_n(R_a(\psi_1,\hdots,\psi_{2^{n-1}}))}\qwx[-1]&\qw\\}}\end{eqnarray} can be written as
    
    \begin{eqnarray}\label{multidec}
{\Qcircuit @C=1em @R=.7em {
    &\lstick{1}&\qw& \qw &\ctrl{4} &\qw& \qw &\ctrl{4} &\qw \\
    &\lstick{2}&\qw&\gate{ }&\qw&\qw&\gate{ }&\qw&\qw\\
    &\lstick{\vdots}&\qw&\gate{ }\qwx[-1]&\qw&\qw&\gate{ }\qwx[-1]&\qw&\qw\\
    &\lstick{n-1}&\qw&\gate{ }\qwx[-1]&\qw&\qw&\gate{ }\qwx[-1]&\qw&\qw\\
    &\lstick{n}&\qw&\gate{F_{n-1}(R_a(\theta_1,\hdots,\theta_{2^{n-2}}))}\qwx[-1]&\targ&\qw&\gate{F_{n-1}(R_a(\phi_1,\hdots,\phi_{2^{n-2}}))}\qwx[-1]&\targ&\qw\\}}\end{eqnarray}  where $$\psi_k=\begin{cases}
        \theta_j+\phi_j  \mbox{ where } 1\leq j\leq 2^{n-2}, k=j\\
        \theta_j-\phi_j \mbox{ where } 1\leq j\leq 2^{n-2}, k=j+2^{n-2}.
    \end{cases}$$

\begin{lemma}\label{mzyz}
 The quantum circuits in equation (\ref{multi1}) and equation (\ref{multidec}) are equivalent.  
\end{lemma}

\pf  The proof is computational and easy to verify. \hfill{$\square$} 

Now with the help of the multi-controlled rotation gates, we consider writing $2$-sparse block diagonal matrix of the form \begin{eqnarray}\label{blkdg}
\left[ 
\begin{array}{cccc} 
      U_1(\alpha_1,\beta_1,\gamma_1) &   & & \\ 
       &  U_2(\alpha_2,\beta_2,\gamma_2) & & \\
       &  & \ddots &   \\
        & & &U_{2^{k-1}}(\alpha_{2^{n-1}},\beta_{2^{n-1}},\gamma_{2^{n-1}})
    \end{array} 
    \right] \end{eqnarray} in terms of the proposed basis elements, where $U_{j}(\Theta_j) \in\Sf\Uf(2),\Theta_j:=(\alpha_j,\beta_j,\gamma_j),$ $1\leq j\leq 2^{n-1}$ is a $2\times 2$ special unitary matrix such that \begin{equation}\label{def:uj}U_{j}(\Theta_j)=\bmatrix{e^{\iota  (\alpha_j+\beta_j)}\cos{\gamma_j}&e^{\iota  (\alpha_j-\beta_j)}\sin{\gamma_j}\\-e^{-\iota  (\alpha_j-\beta_j)}\sin{\gamma_j}&e^{-\iota  (\alpha_j+\beta_j)}\cos{\gamma_j}}.\end{equation}

Since any $2\times 2$ special unitary matrix has a $ZYZ$ decomposition, the matrices in equation (\ref{blkdg}) have circuit from using the multi-controlled rotation gates as 
\small\begin{eqnarray}\label{mzyz2qb}
   \hspace{-0.7cm}{\Qcircuit @C=0.6em @R=.7em {
    &\lstick{1}&\qw& \gate{} &\qw& \gate{} &\qw& \gate{} &\qw\\
    &\lstick{\vdots}&\qw& \gate{}\qwx[-1]&\qw& \gate{}\qwx[-1]&\qw& \gate{}\qwx[-1]&\qw \\
    &\lstick{n-1}&\qw&\gate{ }\qwx[-1]&\qw&\gate{ }\qwx[-1]&\qw&\gate{ }\qwx[-1]&\qw\\
    &\lstick{n}&\qw&\gate{F_n(R_z(\alpha_1,\hdots,\alpha_{2^{n-1}}))}\qwx[-1]&\qw&\gate{F_n(R_y(\gamma_1,\hdots,\gamma_{2^{n-1}}))}\qwx[-1]&\qw&\gate{F_n(R_z(\beta_1,\hdots,\beta_{2^{n-1}}))}\qwx[-1]&\qw\\}}\end{eqnarray}\normalsize which we denote as $M_nZYZ$.    
    
Now, we introduce a handy notation for extracting diagonal entry of a diagonal matrix. Define \begin{equation}\label{eta} \eta^{(j)}_M=M_{jj}, 1\leq j\leq k\end{equation} for any diagonal matrix $M\in\{1,-1\}^{k\times k}.$
    
\begin{lemma}\label{Hadamarddiag}
The set of vectors $\left\{\bmatrix{\eta^{(1)}_{\chi^{-1}_{n}(k)\otimes \sigma_3} & \eta^{(3)}_{\chi^{-1}_{n}(k)\otimes \sigma_3} & \hdots & \eta^{(2^{n+1}-1)}_{\chi^{-1}_{n}(k)\otimes \sigma_3}}^T \,|\, 0\leq k\leq 2^{n}-1\right\}$ is equal to the set of column vectors of the matrix $2^{n/2}H_{n}$  where $\eta$ is defined in  equation (\ref{eta}) and $\chi$ is defined in equation (\ref{definition2}).
\end{lemma}

\pf The proof follows from the fact that $$\bmatrix{\eta^{(1)}_{\chi^{-1}_{n}(k)\otimes \sigma_3}& \eta^{(3)}_{\chi^{-1}_{n}(k)\otimes \sigma_3}& \hdots & \eta^{(2^{n+1}-1)}_{\chi^{-1}_{n}(k)\otimes \sigma_3}}^T =\diag(\chi^{-1}_{n}(k)),$$ where $0\leq k\leq 2^{n}-1.$ \hfill{$\square$}

\begin{theorem}\label{2mult}
The unitary matrix corresponding to an $M_nZYZ$ given by equation (\ref{blkdg}) can be written as  $$\left( \prod_{j\in\mathcal{J}} \exp(\iota t_j \chi_{n-1}^{-1}(j)) \otimes \sigma_3) \right)\left(\prod_{j=1}^{2^{n-1}}\exp{(\iota \theta_{4j^2-2j}U^{(2^n)}_{4j^2-2j})}\right) \left(\prod_{j\in\mathcal{J}} \exp(\iota t'_j \chi_{n-1}^{-1}(j)) \otimes \sigma_3)\right)$$
 where $\theta_{4j^2-2j}=\gamma_{j}\in \R, 1\leq j \leq  2^{n-1}, t_j,t'_j\in \mathbb{R}.$ 
\end{theorem}

\pf We know that $R_y(\theta)=\exp(\iota \theta\sigma_2)=\cos\theta I_2+i\sin\theta\sigma_2=\bmatrix{\cos\theta & \sin\theta \\ - \sin\theta & \cos \theta}.$ From the proposed traceless basis elements, we have $$U_j^{(2^n)}\in \left\{\bmatrix{D_1&0&0 \\ 0 & \sigma_2&0\\ 0&0&D_2}, \bmatrix{\sigma_2&0\\0&D}, \bmatrix{D&0\\0&\sigma_2}\right\}$$ when $j\in \mathcal{J}_{\sigma_2}=\{l^2-l : 2\leq l\leq 2^n\}$ and $D_1,D_2,D$ are diagonal matrices with entries from $\{1,-1\}.$ In particular, the $i$-th diagonal entry of $D,D_1,$ and $D_2$ is $1$ if $i$ is even and $-1$ otherwise. Further, choosing $j=4l^2-2l,$ $1\leq l\leq 2^{n-1}$ it is evident that the block diagonal matrices $U^{(2^n)}_j$ will have non-overlapping positions of $\sigma_2$ in the diagonal. 

Thus we obtain $$\prod_{l=1}^{2^{n-1}} \exp(\iota \theta_{4l^2-2l}U^{(2^n)}_{4l^2-2l}) =\bmatrix{V_2&&& \\ &V_{4}&& \\ &&\ddots& \\ &&&V_{2^n}},$$ where 
\begin{eqnarray}
V_{2l} &=& \left(\prod_{j<l} \bmatrix{e^{\iota \theta_{4j^2-2j}}&0\\0& e^{-\iota \theta_{4j^2-2j}}}\right) \bmatrix{\cos{\theta_{4l^2-2l}} & \sin{\theta_{4l^2-2l}} \\ \sin{\theta_{4l^2-2l}} & \cos{\theta_{4l^2-2l}}} \left(\prod_{j>l} \bmatrix{e^{\iota \theta_{4j^2-2j}}&0\\0& e^{-\iota \theta_{4j^2-2j}}}\right) \nonumber \\
&=& \bmatrix{e^{\iota \sum_{j\neq l}\theta_{4j^2-2j}} \cos \theta_{4l^2-2l} & e^{\iota \left(\sum_{j<l} \theta_{4j^2-2j} - \sum_{j>l} \theta_{4j^2-2j}\right)}\sin \theta_{4l^2-2l} \\  -e^{-\iota \left(\sum_{j<l} \theta_{4j^2-2j} - \sum_{j>l} \theta_{4j^2-2j}\right)}\sin \theta_{4l^2-2l} & e^{-\iota \sum_{j\neq l}\theta_{4j^2-2j}} \cos \theta_{4l^2-2l}}, \nonumber
\end{eqnarray} $1\leq l\leq 2^{n-1}.$

Now setting $\theta_{4l^2-2l}=\gamma_l,$ $1\leq l\leq 2^{n-1}$ for the representation of a $M_nZYZ$ unitary matrix given by equation (\ref{blkdg}) with each diagonal block given by equation (\ref{def:uj}) through the proposed basis elements, the Next, is to find parameters that can provide the values of the remaining parameters $\alpha_l,\beta_l.$ Setting $x_l=\sum_{j\neq l}\theta_{4j^2-2j}$ and $y_l=\sum_{j<l} \theta_{4j^2-2j} - \sum_{j>l} \theta_{4j^2-2j},$ we obtain $$V_{2l}=\bmatrix{e^{\iota x_l}\cos\gamma_l & e^{\iota y_l}\sin\gamma_l \\ -e^{-\iota y_l}\sin\gamma_l & e^{-\iota x_l}\cos\gamma_l}.$$ Now let $A_{2l}=\bmatrix{e^{\iota a_l}&0\\0&e^{-\iota a_l}}$ and $B_{2l}=\bmatrix{e^{\iota b_l}&0\\ 0&e^{-\iota b_l}},$ $a_l,b_l\in \R$ such that $$U_l(\alpha_l,\beta_l,\gamma_l)=A_{2l}V_{2l}B_{2l}=\bmatrix{e^{\iota  (\alpha_l+\beta_l)}\cos{\gamma_l}&e^{\iota  (\alpha_l-\beta_l)}\sin{\gamma_l}\\-e^{-\iota  (\alpha_l-\beta_l)}\sin{\gamma_l}&e^{-\iota  (\alpha_l+\beta_l)}\cos{\gamma_l}},$$ and $U=AVB$, where $$A=\bmatrix{A_2&& \\ &\ddots& \\ &&A_{2^{n}}},  V= \bmatrix{V_2&& \\ &\ddots& \\ &&V_{2^n}}, B=\bmatrix{B_2&& \\ &\ddots& \\ &&B_{2^{n}}}.$$ 

Then setting $$A_{2l}=\bmatrix{\exp{(\iota  \sum_{j=0}^{2^{n-1}-1}\eta^{(2l-1)}_{\chi^{-1}_{n-1}(j)\otimes \sigma_3}t_j)} & 0 \\ 0 & \exp{-(i \sum_{j=0}^{2^{n-1}-1}\eta^{(2l-1)}_{{\chi^{-1}_{n-1}(j)}\otimes \sigma_3}t_j)}}$$ where $$B_{2l}=\bmatrix{\exp{(\iota  \sum_{j=0}^{2^{n-1}-1}\eta^{(2l-1)}_{\chi^{-1}_{n-1}(j)\otimes \sigma_3}t'_j)} & 0 \\ 0 & \exp{-(i \sum_{j=0}^{2^{n-1}-1}\eta^{(2l-1)}_{{\chi^{-1}_{n-1}(j)}\otimes \sigma_3}t'_j)}}$$ This is obtained by applying all $l$-th blocks of $exp{(i\otimes_{k=1}^{n-1}A_{k}\otimes \sigma_3)}$ where $A_{k}\in \{I, \sigma_3\}$ so that $A_{2l}, B_{2l}$ are of the desired forms. Then equating the entries of both sides of the equation $U=AVA$ provides a system of linear equation of the form $Hx=b,$ where 
$b_{2l-1}=-(x_l-\alpha_l-\beta_l)$ and $b_{2l}=-(y_l-\beta_l+\alpha_l)$, and $\eta$ defined in equation (\ref{eta}), $1\leq l\leq 2^{n-1}$ and $x=\bmatrix{t_0&t_1&\hdots & t_{2^{n-1}-1}&t'_0&\hdots&t'_{2^{n-1}-1}}^T$ and $$H=\bmatrix{\eta^{(1)}_{\chi^{-1}_{n-1}(0)\otimes \sigma_3} & \hdots & \eta^{(1)}_{\chi^{-1}_{n-1}(2^{n-1}-1)\otimes \sigma_3} & \eta^{(1)}_{\chi^{-1}_{n-1}(0)\otimes \sigma_3} & \hdots & \eta^{(1)}_{\chi^{-1}_{n-1}(2^{n-1}-1)\otimes \sigma_3}\\\eta^{(1)}_{\chi^{-1}_{n-1}(0)\otimes \sigma_3} & \hdots & \eta^{(1)}_{\chi^{-1}_{n-1}(2^{n-1}-1)\otimes \sigma_3} & -\eta^{(1)}_{\chi^{-1}_{n-1}(0)\otimes \sigma_3} & \hdots & -\eta^{(1)}_{\chi^{-1}_{n-1}(2^{n-1}-1)\otimes \sigma_3}\\
\vdots &\vdots&\vdots&\vdots&\vdots&\vdots\\
\eta^{(2l-1)}_{\chi^{-1}_{n-1}(0)\otimes \sigma_3} & \hdots & \eta^{(2l-1)}_{\chi^{-1}_{n-1}(2^{n-1}-1)\otimes \sigma_3} & \eta^{(2l-1)}_{\chi^{-1}_{n-1}(0)\otimes \sigma_3} & \hdots & \eta^{(2l-1)}_{\chi^{-1}_{n-1}(2^{n-1}-1)\otimes \sigma_3}\\\eta^{(2l-1)}_{\chi^{-1}_{n-1}(0)\otimes \sigma_3} & \hdots & \eta^{(2l-1)}_{\chi^{-1}_{n-1}(2^{n-1}-1)\otimes \sigma_3} & -\eta^{(2l-1)}_{\chi^{-1}_{n-1}(0)\otimes \sigma_3} & \hdots & -\eta^{(2l-1)}_{\chi^{-1}_{n-1}(2^{n-1}-1)\otimes \sigma_3}\\ \vdots &\vdots&\vdots&\vdots&\vdots&\vdots\\ \eta^{(2^n-1)}_{\chi^{-1}_{n-1}(0)\otimes \sigma_3} & \hdots & \eta^{(2^n-1)}_{\chi^{-1}_{n-1}(2^{n-1}-1)\otimes \sigma_3} & \eta^{(2^n-1)}_{\chi^{-1}_{n-1}(0)\otimes \sigma_3} & \hdots & \eta^{(2^n-1)}_{\chi^{-1}_{n-1}(2^{n-1}-1)\otimes \sigma_3}\\\eta^{(2^n-1)}_{\chi^{-1}_{n-1}(0)\otimes \sigma_3} & \hdots & \eta^{(2^n-1)}_{\chi^{-1}_{n-1}(2^{n-1}-1)\otimes \sigma_3} & -\eta^{(2^n-1)}_{\chi^{-1}_{n-1}(0)\otimes \sigma_3} & \hdots & -\eta^{(2^n-1)}_{\chi^{-1}_{n-1}(2^{n-1}-1)\otimes \sigma_3}}_{2^n\times 2^n}$$

Now $H$ is nonsingular since $H$ can also be written in the following form.
 $$H=P\bmatrix{\eta^{(1)}_{\chi^{-1}_{n-1}(0)\otimes \sigma_3} & \hdots & \eta^{(1)}_{\chi^{-1}_{n-1}(2^{n-1}-1)\otimes \sigma_3} & \eta^{(1)}_{\chi^{-1}_{n-1}(0)\otimes \sigma_3} & \hdots & \eta^{(1)}_{\chi^{-1}_{n-1}(2^{n-1}-1)\otimes \sigma_3}\\
\eta^{(3)}_{\chi^{-1}_{n-1}(0)\otimes \sigma_3} & \hdots & \eta^{(3)}_{\chi^{-1}_{n-1}(2^{n-1}-1)\otimes \sigma_3} & \eta^{(3)}_{\chi^{-1}_{n-1}(0)\otimes \sigma_3} & \hdots & \eta^{(3)}_{\chi^{-1}_{n-1}(2^{n-1}-1)\otimes \sigma_3}\\
\vdots &\vdots&\vdots&\vdots&\vdots&\vdots\\
\eta^{(2l-1)}_{\chi^{-1}_{n-1}(0)\otimes \sigma_3} & \hdots & \eta^{(2l-1)}_{\chi^{-1}_{n-1}(2^{n-1}-1)\otimes \sigma_3} & \eta^{(2l-1)}_{\chi^{-1}_{n-1}(0)\otimes \sigma_3} & \hdots & \eta^{(2l-1)}_{\chi^{-1}_{n-1}(2^{n-1}-1)\otimes \sigma_3}\\\vdots &\vdots&\vdots&\vdots&\vdots&\vdots\\
\eta^{(2^n-1)}_{\chi^{-1}_{n-1}(0)\otimes \sigma_3} & \hdots & \eta^{(2^n-1)}_{\chi^{-1}_{n-1}(2^{n-1}-1)\otimes \sigma_3} & \eta^{(2^n-1)}_{\chi^{-1}_{n-1}(0)\otimes \sigma_3} & \hdots & \eta^{(2^n-1)}_{\chi^{-1}_{n-1}(2^{n-1}-1)\otimes \sigma_3}\\\eta^{(1)}_{\chi^{-1}_{n-1}(0)\otimes \sigma_3} & \hdots & \eta^{(1)}_{\chi^{-1}_{n-1}(2^{n-1}-1)\otimes \sigma_3} & -\eta^{(1)}_{\chi^{-1}_{n-1}(0)\otimes \sigma_3} & \hdots & -\eta^{(1)}_{\chi^{-1}_{n-1}(2^{n-1}-1)\otimes \sigma_3}\\\eta^{(3)}_{\chi^{-1}_{n-1}(0)\otimes \sigma_3} & \hdots & \eta^{(3)}_{\chi^{-1}_{n-1}(2^{n-1}-1)\otimes \sigma_3} & -\eta^{(3)}_{\chi^{-1}_{n-1}(0)\otimes \sigma_3} & \hdots & -\eta^{(3)}_{\chi^{-1}_{n-1}(2^{n-1}-1)\otimes \sigma_3}\\\vdots &\vdots&\vdots&\vdots&\vdots&\vdots\\\eta^{(2^n-1)}_{\chi^{-1}_{n-1}(0)\otimes \sigma_3} & \hdots & \eta^{(2^n-1)}_{\chi^{-1}_{n-1}(2^{n-1}-1)\otimes \sigma_3} & -\eta^{(2^n-1)}_{\chi^{-1}_{n-1}(0)\otimes \sigma_3} & \hdots & -\eta^{(2^n-1)}_{\chi^{-1}_{n-1}(2^{n-1}-1)\otimes \sigma_3}}$$

Note that $$A=P\left[ 
    \begin{array}{cc} 
      2^{(n-1)/2}H_{n-1} & 2^{(n-1)/2}H_{n-1} \\ 
      2^{(n-1)/2}H_{n-1} & -2^{(n-1)/2}H_{n-1} 
    \end{array} 
    \right]^T=2^{n/2}H_n$$ follows from Lemma \ref{Hadamarddiag} and $(\diag(\otimes_{k=1}^{n-1}M_k \otimes \sigma_3)=2^{(n-1)/2}H_{2^{n-1}}$, 
 where $P$ is a permutation matrix. This completes the proof. \hfill{$\square$}

\begin{remark}
\begin{enumerate}
    \item[(a)] As Givens rotation matrices, which are $2$-level unitary matrices, can be used to construct any unitary matrix through matrix multiplication \cite{golub2013}, it follows that any unitary matrix can be expressed as a product of exponentials of the RBB elements. It is worth noting that computing the exponentials of RBB elements can be performed in $O(1)$ time. Therefore, if the Givens rotation corresponding to a given unitary matrix is known, expressing that matrix in terms of the proposed basis elements becomes a straightforward task.
    
    \item[(b)] It can be noted that using Pauli  string basis to form matrices of dimensions $2^n$ is not ideal when compared to the proposed basis, due to two main reasons. Firstly, computing the exponentials of Pauli string matrices is a difficult task because the fundamental Pauli matrices do not commute. Secondly, in the worst-case scenario, generating a Pauli string for an $n$-qubit system would require $O(n2^{2n})$ operations using generic Kronecker product algorithms \cite{VidalRomero2023}. On the contrary, the construction of the proposed basis matrices do not require any operation as the construction is completely prescribed by the pattern of the non-zero entries of the basis elements. 
\end{enumerate}    
\end{remark}

  \subsection{Algorithmic approximation of special unitary matrices} 
  We can utilize Theorem \ref{thm:span} to find a value $\theta>0$ such that the set $\left\{\prod_{j=1}^{d^2} \exp\left({i\theta_j B^{(d)}_j}\right)\right\},$ where $(\theta_1,\hdots,\theta_{d^2})\in I^{d^2}_\theta$, generates the unitary group $\Uf(d)$. As a result, any unitary matrix $U$ up to permutation of indices of the basis elements can be represented as 
\begin{equation}\label{eqn:uapprox} U=\underbrace{\left(\prod_{j=1}^{d^2} \exp\left({i\theta_j B^{(d)}_j}\right)\right)\hdots \left(\prod_{j=1}^{d^2} \exp\left({i\theta_j B^{(d)}_j}\right)\right)}_{L\, \mbox{times}} := \prod_{l=1}^L \left(\prod_{j=1}^{d^2} \exp\left({i\theta_{lj}B_{j}^{(d)}}\right)\right)\end{equation} 
for some positive integer $L$, which we call the number of layers or iterations for approximating $U$. However, determining the appropriate value of $L$ for a given $U\in \Uf(d)$ is challenging in practice. Further, the ordering of the basis elements in Corollary \ref{cor:basis} given by $\boldsymbol{\mathcal{U}}^{(2^n)}$ is fixed to ensure that the recursive construction method works.

Indeed, we propose to find a parametric representation of a given unitary through solving the following optimization problem  $$\min_{\theta_{lj}\in I^{Kd^2}_\theta}\left\|U-\prod_{l=1}^L \left(\prod_{j=1}^{d^2} \exp\left({i\theta_{lj}B_{j}^{(d)}}\right)\right)\right\|$$ for some $\theta>0,$ where $\|\cdot\|$ is a desired matrix norm. In this section, we explore approximating unitaries that are dominant in quantum computing and perform an accuracy analysis of this approximation corresponding to Frobenius norm. The optimization problem is solved using Nelder-Mead method in the numerical simulation. We also explore whether altering the ordering impact the accuracy of approximating a unitary $U$ through parameter optimization.

Employing the basis described in Theorem \ref{thm:basis2}, the Algorithm \ref{algo1} describes how to approximate a unitary $U\in\Sf\Uf(d).$

\begin{algorithm}[H]
\caption{Approximating $d\times d$ special unitary matrix }\label{algo1}
\textbf{Provided:} Consider the basis $\boldsymbol{\mathcal{B}}^{(d)}$ of $\Sf\Uf(d)$ from Theorem \ref{thm:basis2} and set $$\psi(a_1,a_2,\hdots,a_{d^2-1})=\prod_{j=1}^{d^2-1}\exp(\iota  a_j B_{j}^{(d)})$$
\small
\textbf{Input:} $U_1\in \Sf\Uf(d), d^2-1$ real parameters $(a_1^{(1)},\hdots, a_{d^2-1}^{(1)}),$ $\epsilon >0$ 

\textbf{Output:} A unitary matrix $A$ such that $\|U-A\|_F\leq \epsilon$ where $$A=\prod_{l=1}^L \left(\prod_{j=1}^{d^2-1}\exp(\iota  a_j^{(l)}B_{j}^{(d)})\right)$$

\small\begin{algorithmic}

\Procedure{}{Unitary Matrix $U$}      
\State $A\rightarrow I$
\For{$t=1;t++$}
\State Use optimization methods like Nelder-Mead/Powell's or Gradient descent to determine $U_t=\psi(a_1^{(t)},\hdots,a_{n^2-1}^{(t)})$ such that $\left\|U-U_t\right\|_F$ is minimum. Set $\|U-U_t\|=\epsilon_t. $ 
\If{$\epsilon_t\leq \epsilon$}
\State \textbf{Break}
\State $A \rightarrow A\psi(a_1^{(t)},a_2^{(t)},\hdots,a_{n^2-1}^{(t)})$
\Else
\State $U_{t+1}\rightarrow U_{t}A^*$\\
\EndIf
\State \textbf{End}
\EndFor
\State \textbf{End}
\State \textbf{End Procedure}
\EndProcedure
\end{algorithmic}
\end{algorithm}\normalsize

One drawback of Algorithm \ref{algo1} lies in the fact that the optimization may end at a local minima since the objective function is not convex. Further, the initial condition i.e. the choice of of the parameters involved can have adverse effect on the efficiency of the algorithm. A procedure to decide on the choice of the parameters can possibly be overcome by generating several values of parameters in the initial stage of the algorithm and then apply the optimization method. For $n$-qubit system the algorithm faces another significant problem for implementation of the unitary matrices through elementary gates, since the ultimate goal is to implement any unitary through strings of elementary gates. Indeed, for unitary matrices of order $d=2^n,$ the problem with Algorithm \ref{algo1}, lies in the fact that in order to construct a quantum circuit for the proposed ordering of the basis elements while approximating any unitary from $\Sf\Uf(2^n)$, the number of $\cnot$ gates required for a single iteration becomes $O(2^{3n})$ as follows from equation (\ref{eqn:uapprox}) setting $L=1$. This is due to the fact all non-diagonal RBB matrices generate $2$-level unitary matrices and  a single $2$-level unitary matrix requires at least $2^{n-1}$ $\cnot$ gates from this ordering of the basis elements and there are $2^{2n}-2^n$ non-diagonal basis matrices. 


Thus the question is: how to choose a suitable ordering of the basis elements? One motivation for a suitable choice is to reduce the number of $\cnot$ gates in  a quantum circuit implementation of a given unitary matrix using equation (\ref{eqn:uapprox}). First we introduce two functions through which we like to call the proposed basis elements of particular index. For a given integer $d\geq 2,$ we define the functions: 
\begin{eqnarray}\label{definition}
\hspace{-0.65cm}\begin{cases}
f:\mathbb{N}\times \mathbb{Z}\rightarrow \mathbb{Z} \mbox{ such that }
f(n,k):=f_n(k)=(n-1)^2+(n-1)+(k \mbox{ mod } (n-1))\\
h:\mathbb{N}\times \mathbb{Z}\rightarrow \mathbb{Z} \mbox{ such that } h(n,k):=h_n(k)=(n-1)^2+(k \mbox{ mod } (n-1)).\\
\end{cases}\end{eqnarray}

Now in the next section, we propose a new ordering for the SRBB that can give an optimal number of $\cnot$ gates in the corresponding quantum circuit implementation using equation (\ref{eqn:uapprox}), setting $L=1.$

\section{Approximation of $n$-qubit unitaries}\label{Sec:4}
First observe that a SRBB element $U^{(2^n)}_j\in \boldsymbol{\mathcal{U}}^{(2^n)},$ $\exp(\iota  \theta U^{(2^n)}_{j}), j=h_q(p), p<q,q\in \{2,\hdots,2^n\}$, can be written as $$\left(\prod_{l=0}^{2^{n-1}-1} \exp(\iota  t_{l}(\chi_{n-1}^{-1}(l)\otimes \sigma_3)\right) \exp{(\iota  \theta U^{(2^n)}_{j'})} \left(\prod_{l=0}^{2^{n-1}-1} \exp(\iota  t'_{l}(\chi_{n-1}^{-1}(l)\otimes \sigma_3)\right) $$ where $j'=f_q(p), p<q,q\in \{2,\hdots,2^n\}$ for some $t_l,t'_l\in \mathbb{R}$. Moreover, we would like to consider the ordering of the SRB such that products of  the exponentials of certain non-diagonal SRBB elements in that order should generate $M_nZYZ$ type matrix or permutation of a $M_nZYZ$ type matrix. For example, note from Theorem \ref{2mult} that in the original ordering of the non-diagonal SRBB basis matrices with indices $4j^2-2j$ and diagonal SRB matrices, the matrix  $$\left(\prod_{l=0}^{2^{n-1}-1} \exp(\iota  t_{l}(\chi_{n-1}^{-1}(l)\otimes \sigma_3)\right) \left(\prod_{j=1}^{2^{n-1}}\theta_{4j^2-2j}U^{(2^n)}_{4j^2-2j}\right) \left(\prod_{l=0}^{2^{n-1}-1} \exp(\iota  t'_{l}(\chi_{n-1}^{-1}(l)\otimes \sigma_3)\right)$$ is a $M_nZYZ$ type matrix. Besides, it is well known that quantum circuit for $M_nZYZ$ is prevalent in literature \cite{krol2022}.

From Corollary \ref{cor:basis}, note that any non-diagonal element of the SRBB is given by $U_j^{(2^n)}=PMP,$ where $M$ is a block diagonal matrix with a maximum $3$ blocks, two of which are diagonal matrices and one block is $\sigma\in\{\sigma_1,\sigma_3\},$ and $P=P_{(\alpha,\beta)}$ is a transposition with $0<\alpha\leq \beta\leq 2^n.$ Moreover, the permutation matrix $P_{(\alpha,\beta)}$ is uniquely identified with the SRBB element index $j,$ except when it is identity. We shall see now another interesting aspect of the SRBB elements. First we consider the following example.

\begin{example} The exponentials of SRBB elements for $\C^{4\times 4}$ are as follows.
\footnotesize\begin{eqnarray*}
&& \exp(\iota \theta_1 U_1^{(4)}) = \bmatrix{\cos{\theta_1} & \iota \sin{\theta_1} & 0 & 0\\ \iota \sin{\theta_1} &  \cos{\theta_1}  & 0 & 0\\0& 0 & e^{\iota  \theta_1} &0\\0&0&0&e^{-\iota  \theta_1}}, \,\, \exp(\iota  \theta_2 U_2^{(4)})=\bmatrix{\cos{\theta_2} &  \sin{\theta_1} & 0 & 0\\-\sin{\theta_2} &  \cos{\theta_2}  & 0 & 0\\0& 0 & e^{\iota  \theta_2} &0\\0&0&0&e^{-\iota  \theta_2}} \\
&& \exp(\iota  \theta_3 U_3^{(4)})=\bmatrix{e^{\iota  \theta_3} & 0 & 0 &0\\0&e^{-\iota  \theta_3} &0&0\\0&0&e^{\iota  \theta_3} &0\\0&0&0&e^{-\iota  \theta_3} }, \,\, \exp(\iota  \theta_4 U_4^{(4)})=\bmatrix{e^{\iota  \theta_4}&0&0&0\\0&\cos{\theta_4}& \iota \sin{\theta_4}&0\\0& \iota \sin{\theta_4}&\cos{\theta_4}&0\\0&0&0&e^{-\iota  \theta_4}} \\
&&  \exp(\iota  \theta_5 U_5^{(4)})=\bmatrix{\cos{\theta_4}&0& \iota \sin{\theta_5}&0\\0&e^{\iota  \theta_5}&0&0\\ \iota \sin{\theta_5}&0&\cos{\theta_5}&0\\0&0&0&e^{-\iota  \theta_5}}, \,\, \exp(\iota  \theta_6 U_6^{(4)})=\bmatrix{e^{\iota  \theta_6}&0&0&0\\0&\cos{\theta_6}&\sin{\theta_6}&0\\0&-\sin{\theta_6}&\cos{\theta_6}&0\\0&0&0&e^{-\iota  \theta_6}},\\
&& \exp(\iota  \theta_7 U_7^{(4)})=\bmatrix{\cos{\theta_4}&0&\sin{\theta_5}&0\\0&e^{\iota  \theta_7}&0&0\\ -\sin{\theta_7}&0&\cos{\theta_7}&0\\0&0&0&e^{-\iota  \theta_7}}, \,\, \exp(\iota  \theta_8 U_8^{(4)})=\bmatrix{e^{\iota  \theta_8 }&0&0&0\\0&e^{\iota  \theta_8 }&0&0\\0&0&e^{-\iota  \theta_8 }&0\\0&0&0&e^{-\iota  \theta_8 }},\\
&&\exp(\iota  \theta_9 U_{9}^{(4)})=\bmatrix{e^{\iota  \theta_9}&0&0&0\\0&e^{- i \theta_9}&0&0\\0&0&\cos{\theta_9}& \iota \sin{\theta_9}\\0&0& \iota \sin{\theta_9}&\cos{\theta_9}}, \,\, \exp(\iota  \theta_{10} U_{10}^{(4)})=\bmatrix{\cos{\theta_{10}}&0&0& \iota \sin{\theta_{10}}\\0&e^{-\iota  \theta_{10}}&0&0\\0&0&e^{\iota  \theta_{10}}&0\\ \iota \sin{\theta_{10}}&0&0&\cos{\theta_{10}}},\end{eqnarray*}\begin{eqnarray*}
&& \exp(\iota  \theta_{11} U_{11}^{(4)})=\bmatrix{e^{\iota  \theta_{11}}&0&0&0\\0&\cos{\theta_{11}}&0& \iota \sin{\theta_{11}}\\0&0&e^{-\iota  \theta_{11}}&0\\0& \iota \sin{\theta_{11}}&0&\cos{\theta_{11}}}, \,\, \exp(\iota  \theta_{12} U_{12}^{(4)})=\bmatrix{e^{\iota  \theta_{12}}&0&0&0\\0&e^{- i \theta_{12}}&0&0\\0&0&\cos{\theta_{12}}&\sin{\theta_{12}}\\0&0&-\sin{\theta_{12}}&\cos{\theta_{12}}},\\
&& \exp(\iota  \theta_{13} U_{13}^{(4)})=\bmatrix{\cos{\theta_{13}}&0&0& \sin{\theta_{13}}\\0&e^{-\iota  \theta_{13}}&0&0\\0&0&e^{\iota  \theta_{13}}&0\\-\sin{\theta_{13}}&0&0&\cos{\theta_{13}}}, \,\, \exp(\iota  \theta_{14} U_{14}^{(4)})=\bmatrix{e^{\iota  \theta_{14}}&0&0&0\\0&\cos{\theta_{14}}&0& \sin{\theta_{14}}\\0&0&e^{-\iota  \theta_{14}}&0\\0& -\sin{\theta_{14}}&0&\cos{\theta_{14}}},\\
&&  \exp(\iota  \theta_{15}U_{15}^{(4)})=\bmatrix{e^{\iota  \theta_{15}}&0&0&0\\0&e^{-\iota  \theta_{15}}&0&0\\0&0&e^{-\iota  \theta_{15}}&0\\0&0&0&e^{\iota  \theta_{15}}}
\end{eqnarray*}\normalsize 
\end{example}
Then, note that $\exp{(\iota  \theta_1U^{(4)}_1)}\exp{(\iota  \theta_2U^{(4)}_2)}\exp{(\iota  \theta_1U^{(4)}_9)}\exp{(\iota  \theta_{12}U^{(4)}_{12})}$ forms a $M_2ZYZ$ matrix. Further, the product \begin{eqnarray*}
    P_{(2,4)}\exp{(\iota  \theta_4 U^{(4)}_4)}\exp{(\iota  \theta_6 U^{(4)}_6)}\exp{(\iota  \theta_{10} U^{(4)}_{10})}\exp{(\iota  \theta_{13} U^{(4)}_{13})}P_{(2,4)}
\end{eqnarray*} is of the form $M_2ZYZ$, and \begin{eqnarray*}
    P_{(2,3)}\exp{(\iota  \theta_5 U^{(4)}_5)}\exp{(\iota  \theta_7 U^{(4)}_7)}\exp{(\iota  \theta_{11} U^{(4)}_{11})}\exp{(\iota  \theta_{14} U^{(4)}_{14})}P_{(2,3)}
\end{eqnarray*} is a block diagonal matrix. Thus, we conclude that product of exponentials of certain non-diagonal SRBB elements is permutation similar to either a $M_2ZYZ$ type matrix or a block-diagonal matrix.\\

Now, we show that the above observation is true for non-diagonal SRBB elements of $\Sf\Uf(2^n).$ Indeed, for a pair $1\leq \alpha<\beta\leq 2^n$ with $\beta$ is even and $\alpha$ is odd, we have \small\begin{equation}\label{eqn:blockx}\hspace{-1.5cm}\exp\left(i\theta U_{f_\beta(\alpha)}^{(2^n)}\right)=\bmatrix{\exp\left(i\theta D_{\alpha-1}\right) & &&& \\  & \cos\theta&& \sin\theta & \\ \vdots&& \exp\left(i\theta D_{\beta-\alpha-1}\right) && \\ & -\sin\theta&&\cos\theta& \\ &&&& \exp\left(i\theta D_{2^n-\beta}\right)}\end{equation}\normalsize where $D_x$ is a diagonal matrix of order $x$ with $l$-th diagonal entry is $1$ if $l$ is odd and it is $1$  if $l$ is even. Then clearly $P_{(\alpha+1,\beta)}\exp\left(i \theta U_{f_\beta(\alpha)}^{(2^n)}\right)P_{(\alpha+1,\beta)}$ gives a $M_nZYZ$ type matrix. Similarly, if $\beta$ is odd and $\alpha$ is even then $P_{(\alpha,\beta+1)}\exp\left(i \theta U_{f_\beta(\alpha)}^{(2^n)}\right) P_{(\alpha,\beta+1)}$ is a $M_nZYZ$ type matrix. Next, if $\alpha$ and $\beta$ are both odd then for the non-diagonal basis element $\exp\left(i \theta U_{f_\beta(\alpha)}^{(2^n)}\right)$ will have same pattern as equation (\ref{eqn:blockx}) but 

$P_{(\alpha+1,\beta)}\exp\left(\iota \theta U_{f_\beta(\alpha)}^{(2^n)}\right)P_{(\alpha+1,\beta)}$ will be a block-diagonal matrix consists of blocks are of size $2$ belonging to $\uf(2)$ with at least one block of the form $\bmatrix{\exp(\iota \theta)&0\\ 0& \exp(\iota \theta)}.$ Similarly, if $\alpha, \beta$ are even then 

$P_{(\alpha,\beta-1)}\exp\left(i \theta U_{f_q(p)}^{(2^n)}\right)P_{(\alpha,\beta-1)}$ is a special unitary block diagonal matrix with at least one block is from $\Uf(2).$ Similar observations also hold for the function $h.$

Finally observe that all the transpositions $P_{(\alpha,\beta)}$ whose pre and pro multiplication make a matrix $U_j^{(2^n)}\in \boldsymbol{\mathcal{U}}^{(2^n)}$ a matrix of type $M_nZYZ$ or a special unitary block diagonal matrix, have the values of $\alpha,\beta$ both to be even or $\alpha$ is even and $\beta$ is odd, where $1\leq \alpha<\beta\leq 2^n.$ Thus we consider two sets of permutation matrices \begin{eqnarray*}
\mathcal{P}_{2^n, \mbox{even}} &=& \{P_{(\alpha,\beta)}\in \mathcal{P}_{2^n} \, | \, \alpha< \beta \mbox{ and } \alpha, \beta \,\, \mbox{are even}\} \\
\mathcal{P}_{2^n, \mbox{odd}} &=& \{P_{(\alpha,\beta)} \in \mathcal{P}_{2^n} \,| \,  \alpha< \beta\mbox{ and }\alpha \, \, \mbox{is even}, \, \beta \,\, \mbox{is odd}\}
\end{eqnarray*} which we will use in order to approximate a unitary matrix as a product of $M_nZYZ$ or unitary block diagonal matrices and its permutations. Then it follows that $\left|\mathcal{P}_{2^n, \mbox{even}}\right|=2^{2n-3}-2^{n-2}=\left|\mathcal{P}_{2^n, \mbox{odd}}\right|.$ 

We will see in Theorem \ref{even} that the product of exponentials of certain SRBB elements create a matrix which is permutation similar to a $M_nZYZ$ matrix or a special unitary block diagonal matrix, and the corresponding permutation matrix is a product of  $2^{n-2}$ disjoint transpositions either from $\mathcal{P}_{2^n, \mbox{even}}$ or $\mathcal{P}_{2^n, \mbox{odd}}$ and total number of such permutation matrices is $(2^{n-1}-1)$ which makes the total number of elements in $\mathcal{P}_{2^n, \mbox{even}}$ and $\mathcal{P}_{2^n, \mbox{odd}}$ to be $2^{n-2}\times (2^{n-1}-1)=2^{2n-3}-2^{n-2}.$ {Let $T^{e}_x$ and $T^{o}_x$ be sets of $2^{n-2}$ disjoint transpositions from $\mathcal{P}_{2^n, \mbox{even}}$ and $\mathcal{P}_{2^n, \mbox{odd}}$ respectively, $1\leq x\leq 2^{n-1}-1$ such that $\cupdot_x T^e_x=\mathcal{P}_{2^n, \mbox{even}}$ and $\cupdot_x T^o_x=\mathcal{P}_{2^n, \mbox{odd}},$ where $\cupdot$ denotes disjoint union i.e. the variable $x$ determines set of selected disjoint transpositions. Define \begin{equation}\label{eqn:pieox}\Pi \mathsf{T}^e_{n,x}=\prod_{(\alpha,\beta)\in T^e_x} P_{(\alpha,\beta)} \,\, \mbox{and} \,\, \Pi \mathsf{T}^o_{n,x}=\prod_{(\alpha,\beta)\in T^o_x} P_{(\alpha,\beta)},\end{equation} $1\leq x\leq 2^{n-1}-1.$ In the later part of the paper, we shall provide an explicit quantum circuit construction and definition of $\Pi \mathsf{T}^g_{n,x},\alpha,\beta$ where $g\in\{e,o\},1\leq x\leq 2^{n-1}-1$.i.e. calculating $\Pi \mathsf{T}^g_{n,x}$ depending on $x$. (See equation (\ref{permusch1}).)}

The motivation behind creating unitary block diagonal or $M_nZYZ$ matrices lies in the fact that the quantum circuits for such matrices are easy to implement. The quantum circuit for $M_nZYZ$ matrices can be found in \cite{krol2022}. We shall see later that adding a few $\cnot$ and $R_z$ gates it is possible to define a circuit for a block diagonal matrix with $2\times 2$ blocks from a circuit that represents a $M_nZYZ$ matrix.

Now, we prove certain results which will be used to approximate a unitary matrix through product of exponentials of SRBB elements in certain order.

\begin{lemma}\label{blockdiagonal2}
    A block diagonal matrix $U\in \Sf\Uf(2^n)$ consisting of $2\times 2$ blocks is of the form $$\left(\prod_{t=2}^{2^n} \exp\left(i \theta_{t^2-1} U^{(2^n)}_{t^2-1}\right)\right) \left(\prod_{j=1}^{2^{n-1}}\exp\left(i\theta_{4j^2-2j} U^{(2^n)}_{4j^2-2j}\right)\right) \left(\prod_{t=2}^{2^n} \exp\left(i \theta_{t^2-1} U^{(2^n)}_{t^2-1}\right)\right)$$ where $\theta_{4j^2-2j}\in \mathbb{R},1\leq j\leq 2^{n-1}, \theta_{t^2-1},\theta'_{t^2-1}$ are obtained from Theorem \ref{2levSun}  and Theorem \ref{2mult}.
\end{lemma}
\pf  The matrices $U^{(2^n)}_{t^2-1},2\leq t\leq 2^{n}$ are used to construct diagonal unitary matrices. Thus the proof follows from how to get a $M_nZYZ$ matrix and finally we apply the same procedure used in theorem \ref{2levSun}. \hfill{$\square$}

\begin{lemma}\label{lemblkdg}
Suppose $U(\boldsymbol{\alpha}, \boldsymbol{\beta},\boldsymbol{\gamma})$ is a unitary matrix given by equation (\ref{blkdg}), where $\boldsymbol{\alpha}=\{\alpha_j\}_{j=1}^{2^{n-1}}$, $\boldsymbol{\beta}=\{\beta_j\}_{j=1}^{2^{n-1}}$, $\boldsymbol{\gamma}=\{\gamma_j\}_{j=1}^{2^{n-1}}.$ Then $$\left(\prod_{p=0}^{2^{n-1}-1} \exp\left(i t_{p}(\chi_{n-1}^{-1}(p)\otimes \sigma_3)\right) \right) U(\boldsymbol{\alpha}, \boldsymbol{\beta},\boldsymbol{\gamma}) \left(\prod_{p=0}^{2^{n-1}-1} \exp\left(i t'_{p}(\chi_{n-1}^{-1}(p)\otimes \sigma_3\right)\right)=\Tilde{U}(\widetilde{\boldsymbol{\alpha}}, \widetilde{\boldsymbol{\beta}}, \widetilde{\boldsymbol{\gamma}}),$$ which is of the form $M_nZYZ,$ where $\widetilde{\boldsymbol{\alpha}}=\{\widetilde{\alpha}_j\}_{j=1}^{2^{n-1}}$, $\widetilde{\boldsymbol{\beta}}=\{\widetilde{\beta}_j\}_{j=1}^{2^{n-1}}$, $\widetilde{\boldsymbol{\gamma}}=\{\widetilde{\gamma}_j\}_{j=1}^{2^{n-1}}$ with \begin{eqnarray*}
\widetilde{\alpha}_j &=& (\alpha_j+(\sum_{p=0}^{2^{n-1}-1}\eta^{(2j-1)}_{\chi^{-1}_{n-1}(p)\otimes Z}t_p+\sum_{p=0}^{2^{n-1}-1}\eta^{(2j-1)}_{\chi^{-1}_{n-1}(p)\otimes Z}t'_p)) \\
\widetilde{\beta}_j &=& (\beta_j+(\sum_{p=0}^{2^{n-1}-1}\eta^{(2j-1)}_{\chi^{-1}_{n-1}(p)\otimes Z}t_p-\sum_{p=0}^{2^{n-1}-1}\eta^{(2j-1)}_{\chi^{-1}_{n-1}(p)\otimes Z}t'_p)),
\end{eqnarray*} $\widetilde{\gamma}_j=\gamma_j,$ $1\leq j\leq 2^{n-1},$ $t_p,t'_p\in \R.$
\end{lemma}

\pf The proof follows adapting a similar procedure as described in Theorem \ref{2mult} \hfill{$\square$}.

 \begin{lemma}[\textbf{Product of exponentials of certain basis elements makes a $M_nZYZ$ matrix}]\label{lemmamult} The matrix
      \begin{eqnarray*}
  \prod_{q=1}^{2^{n-1}} \left(\exp{(\iota \theta_{h_{2q}(0)}U^{(2^n)}_{h_{2q}(0)}}\right) \left(\exp{(\iota \theta_{f_{2q}(0)}U^{(2^n)}_{f_{2q}(0)}}\right)= \prod_{j=1}^{2^{n-1}} \left(\exp{(\iota \theta_{(2j-1)^2}U^{(2^n)}_{(2j-1)^2})}\right) \left(\exp{(\iota \theta_{4j^2-2j}U^{(2^n)}_{4j^2-2j})}\right)
      \end{eqnarray*} is of the form of a $M_nZYZ$ matrix. Besides, this matrix can also be expressed alternatively as 
       $$\left(\prod_{p=0}^{2^{n-1}-1} \exp(\iota  t_{p}(\chi_{n-1}^{-1}(p)\otimes \sigma_3)\right) \left(\prod_{j=1}^{2^{n-1}} \exp{(\iota \phi_{4j^2-2j}U^{(2^n)}_{4j^2-2j})}\right) \left(\prod_{p=1}^{2^{n-1}} \exp(\iota  t'_{p}(\chi_{n-1}^{-1}(p)\otimes \sigma_3)\right),$$
 where $\theta_{4j^2-2j},\theta_{(2j-1)^2}\in \R, 1\leq j \leq  2^{n-1},$ $${\phi_{4j^2-2j}}=\arccos{\sqrt{(\cos{\theta_{(2j-1)^2}}\cos{\theta_{(4j^2-2j)}})^2+(\sin{\theta_{(2j-1)^2}}\sin{\theta_{(4j^2-2j)}})^2}},$$ and  $t_p, t'_p, $ $0\leq p\leq 2^{n-1}-1$ can be obtained by solving a linear system.
 \end{lemma}
 
\pf The first identity follows from writing the values of $h_{2q}(0)$ and $f_{2q}(0)$. From the definition of $U^{(2^n)}_j,$ observe that  
\begin{eqnarray*}
\hspace{-0.75cm}\exp\left(i\theta_{4j^2-2j}U^{(2^n)}_{4j^2-2j}\right) &=& \left[\begin{array}{c|c|c} 
      D_{{(2j-2)}\times {(2j-2)}}& 0 &0\\ 
      \hline 
       0&\bmatrix{\cos{\theta_{4j^2-2j}} & \sin{\theta_{4j^2-2j}}\\-\sin{\theta_{4j^2-2j}} & \cos{\theta_{4j^2-2j}}} &0\\
       \hline
      0& 0 & D_{(2^n-2j)\times (2^n-2j)} 
    \end{array} 
    \right] \\
    \exp{(\iota  \theta_{(2j-1)^2}B^{(2^n)}_{(2j-1)^2})} &=& \left[\begin{array}{c|c|c} 
      D_{(2j-2)\times (2j-2)}& 0 &0\\ 
      \hline 
       0&\bmatrix{\cos{\theta_{(2j-1)^2}} &\iota \sin{\theta_{(2j-1)^2}}\\ \iota \sin{\theta_{(2j-1)^2}} & \cos{\theta_{(2j-1)^2}}} &0\\
       \hline
      0& 0 & D_{(2^n-2j)\times (2^n-2j)}
    \end{array} 
    \right]
\end{eqnarray*} where $D$ is a diagonal matrix with $k$-th diagonal entry is given by $$D_{kk}=\begin{cases}
        \exp{(\iota  \theta_{4j^2-2j})} ,\mbox{if} \,\, k \,\, \mbox{is odd}\\
        \exp{(-i \theta_{4j^2-2j})},\mbox{if} \,\, k \,\, \mbox{is even}
    \end{cases},$$ $1\leq k\leq 2^{n-1}.$

 Now,
 \small\begin{eqnarray*}
 && \bmatrix{\cos{\theta_{(2j-1)^2}} & \iota \sin{\theta_{(2j-1)^2}}\\ \iota \sin{\theta_{(2j-1)^2}} &\cos{\theta_{(2j-1)^2}}}\bmatrix{\cos{\theta_{4j^2-2j}} & \sin{\theta_{4j^2-2j}}\\ -\sin{\theta_{4j^2-2j}} &\cos{\theta_{4j^2-2j}}} \\
 &=& \bmatrix{\cos{\theta_{(2j-1)^2}}\cos{\theta_{4j^2-2j}}-\iota \sin{\theta_{(2j-1)^2}}\sin{\theta_{4j^2-2j}}  & \sin{\theta_{4j^2-2j}}\cos{\theta_{(2j-1)^2}}+i \cos{\theta_{4j^2-2j}}\sin{\theta_{(2j-1)^2}}\\ -\sin{\theta_{4j^2-2j}}\cos{\theta_{(2j-1)^2}}+i \cos{\theta_{4j^2-2j}}\sin{\theta_{(2j-1)^2}}&\cos{\theta_{(2j-1)^2}}\cos{\theta_{4j^2-2j}}+\iota \sin{\theta_{(2j-1)^2}}\sin{\theta_{4j^2-2j}}} \\&=& \bmatrix{\cos{\phi_{4j^2-2j}}\exp{(-i \beta_{4j^2-2j})} & \sin{\phi_{4j^2-2j}}\exp{(\iota  \gamma_{4j^2-2j})}\\-\sin{\phi_{4j^2-2j}}\exp{(-i \gamma_{4j^2-2j})} &\cos{\phi_{4j^2-2j}}\exp{(\iota\beta_{4j^2-2j})}}
 \end{eqnarray*}\normalsize Then, utilizing \small\begin{eqnarray*} \hspace{-1.5cm}{(\cos{\theta_{(2j-1)^2}}\cos{\theta_{4j^2-2j}})^2+(\sin{\theta_{(2j-1)^2}}\sin{\theta_{4j^2-2j}})^2+(\sin{\theta_{4j^2-2j}}\cos{\theta_{(2j-1)^2}})^2+(\cos{\theta_{4j^2-2j}}\sin{\theta_{(2j-1)^2}})^2}=1
 \end{eqnarray*}\normalsize with 
 \begin{eqnarray*}
 \cos{\phi_{4j^2-2j}} &=& \sqrt{(\cos{\theta_{(2j-1)^2}}\cos{\theta_{4j^2-2j}})^2+(\sin{\theta_{(2j-1)^2}}\sin{\theta_{4j^2-2j}})^2}, \\
 \sin{\phi_{4j^2-2j}} &=& \sqrt{(\sin{\theta_{4j^2-2j}}\cos{\theta_{(2j-1)^2}})^2+(\cos{\theta_{4j^2-2j}}\sin{\theta_{(2j-1)^2}})^2}, \\
 \beta_{4j^2-2j} &=& \arcsin{\frac{ \sin{\theta_{(2j-1)^2}}\sin{\theta_{4j^2-2j}} }{\cos{\phi_{4j^2-2j}}}}, \\
 \gamma_{4j^2-2j} &=& \arcsin{\frac{\cos{\theta_{4j^2-2j}}\sin{\theta_{(2j-1)^2}}}{\sin{\phi_{4j^2-2j}}}}.
 \end{eqnarray*}
 
Therefore, $$\prod_{j=1}^{2^{n-1}} \exp\left(i \theta_{(2j-1)^2}U^{(2^n)}_{(2j-1)^2}\right) \exp\left(i \theta_{4j^2-2j}B^{(2^n)}_{4j^2-2j}\right) :=V =\left[ 
\begin{array}{c|c|c|c} 
      V_2 &  0 &0 & 0\\ 
      \hline 
       0 &  V_{4} &0 & 0\\ 
       \hline
       0 &  0&\ddots &0\\
        \hline
       0 & 0& 0&V_{2^{n}}
    \end{array} 
    \right]$$
where \begin{eqnarray*}
V_{{2j}} &=& \bmatrix{e^{\iota  (\sum_{l<j}(\theta_{4l^2-4l}+\theta_{(2l-1)^2}))}&0\\0&e^{(-i \sum_{l<j}(\theta_{4l^2-4i}+\theta_{(2l-1)^2}))}}\\&& \bmatrix{e^{(-i \beta_{4j^2-2j})}\cos{\phi_{4j^2-2j}} & e^{(i \gamma_{4j^2-2j})}\sin{\phi_{4j^2-2j}}\\-e^{(-\iota\gamma_{4j^2-2j})}\sin{\phi_{4j^2-2j}} &e^{(i \beta_{4j^2-2j})}\cos{\phi_{4j^2-2j}}} \\
&& \bmatrix{e^{(i (\sum_{l=j+1}^{2^{n-1}}(\theta_{4l^2-4l}+\theta_{(2l-1)^2})))} & 0\\0 & e^{(-i (\sum_{l=j+1}^{2^{n-1}}(\theta_{4l^2-4l}+\theta_{(2l-1)^2})))}}
\end{eqnarray*}\normalsize

Finally, from Theorem \ref{2mult}, we have $$\prod_{j=1}^{2^{n-1}}\exp\left(i \phi_{4j^2-2j}U^{(2^n)}_{4j^2-2j}\right)=\left[ 
\begin{array}{c|c|c|c} 
      W_2 &  0 &0 & 0\\ 
      \hline 
       0 &  0&\ddots &0\\
        \hline
       0 & 0& 0&W_{2^{n}}
    \end{array} 
    \right]$$ with $${W_{2j}}=\bmatrix{e^{\iota \sum_{l\neq j}\phi_{4l^2-2l}}\cos{\phi_{4j^2-2j}} & e^{\iota  (\sum_{l<j}\phi_{4l^2-2l}-\sum_{l>j} \phi_{4l^2-2l})}\sin{\phi_{4j^2-2j}}\\-e^{-\iota  (\sum_{l<j}\phi_{4l^2-2l}-\sum_{l>j}\phi_{4l^2-2l})}\sin{\phi_{4j^2-2j}} & e^{-\iota  (\sum_{l\neq j}\phi_{4i^2-2i})}\cos{\phi_{4j^2-2j}}},$$ $1 \leq j\leq 2^{n-1}.$ 
    
    Thus, 
    \begin{eqnarray*}
       \left(\prod_{p=0}^{2^{n-1}-1} \exp\left(i t_{p}(\chi_{n-1}^{-1}(p)\otimes \sigma_3\right)\right) \left(\prod_{j=1}^{2^{n-1}} \exp\left(i \phi_{4j^2-2j}U^{(2^n)}_{4j^2-2j}\right)\right) \left(\prod_{p=0}^{2^{n-1}-1} \exp\left(i t'_{p}(\chi_{n-1}^{-1}(p)\otimes \sigma_3\right)\right)=V,
    \end{eqnarray*}
  which is equivalent to $V_{2j}=M_{2j} {W}_{2j} M'_{2j}$such that \begin{eqnarray*}
   M_{2j} &=& \bmatrix{\exp{(\iota  \sum_{p=0}^{2^{n-1}-1}\eta^{(2j-1)}_{\chi^{-1}_{n-1}(p)\otimes \sigma_3}t_p)} & 0 \\ 0 & \exp{-(i \sum_{p=0}^{2^{n-1}-1}\eta^{(2j-1)}_{{\chi^{-1}_{n-1}(p)}\otimes \sigma_3}t_p)}} \\
    M'_{2j} &=& \bmatrix{\exp{(\iota  \sum_{p=0}^{2^{n-1}-1}\eta^{(2j-1)}_{\chi^{-1}_{n-1}(p)\otimes \sigma_3}t'_p)} & 0 \\ 0 & \exp{-(i \sum_{p=0}^{2^{n-1}-1}\eta^{(2j-1)}_{{\chi^{-1}_{n-1}(p)}\otimes \sigma_3}t'_p)}}
    \end{eqnarray*}
    will hold for certain values of $t_p, t'_p\in\R$ that can be obtained by solving a linear system as described in Theorem \ref{2mult}. This completes the proof. \hfill{$\square$}

\begin{remark}
In the above lemma we see how to select certain basis elements to obtain $M_nZYZ$ matrices, and the number of non-diagonal basis elements used to create such a matrix is $2^n$. Total number of non-diagonal basis elements is $2^{2n}-2^n$. Hence, we need to allocate these matrices into $(2^n-1)$ bundles each of which contains $2^n$ matrix multiplications such that each bundle gives us a matrix which is permutationally similar to $M_nZYZ$ matrices i.e. the matrix multiplication in the bundle is of the form $P\mathcal{U}P^{-1}$ where $\mathcal{U}$ is a $M_nZYZ$ matrix. In order to find what basis we shall use for multiplication and what permutation matrices are allowed, we first look at a theorem that tells us about the effect of permutation matrix on the exponentiation of non-diagonal basis elements. 
\end{remark}

\begin{lemma}\label{lemex}
Let $0<\alpha\leq \beta\leq 2^n$ be a pair of even integers, and $P_{(\alpha,\beta)}$ denote a $2$-cycle permutation on $2^n$ elements. Then,  \small\begin{eqnarray*}
    P_{(\alpha,\beta)}\exp\left(i \theta_{h_{\beta}(\alpha-1)} U^{(2^n)}_{h_{\beta}(\alpha-1)}\right) \exp\left(i \theta_{f_{\beta}(\alpha-1)} U^{(2^n)}_{f_{\beta}(\alpha-1)}\right) \exp\left(i \theta_{h_{\beta-1}(\alpha)} U^{(2^n)}_{h_{\beta-1}(\alpha)}\right) \exp\left(i \theta_{f_{\beta-1}(\alpha)}U^{(2^n)}_{f_{\beta-1}(\alpha)}\right) P_{(\alpha,\beta)}
\end{eqnarray*}\normalsize is a block diagonal matrix of the form $M_nZYZ$ given by $$\left[ 
\begin{array}{c|c|c|c} 
      {U}_2 &  0 &0 &0\\ 
      \hline 
      0&U_4&0 &0\\
      \hline
       0 &  0&\ddots &0\\
        \hline
       0 & 0&0& {U}_{2^{n}}
    \end{array} 
    \right]$$ such that \begin{eqnarray*}
    U_{\alpha} &=& \exp\left(i\theta_{h_{\beta}(\alpha-1)}\sigma_1\right) \exp\left(i \theta_{f_{\beta}(\alpha-1)}\sigma_2\right) \exp\left(i (\theta_{f_{\beta-1}(\alpha)}+\theta_{h_{\beta-1}(\alpha)}) \sigma_3\right) \\
    U_{\beta} &=& \exp\left(i (\theta_{f_{\beta}(\alpha-1)}+\theta_{h_{\beta}(\alpha-1)}) \sigma_3\right) \exp\left(i\theta_{h_{\beta-1}(\alpha)}\sigma_1\right) \exp\left(-i\theta_{f_{\beta-1}(\alpha)}\sigma_2\right) \\
    U_{2l} &=& \exp\left(i(\theta_{f_{\beta-1}(\alpha)}+\theta_{h_{\beta-1}(\alpha)}+\theta_{f_{\beta}(\alpha-1)}+\theta_{h_{\beta}(\alpha-1)})\sigma_3\right), \, l\in \{1,2,\hdots,2^{n-1}\}\setminus\{\alpha/2,\beta/2\}.
    \end{eqnarray*} 
\end{lemma}

\pf Note that 
\small\begin{eqnarray*}
\hspace{-.75cm}\exp\left(i \theta_{h_{\beta}(\alpha-1)} U^{(2^n)}_{h_{\beta}(\alpha-1)}\right) &=& \bmatrix{e^{\iota (\theta_{h_{\beta}(\alpha-1)})D_{\alpha-2}} & &&& \\ & \cos{(\theta_{h_{\beta}(\alpha-1)})} &  & i  \sin{{(\theta_{h_{\beta}(\alpha-1)})}} &  \\  &  & e^{\iota (\theta_{h_{\beta}(\alpha-1)})D_{\beta-\alpha+1}} &  &  \\ & \iota \sin{(\theta_{h_{\beta}(\alpha-1)})} & & \cos{(\theta_{h_{\beta}(\alpha-1)})}  &  \\ &&&& e^{\iota (\theta_{h_{\beta}(\alpha-1)})D_{2^n-\beta}}}, \\\end{eqnarray*}\begin{eqnarray*}
\hspace{-.75cm}\exp\left(i \theta_{f_{\beta}(\alpha-1)} U^{(2^n)}_{f_{\beta}(\alpha-1)}\right) &=& \bmatrix{e^{\iota (\theta_{f_{\beta}(\alpha-1)})D_{\alpha-2}} & &&& \\ & \cos{(\theta_{f_{\beta}(\alpha-1)})} &  &  \sin{{(\theta_{f_{\beta}(\alpha-1)})}} &  \\  &  & e^{\iota (\theta_{f_{\beta}(\alpha-1)})D_{\beta-\alpha+1}} &  &  \\ & -\sin{(\theta_{f_{\beta}(\alpha-1)})} & & \cos{(\theta_{f_{\beta}(\alpha-1)})}  &  \\ &&&& e^{\iota (\theta_{f_{\beta}(\alpha-1)})D_{2^n-\beta}}},
\end{eqnarray*}\normalsize where $D_{\alpha-2}, D_{\beta-\alpha+1}$ and $D_{2^n-\beta}$ are diagonal matrices of order $\alpha-2,$ $\beta-\alpha+1$ and $2^n-\beta$ respectively with the $k$-th diagonal entry is $1$ if $k$ is odd and $-1$ otherwise. Therefore, 
\begin{eqnarray*}
P_{(\alpha,\beta)}\exp{(\iota  \theta_{h_{\beta}(\alpha-1)} U^{(2^n)}_{h_{\beta}(\alpha-1)})}\exp{(\iota \theta_{f_{\beta}(\alpha-1)} U^{(2^n)}_{h_{\beta}(\alpha-1)})}P_{(\alpha,\beta)} &=& \left[ 
\begin{array}{c|c|c} 
      V_2 &  0 &0 \\ 
      \hline 
       0 & \ddots &0\\
        \hline
       0 & 0&V_{2^{n}}
    \end{array} 
    \right], \\
P_{(\alpha,\beta)}\exp{(\iota  \theta_{h_{\beta-1}(\alpha)} U^{(2^n)}_{h_{\beta-1}(\alpha)})}\exp{(\iota  \theta_{f_{\beta-1}(\alpha)}B^{(2^n)}_{f_{\beta-1}(\alpha)})})P_{(\alpha,\beta)} &=& \left[ 
\begin{array}{c|c|c} 
      W_2 &  0 &0 \\ 
      \hline 
       0 & \ddots &0\\
        \hline
       0 & 0&W_{2^{n}}
    \end{array} 
    \right],
\end{eqnarray*} where 
\begin{eqnarray*}
V_{\alpha} &=& \exp{(\iota \theta_{h_{\beta}(\alpha-1)}\sigma_1)}\exp{(\iota \theta_{f_{\beta}(\alpha-1)}\sigma_2)} \\
V_{2l} &=& \exp{(\iota \theta_{h_{\beta}(\alpha-1)}\sigma_3)}\exp{(\iota \theta_{f_{\beta}(\alpha-1)}\sigma_3)},\forall l\in \{1,2,\hdots,2^{n-1}\}\setminus \{\alpha/2\} \\
W_{\beta} &=& \exp{(\iota \theta_{h_{\beta-1}(\alpha)}\sigma_1)}\exp{(-i\theta_{f_{\beta-1}(\alpha)}\sigma_2)} \\
W_{2l} &=& \exp{(\iota \theta_{h_{\beta-1}(\alpha)}\sigma_3)}\exp{(\iota \theta_{f_{\beta-1}(\alpha)}\sigma_3)},\forall l\in \{1,2,\hdots,2^{n-1}\}\setminus \{\beta/2\}.
\end{eqnarray*}
Thus the desired result follows. \hfill{$\square$}

\begin{corollary}\label{corinvariant}
 Let $0<\alpha\leq \beta\leq 2^n$ and $0<\delta\leq \gamma\leq 2^n$ be two distinct pairs of even integers, and $P_{(\alpha,\beta)},$ $P_{(\delta,\gamma)}$ denote the permutation matrices of order $2^n$ corresponding to the transpositions $(\alpha,\beta)$ and $(\delta,\gamma)$ respectively. Then $P_{(\delta,\gamma)} A P_{(\delta,\gamma)}=A,$ where \begin{eqnarray*}
     \hspace{-0.5cm}A=\exp\left(i \theta_{h_{\beta}(\alpha- 1)} U^{(2^n)}_{h_{\beta}(\alpha- 1)}\right) \exp\left(i \theta_{f_{\beta}(\alpha- 1)} U^{(2^n)}_{f_{\beta}(\alpha- 1)}\right) \exp\left( \theta_{h_{\beta- 1}(\alpha)} U^{(2^n)}_{h_{\beta- 1}(\alpha)}\right) \exp\left(i \theta_{f_{\beta- 1}(\alpha)}U^{(2^n)}_{f_{\beta- 1}(\alpha)}\right)
 \end{eqnarray*} 
  \end{corollary}
\pf The proof is computational and follows from Theorem \ref{lemex}. \hfill{$\square$}

\begin{theorem}[\textbf{Product of exponentials of certain basis elements make a matrix permutationally similar to a $M_nZYZ$ matrix}]\label{even}
Let $P=\prod_{j=1}^{2^{n-2}}P_{(\alpha_j,\beta_j)}$ be a product of $2^{n-2}$ permutation matrices of order $2^n$ corresponding to transposition $(\alpha_j,\beta_j)$, where $0<\alpha_j\leq \beta_j\leq 2^n, 1\leq j\leq 2^{n-2}$ are distinct pairs of even integers. Then \begin{eqnarray}   && P\left[ \prod_{j=1}^{2^{n-2}} \exp\left(i \theta_{h_{\beta_j}(\alpha_j-1)} U^{(2^n)}_{h_{\beta_j}(\alpha_j-1)}\right) \exp\left(i \theta_{f_{\beta_j}(\alpha_j-1)} U^{(2^n)}_{f_{\beta_j}(\alpha_j-1)}\right) \right. \nonumber \\
&& \left. \exp\left(i \theta_{h_{\beta_j-1}(\alpha_j)} U^{(2^n)}_{h_{\beta_j-1}(\alpha_j))}\right) \exp\left(i \theta_{f_{\beta_j-1}(\alpha_j)}U^{(2^n)}_{f_{\beta_j-1}(\alpha_j))}\right) \right]P, \nonumber \end{eqnarray} is a block diagonal matrix of the form $M_nZYZ$, where the diagonal blocks are given by 
\small\begin{eqnarray*}
U_{\alpha_j} &=& \exp\left(i \sum_{m<j}(\theta_{h_{\beta_m}(\alpha_m-1)}+\theta_{f_{\beta_m}(\alpha_m-1)}+\theta_{h_{\beta_m-1}(\alpha_m)}+\theta_{f_{\beta_m-1}(\alpha_m)})\sigma_3\right) \exp\left(i \theta_{h_{\beta_j}(\alpha_j-1)}\sigma_1\right) \\
&& \exp\left(i \theta_{f_{\beta_j}(\alpha_j-1)}\sigma_2\right) \exp\left(i (\theta_{f_{\beta_j-1}(\alpha_j)}+\theta_{h_{\beta_j-1}(\alpha_j)}) \sigma_3\right) \\
&& \exp\left(i\sum_{l=j+1}^{2^{n-2}}(\theta_{h_{\beta_l}(\alpha_l-1)}+\theta_{f_{\beta_l}(\alpha_l-1)}+\theta_{h_{\beta_l-1}(\alpha_l)}+\theta_{f_{\beta_l-1}(\alpha_l)})\sigma_3\right) \\
U_{\beta_j} &=& \exp \left(i \sum_{m<j}(\theta_{h_{\beta_m-1}(\alpha_m)}+\theta_{f_{\beta_m-1}(\alpha_m)}+\theta_{h_{\beta_m}(\alpha_m-1)}+\theta_{f_{\beta_m}(\alpha_m-1)})\sigma_3\right) \exp \left(i\theta_{h_{\beta_j-1}(\alpha_j)}\sigma_1\right)\\
&& \exp \left(-i \theta_{f_{\beta_j-1}(\alpha_j)}\sigma_2\right) \exp \left(i (\theta_{f_{\beta_j}(\alpha_j-1)}+\theta_{h_{\beta_j}(\alpha_j-1)}) \sigma_3\right) \\ && \exp\left(i\sum_{l=j+1}^{2^{n-2}}(\theta_{h_{\beta_l}(\alpha_l-1)}+\theta_{f_{\beta_l}(\alpha_l-1)}+\theta_{h_{\beta_l-1}(\alpha_l)}+\theta_{f_{\beta_l-1}(\alpha_l)})\sigma_3\right),
\end{eqnarray*}\normalsize $1\leq j\leq2^{n-2}.$
\end{theorem}

\pf From Lemma \ref{corinvariant}, \begin{eqnarray*} \prod_{j=1}^{2^{n-2}} && P\left[\exp\left(i \theta_{h_{\beta_j}(\alpha_j-1)} U^{(2^n)}_{h_{\beta_j}(\alpha_j-1)}\right)  
 \exp\left(i \theta_{f_{\beta_j}(\alpha_j-1)} U^{(2^n)}_{f_{\beta_j}(\alpha_j-1)}\right) \right.\\ && \left. \exp\left(i \theta_{h_{\beta_j-1}(\alpha_j)} U^{(2^n)}_{h_{\beta_j-1}(\alpha_j)}\right) 
  \exp\left(i \theta_{f_{\beta_j-1}(\alpha_j)} U^{(2^n)}_{f_{\beta_j-1}(\alpha_j)}\right)\right]P
  \\ = && \prod_{j=1}^{2^{n-2}}  P_{(\alpha_j,\beta_j)}\left[\exp\left(i \theta_{h_{\beta_j}(\alpha_j-1)} U^{(2^n)}_{h_{\beta_j}(\alpha_j-1)}\right)  
 \exp\left(i \theta_{f_{\beta_j}(\alpha_j-1)} U^{(2^n)}_{f_{\beta_j}(\alpha_j-1)}\right) \right.\\ && \left. \exp\left(i \theta_{h_{\beta_j-1}(\alpha_j)} U^{(2^n)}_{h_{\beta_j-1}(\alpha_j)}\right) 
  \exp\left(i \theta_{f_{\beta_j-1}(\alpha_j)} U^{(2^n)}_{f_{\beta_j-1}(\alpha_j)}\right)\right]P_{(\alpha_j,\beta_j)} \end{eqnarray*} 
  Then the proof follows from  Lemma \ref{lemblkdg}, Lemma \ref{lemex}  and Theorem \ref{2mult}. \hfill{$\square$}

\begin{remark}
The theorem above deals with the cases when the product is  permutationally similar to $M_nZYZ$ matrices where the permutation matrix $P$ is a product of $2^{n-2}$ disjoint transpositions of the form $P_{(2m,2n)},m<n$.
\end{remark}

\begin{lemma}\label{lemexodd}
Let $1\leq \alpha<\beta\leq 2^n$ with $\alpha$ is even and $\beta$ is odd, and $P_{(\alpha,\beta)}$ denote the permutation matrix corresponding to the transposition $(\alpha,\beta)$. Then \begin{eqnarray} && P_{(\alpha,\beta)} \left[\exp\left(i \theta_{h_{\beta}(\alpha-1)} U^{(2^n)}_{h_{\beta}(\alpha-1)}\right) \exp\left(i \theta_{f_{\beta}(\alpha-1)} U^{(2^n)}_{f_{\beta}(\alpha-1)}\right) \right. \nonumber \\ 
&& \left. \exp\left(i \theta_{h_{\beta+1}(\alpha)} U^{(2^n)}_{h_{\beta+1}(\alpha)}\right) \exp\left(i \theta_{f_{\beta+1}(\alpha)} U^{(2^n)}_{f_{\beta+1}(\alpha)}\right)\right] P_{(\alpha,\beta)} \nonumber \end{eqnarray} is a block diagonal matrix $U=\left[ 
\begin{array}{c|c|c} 
      {U}_2 &  0 &0 \\ 
      \hline 
       0 &  \ddots &0\\
        \hline
       0 & 0& {U}_{2^{n}}
    \end{array} 
    \right]\in \Sf\Uf(2^n)$ where
    \begin{eqnarray*}
    U_{\alpha} &=& \exp\left(i\theta_{h_{\beta}(\alpha-1)}\sigma_1\right) \exp\left(i \theta_{f_{\beta}(\alpha-1)}\sigma_2\right) \exp\left(i (\theta_{f_{\beta+1}(\alpha)}+\theta_{h_{\beta+1}(\alpha)}) \sigma_3\right) \\ 
    U_{\beta+1} &=& \exp\left(-i (\theta_{f_{\beta}(\alpha-1)}+\theta_{h_{\beta}(\alpha-1)}) I_2\right) \exp\left(i \theta_{h_{\beta+1}(\alpha)}\sigma_1\right) \exp\left(i \theta_{f_{\beta+1}(\alpha)}\sigma_2\right) \\
    U_{\beta-1} &=& \exp\left(i (\theta_{f_{\beta}(\alpha-1)}+\theta_{h_{\beta}(\alpha-1)}) I_2\right) \exp\left(i (\theta_{f_{\beta+1}(\alpha)}+\theta_{h_{\beta+1}(\alpha)}) \sigma_3\right) \\
    U_{2l} &=& \exp\left(i (\theta_{f_{\beta+1}(\alpha)}+\theta_{h_{\beta+1}(\alpha)}+\theta_{f_{\beta}(\alpha-1)}+\theta_{h_{\beta}(\alpha-1)})\sigma_3\right),\end{eqnarray*} where $l\in \{1,2,\hdots,2^{n-1}\}\setminus\{(\alpha+1)/2,(\beta-1)/2),(\beta+1)/2\}.$
\end{lemma}
\pf The proof follows  similar to the proof of Lemma \ref{lemex}. \hfill{$\square$}

\begin{remark}
It is easy to see that the matrix $U$ in the above lemma is not in the form $M_nZYZ$ but a special unitary block diagonal matrix consisting of $2\times 2$ unitary blocks. However, for $n=2$, the matrix is indeed of the type $M_nZYZ$ (It has been pointed out in \cite{Belli2024arx,Belli2024conf}). It is also to be noted that any matrix of the $M_nZYZ$ type is contained in the set of block diagonal special unitary matrices consisting of $2\times 2$ unitary blocks. So the inference of the theorem does not change.   
\end{remark}

\begin{theorem}\label{odd}
Let $P=\prod_{j=1}^{2^{n-2}}P_{(\alpha_j,\beta_j)}$ be the product of $2^{n-2}$ disjoint transpositions where $1<\alpha_j<\beta_j\leq 2^{n}$ with $\alpha_j$ is even and $\beta_j$ is odd, $1\leq j\leq 2^{n-2}$. Then  \begin{eqnarray*}&& P\left[\prod_{j=1}^{2^{n-2}} \exp\left(i \theta_{h_{\beta_j}(\alpha_j-1)} U^{(2^n)}_{h_{\beta_j}(\alpha_j-1)}\right) \exp\left(i \theta_{f_{\beta_j}(\alpha_j-1)} U^{(2^n)}_{f_{\beta_j}(\alpha_j-1)}\right) \exp\left(i \theta_{h_{\beta_j+1}(\alpha_j)} U^{(2^n)}_{h_{\beta_j+1}(\alpha_j)}\right)\right.\\ && \left. \exp\left(i \theta_{f_{\beta_j+1}(\alpha_j)} U^{(2^n)}_{f_{\beta_j-1}(\alpha_j)}\right)\right]P\end{eqnarray*} is equal to $U=\left[ 
\begin{array}{c|c|c} 
      {U}_2 &  0 &0 \\ 
      \hline 
       0 &  \ddots &0\\
        \hline
       0 & 0& {U}_{2^{n}}
    \end{array}\ 
    \right]\in \Sf\Uf(2^n),$ where 
   \small\begin{eqnarray*}
        U_{\alpha_j}=\begin{cases}
        \exp\left(i \sum_{m<j}(\theta_{h_{\beta_m}(\alpha_m-1)}+\theta_{f_{\beta_m}(\alpha_m-1)}+\theta_{h_{\beta_m+1}(\alpha_m)}+\theta_{f_{\beta_m+1}(\alpha_m)})\sigma_3\right)\\
        \exp\left(i \theta_{h_{\beta_j}(\alpha_j-1)}\sigma_1\right) \exp\left(i\theta_{f_{\beta_j}(\alpha_j-1)}\sigma_2\right) \exp\left(i (\theta_{f_{\beta_j+1}(\alpha_j)}+\theta_{h_{\beta_j+1}(\alpha_j)}) \sigma_3\right) \\ 
        \exp\left(\iota\sum_{l=j+1,l\neq k}^{2^{n-2}}(\theta_{h_{\beta_l}(\alpha_l-1)}+\theta_{f_{\beta_l}(\alpha_l-1)}+\theta_{h_{\beta_l+1}(\alpha_l)}+\theta_{f_{\beta_l+1}(\alpha_l)})\sigma_3\right)\\
        \exp\left(i (\theta_{f_{\beta_{k}}(\alpha_{k}-1)}+\theta_{h_{\beta_{k}}(\alpha_{k}-1)}) I_2\right)  \mbox{ if } \beta_{k}-1=\alpha_j,k\geq j \\\\
        \exp\left(i (\theta_{f_{\beta_{k}}(\alpha_{k}-1)}+\theta_{h_{\beta_{k}}(\alpha_{k}-1)}) I_2\right) \\ 
        \exp\left(i \sum_{0<m\neq k<j}(\theta_{h_{\beta_m}(\alpha_m-1)}+\theta_{f_{\beta_m}(\alpha_m-1)}+\theta_{h_{\beta_m+1}(\alpha_m)}+\theta_{f_{\beta_m+1}(\alpha_m)})\sigma_3\right)\\
        \exp\left(i \theta_{h_{\beta_j}(\alpha_j-1)}\sigma_1\right) \exp\left(i\theta_{f_{\beta_i}(\alpha_i-1)}\sigma_2\right) \exp\left(i (\theta_{f_{\beta_i+1}(\alpha_i)}+\theta_{h_{\beta_i+1}(\alpha_i)}) \sigma_3\right)\\
        \exp\left(i\sum_{l=j+1}^{2^{n-2}}(\theta_{h_{\beta_l}(\alpha_l-1)}+\theta_{f_{\beta_l}(\alpha_l-1)}+\theta_{h_{\beta_l+1}(\alpha_l)}+\theta_{f_{\beta_l+1}(\alpha_l)})\sigma_3\right)\mbox{ if } \beta_{k}-1=\alpha_j,k\leq j \\\\
        \exp\left(i \sum_{0<m\neq k<j}(\theta_{h_{\beta_m}(\alpha_m-1)}+\theta_{f_{\beta_m}(\alpha_m-1)}+\theta_{h_{\beta_m+1}(\alpha_m)}+\theta_{f_{\beta_m+1}(\alpha_m)})\sigma_3\right)\\
        \exp\left(i \theta_{h_{\beta_j}(\alpha_j-1)}\sigma_1\right) \exp\left(i\theta_{f_{\beta_i}(\alpha_i-1)}\sigma_2\right) \exp\left(i (\theta_{f_{\beta_i+1}(\alpha_i)}+\theta_{h_{\beta_i+1}(\alpha_i)}) \sigma_3\right)\\
        \exp\left(i\sum_{l=j+1}^{2^{n-2}}(\theta_{h_{\beta_l}(\alpha_l-1)}+\theta_{f_{\beta_l}(\alpha_l-1)}+\theta_{h_{\beta_l+1}(\alpha_l)}+\theta_{f_{\beta_l+1}(\alpha_l)})\sigma_3\right), \,\, \mbox{otherwise}
    \end{cases} \end{eqnarray*}\normalsize and
    
    \small\begin{eqnarray*}
        U_{\beta_j+1}=\begin{cases}
        \exp\left(i (\theta_{f_{\beta_{k}}(\alpha_{k}-1)}+\theta_{h_{\beta_{k}}(\alpha_{k}-1)}) I_2\right) \exp\left(i\sum_{0<m\neq k<j}(\theta_{h_{\beta_m+1}(\alpha_m)}+\theta_{f_{\beta_m+1}(\alpha_m)}+\theta_{h_{\beta_m}(\alpha_m-1)}+\right. \\ 
        \left. \theta_{f_{\beta_m}(\alpha_m-1)})\sigma_3\right)
        \exp\left(-i (\theta_{f_{\beta_j}(\alpha_j-1)}+\theta_{h_{\beta_j}(\alpha_j-1)}) I_2\right) 
         \exp\left(i \theta_{h_{\beta_j+1}(\alpha_j)}\sigma_1\right) \exp\left(i\theta_{f_{\beta_j+1}(\alpha_j)}\sigma_2\right) 
         \\ \exp\left(i (\theta_{f_{\beta_j}(\alpha_j-1)}+\theta_{h_{\beta_j}(\alpha_j-1)}) \sigma_3\right)\\
        \exp\left(i \sum_{l=j+1}^{2^{n-2}}(\theta_{h_{\beta_l}(\alpha_l-1)}+\theta_{f_{\beta_l}(\alpha_l-1)}+\theta_{h_{\beta_l+1}(\alpha_l)}+\theta_{f_{\beta_l+1}(\alpha_l)})\sigma_3\right) \mbox{ if } \beta_{k}-1=\beta_j+1,k<j \vspace{0.5cm}\\
        \exp\left(i \sum_{0<m<j}(\theta_{h_{\beta_m+1}(\alpha_m)}+\theta_{f_{\beta_m+1}(\alpha_m)}+\theta_{h_{\beta_m}(\alpha_m-1)}+\theta_{f_{\beta_m}(\alpha_m-1)})\sigma_3\right)\\
        \exp\left(-i (\theta_{f_{\beta_j}(\alpha_j-1)}+\theta_{h_{\beta_j}(\alpha_j-1)}) I_2\right) \exp\left(i \theta_{h_{\beta_j+1}(\alpha_j)}\sigma_1\right) \exp\left(i \theta_{f_{\beta_j+1}(\alpha_j)}\sigma_2\right) \\
        \exp\left(i (\theta_{f_{\beta_j}(\alpha_j-1)}+\theta_{h_{\beta_j}(\alpha_j-1)}) \sigma_3\right)
        \exp\left(i\sum_{l=j+1}^{2^{n-2}}(\theta_{h_{\beta_l}(\alpha_l-1)}+\theta_{f_{\beta_l}(\alpha_l-1)}+\theta_{h_{\beta_l+1}(\alpha_l)}+\theta_{f_{\beta_l+1}(\alpha_l)})\sigma_3\right) \\
        \exp\left(i (\theta_{f_{\beta_{k}}(\alpha_{k}-1)}+\theta_{h_{\beta_{k}}(\alpha_{k}-1)}) I_2\right) \mbox{ if } \beta_{k}-1=\beta_j+1,k\geq j
 \end{cases} \end{eqnarray*}\normalsize and $1\leq j\leq2^{n-2}$
\end{theorem} 

\pf Follows from \ref{lemexodd} and follows similar to Theorem \ref{even} \hfill{$\square$}

\begin{remark}
\begin{itemize}
    \item[(a)] The theorem above deals with the cases when the product of exponentials of certain SRBB elements is permutationally similar to block-diagonal matrices, where the corresponding permutation matrix is a product of $2^{n-2}$ disjoint transpositions that are of the form $P_{(\alpha,\beta)},$ $\alpha$ is even and $\beta$ is odd. 
    \item[(b)] Besides, from the above theorem we see that when $\alpha$ is even and $\beta$ is odd, the for any transposition $P_{(\delta,\gamma)}$ with $(\delta,\gamma)\neq (\alpha, \beta)$ and $\delta$ is even and $\gamma$ is odd then like Corollary \ref{corinvariant} 
    \begin{eqnarray*}
   &&  P_{(\delta,\gamma)}\left[\exp\left(i \theta_{h_{\beta}(\alpha-1)} U^{(2^n)}_{h_{\beta}(\alpha-1)}\right) \exp\left(i \theta_{f_{\beta}(\alpha-1)} U^{(2^n)}_{f_{\beta}(\alpha+1)}\right) \right.\\ && \left. \exp\left(i \theta_{h_{\beta+1}(\alpha)} U^{(2^n)}_{h_{\beta+1}(\alpha)}\right) \exp\left(i \theta_{f_{\beta+1}(\alpha)} U^{(2^n)}_{f_{\beta+1}(\alpha)}\right)\right]P_{(\delta,\gamma)} \end{eqnarray*} does not give back 
   \begin{eqnarray*}
      \exp\left(i \theta_{h_{\beta}(\alpha-1)} U^{(2^n)}_{h_{\beta}(\alpha-1)}\right) \exp\left(i \theta_{f_{\beta}(\alpha+1)} U^{(2^n)}_{f_{\beta}(\alpha+1)}\right) \exp\left(i \theta_{h_{\beta+1}(\alpha)} U^{(2^n)}_{h_{\beta+1}(\alpha)}\right) \exp\left(i \theta_{f_{\beta+1}(\alpha)} U^{(2^n)}_{f_{\beta+1}(\alpha)}\right)
   \end{eqnarray*} but it gives back 
    \begin{eqnarray*}
    \left[\prod_{t=2}^{2^n} \exp\left(i \theta_{t^2-1} U^{(2^n)}_{t^2-1}\right) \right] \exp\left(i \theta_{h_{\beta}(\alpha-1)} U^{(2^n)}_{h_{\beta}(\alpha-1)}\right) 
     \exp\left(i \theta_{f_{\beta}(\alpha-1)} U^{(2^n)}_{f_{\beta}(\alpha-1)}\right) \exp\left(i \theta_{h_{\beta+1}(\alpha)} U^{(2^n)}_{h_{\beta+1}(\alpha)}\right)  \\
     \exp\left(i \theta_{f_{\beta+1}(\alpha)} U^{(2^n)}_{f_{\beta+1}(\alpha)}\right) \left[\prod_{t=2}^{2^n} \exp\left(i \theta'_{t^2-1}B^{(2^n)}_{t^2-1}\right)\right]
    \end{eqnarray*} for some $\theta'_{t^2-1},\theta_{t^2-1},2\leq t\leq 2^n$. Hence, the product of exponentials of certain SRBB elements scaled with some permutation matrix in Theorem \ref{odd} does not give a $M_nZYZ$ matrix but rather a unitary block-diagonal matrix consisting of $2\times 2$ blocks. However, for $n=2$, the product is indeed of the $M_nZYZ$ type (also see \cite{Belli2024arx,Belli2024conf}). However, as mentioned before, any matrix of the type $M_nZYZ$ is automatically a special unitary block-diagonal matrix consisting of $2\times 2$ blocks. 
\end{itemize}
\end{remark}

Now from equation (\ref{eqn:pieox}), Theorem \ref{even}, and Theorem \ref{odd}, for any $1\leq x\leq 2^{n-1}-1,$ define 
\small\begin{eqnarray}
M^e_x &=& \Pi \mathsf{T}^e_{n,x}  \left[  \prod_{(\alpha,\beta)\in T^e_x} \exp\left(i \theta_{h_{\beta}(\alpha-1)} U^{(2^n)}_{h_{\beta}(\alpha-1)}\right)  \exp\left(i \theta_{f_{\beta}(\alpha-1)} U^{(2^n)}_{f_{\beta}(\alpha-1)}\right) \exp\left(i \theta_{h_{\beta-1}(\alpha)} U^{(2^n)}_{h_{\beta-1}(\alpha)}\right)  \right. \nonumber \\ 
&& \left. \exp\left(i \theta_{f_{\beta-1}(\alpha)} U^{(2^n)}_{f_{\beta-1}(\alpha)}\right) \right] \Pi \mathsf{T}^e_{n,x}, \label{def:mex} \\
M^o_x &=& \Pi \mathsf{T}^o_{n,x}  \left[  \prod_{(\alpha,\beta)\in T^o_x} \exp\left(i \theta_{h_{\beta}(\alpha-1)} U^{(2^n)}_{h_{\beta}(\alpha-1)}\right)  \exp\left(i \theta_{f_{\beta}(\alpha-1)} U^{(2^n)}_{f_{\beta}(\alpha-1)}\right) \exp\left(i \theta_{h_{\beta+1}(\alpha)} U^{(2^n)}_{h_{\beta+1}(\alpha)}\right)  \right. \nonumber \\ 
&& \left. \exp\left(i \theta_{f_{\beta+1}(\alpha)} U^{(2^n)}_{f_{\beta+1}(\alpha)}\right) \right] \Pi \mathsf{T}^o_{n,x}, \label{def:mox}
\end{eqnarray}\normalsize
where $\Pi \mathsf{T}^e_{n,x}$ and $\Pi \mathsf{T}^o_{n,x}$ are defined in equation (\ref{eqn:pieox}). Then it can be seen that $M^o_x\in \Sf\Uf(2^n)$ is a unitary block diagonal matrix with $2\times 2$ blocks and $M^e_x\in \Sf\Uf(2^n)$ is a $M_nZYZ$ matrix, $1\leq x\leq 2^{n-1}-1.$ Now note that for each $x,$ $M^e_x$ and $M^0_x$ include $4\times 2^{n-2}=2^{n}$ non-diagonal SRBB elements, and a total of $2\times (2^{n-1}-1)\times 2^{n}=2^{2n}-2^{n+1}$ SRBB elements. Further, from Lemma \ref{lemmamult}, note that there are $2\times 2^{n-1}=2^n$ non-diagonal SRBB elements whose product gives us a matrix of the form $M_nZYZ.$ Thus the total number of non-diagonal basis elements $2^{2n}-2^n=2^{2n}-2^{n+1}+2^n$ SRBB elements which contribute to unitary block diagonal matrices of matrices of type $M_nZYZ$ which we will now employ to redefine an approximation for any unitary matrix of order $2^n.$

{\bf Approximation of unitary matrices of order $2^n$ with optimal ordering of the SRRB:} Define \begin{eqnarray}
\zeta(\Theta_{\zeta}) &=& \prod_{j=2}^{2^n} \exp\left(i \theta_{j^2-1} U^{(2^n)}_{j^2-1}\right) \label{eqn:thetazeta}\\
\Psi(\Theta_{\psi}) &=&  \left(\prod_{j=1}^{2^{n-1}} \exp\left(i \theta_{(2j-1)^2}U^{(2^n)}_{(2j-1)^2}\right) \exp\left(i \theta_{(4j^2-2j)} U^{(2^n)}_{(4j^2-2j)}\right)\right)\nonumber\\&& \left(\prod_{x=1}^{2^{n-1}-1} \left(\Pi \mathsf{T}^e_{n,x}\right) M^e_x \left(\Pi \mathsf{T}^e_{n,x}\right)\right)\label{eqn:thetapsi}\\
\Phi(\Theta_{\phi}) &=& \left(\prod_{x=1}^{2^{n-1}-1} \left(\Pi \mathsf{T}^o_{n,x}\right) M^o_x \left(\Pi \mathsf{T}^o_{n,x}\right)\right). \label{eqn:thetaphi}
\end{eqnarray} Then note that $\zeta(\Theta_{\zeta})$ is the product of exponentials of all diagonal SRBB elements, $\Psi(\Theta_{\psi})$ is the product of matrices of type $M_nZYZ$ and permutation scaling of $M_nZYZ$ type matrices, and $\Phi(\Theta_{\phi})\in \Sf\Uf(2^n)$ is product of unitary block diagonal matrices, which we will use in the construction of the circuits for these matrices in Section \ref{sec:circuit}. Then we propose a quantum neural network framework \cite{benedetti2019} for approximating a unitary matrix as follows. Given $U\in\Sf\Uf(2^n)$ approximate $U$ as \begin{equation}\label{def:uapprox}
    U = \prod_{l=1}^L \zeta(\Theta_{\zeta}^{(l)})\, \Psi(\Theta_{\psi}^{(l)})\, \Phi(\Theta_{\phi}^{(l)}) 
\end{equation}
where $l$ is called the layer and we call the equation (\ref{def:uapprox}) is called the $L$-layer approximation of $U$ with 
\begin{eqnarray}
\Theta_{\zeta}^{(l)} &=& \left\{\theta_{j^2-1}^{(l)} \,|\, 2\leq j\leq 2^n\right\}, \label{eqn:lthetazeta}\\
\Theta_{\psi}^{(l)} &=& \left\{\theta_{h_{\beta}(\alpha-1)}^{(l)},  \theta_{f_{\beta}(\alpha-1)}^{(l)}, \theta_{h_{\beta-1}(\alpha)}^{(l)}, \theta_{f_{\beta-1}(\alpha)}^{(l)}, \theta_{(2j-1)^2}^{(l)}, \theta_{4j^2-2j}^{(l)} \,|\, 1\leq j\leq 2^{n-1}, \right. \nonumber \\ && \left. (\alpha,\beta)\in T^e_x, 1\leq x\leq 2^{n-1}-1\right\}, \label{eqn:lthetapsi} \\
\Theta_{\phi}^{(l)} &=& \left\{\theta_{h_{\beta}(\alpha-1)}^{(l)},  \theta_{f_{\beta}(\alpha-1)}^{(l)}, \theta_{h_{\beta+1}(\alpha)}^{(l)}, \theta_{f_{\beta+1}(\alpha)}^{(l)} \,|\, (\alpha,\beta)\in T^o_x, 1\leq x\leq 2^{n-1}-1 \right\}. \hspace{2.1em}\label{eqn:lthetaphi}
\end{eqnarray}

It may seem from the equation (\ref{def:uapprox}) that we can change the ordering of making the product of the matrices $\zeta(\Theta_{\zeta}^{(l)}), \Psi(\Theta_{\psi}^{(l)}), \Phi(\Theta_{\phi}^{(l)}),$ which is indeed possible. However, from the perspective of design of quantum circuits for $U$ in order to reduce the count of $\cnot$ gates, this choice of ordering facilitates the nullification of effects of certain $\cnot$ gates while considering this ordering. For instance, see Section \ref{Sec:2qqc}.

\begin{algorithm}[H]
\caption{Modified Algorithm for Approximating $2^n\times 2^n$ special unitary matrix}\label{alg2}
\textbf{Provided:} $U_1\in \Sf\Uf(2^n)$, $U^{(2^n)}_j \in \boldsymbol{\mathcal{U}}^{(2^n)}, 1\leq j\leq 2^{2n}-1,$ $\zeta(\Theta_{\zeta}),$ $\Psi(\Theta_{\psi}),$ $\Phi(\Theta_{\phi})$ given by equation (\ref{eqn:thetazeta}) - (\ref{eqn:thetaphi}).

\textbf{Input:} $\Theta_{\zeta}^{(0)}$, $\Theta_{\psi}^{(0)},$ $\Theta_{\phi}^{(0)},$ $\epsilon>0$ 
 \\
\textbf{Output:} $A=\prod_t \zeta(\Theta^{(t)}_{\zeta}) \Psi(\Theta^{(t)}_{\psi}) \Phi(\Theta^{(t)}_{\phi})$ such that $\|U-A\|_F\leq \epsilon$ 
\begin{algorithmic}

\Procedure{}{Unitary Matrix $U$}       \Comment{}
\State $A\rightarrow I$
\For{$t=1;t++$}
\State Use an optimization method like Nelder-Mead/Powell's or Gradient descent method to find $\Theta_{\zeta}^{(t)}$, $\Theta_{\psi}^{(t)},$ $\Theta_{\phi}^{(t)}$ such that $$\min_{\Theta_{\zeta}^{(t)},\Theta_{\psi}^{(t)},\Theta_{\phi}^{(t)}} \left\|U-\zeta(\Theta^{(t)}_{\zeta}) \Psi(\Theta^{(t)}_{\psi}) \Phi(\Theta^{(t)}_{\phi})\right\|_F=\epsilon_t$$

\If{$\epsilon_t\leq \epsilon$}
\State \textbf{Break}
\State $A \rightarrow A\zeta(\Theta^{(t)}_{\zeta}) \Psi(\Theta^{(t)}_{\psi}) \Phi(\Theta^{(t)}_{\phi})$
\Else
\State $U_{t+1}\rightarrow U_{t}A^*$\\
\EndIf
\State \textbf{End}
\EndFor
\State \textbf{End}
\State \textbf{End Procedure}
\EndProcedure
\end{algorithmic}
\end{algorithm}
\subsection{Numerical simulations}
In this section, we report the performance of the proposed algorithms for approximating unitary matrices through product of exponentials of the proposed RB basis elements in optimal ordering. We have considered several unitary matrices sampled from the Haar distribution and the standard well-known unitaries for two, three and four qubits. Given a target unitary matrix, the initial choice of the parameters can influence the output unitary matrix and since the objective function is non-convex, the optimal approximated values of the parameters may lead to a local minimum. Thus we generate multiple random points from uniform distribution and normal distribution for the set of parameters $\Theta=\{\theta_1,\hdots,\theta_{2^{2n}-1}\},$ where $0 \leq \theta_j\leq 2\pi, 1\leq j\leq 2^{2n}-1$ and execute the proposed algorithms. Finally, we report the error that is least among all those initial parameter values. From our simulations, we sampled $600$ random unitary matrices and we observe that the error lies mostly between $10^{-12}$ to $10^{-15}$ except at a few cases where the error is of the order $10^{-4}.$

We compare our findings with the results found in \cite{krol2022,vidal2004,kraus2001} and see that our method for $2$-qubits is faster as it does not need to perform singular value decomposition. Like \cite{vidal2004,kraus2001}, we don't need to convert the target matrices into magic basis/states and perform Schmidt decomposition in order to check for separable states which is non-trivial and time consuming. We have also seen that employing the modified ordering of the proposed basis elements and decomposing a $2$-qubit gate using the original ordering of the basis elements with $(\prod_{j=1}^{15}\exp{(\iota  \theta_i U^{(4)}_i)})^L$ with $L=1$ and applying Algorithm \ref{alg2}, the error is same. We have performed Algorithm \ref{alg2} on MATLAB and Python 3.0 on a system with $16GB$ RAM, Intel(R) Core(TM) $i5-1035G1$ CPU $@ 1.00GHz 1.19 GHz$ for $2$-qubit and $3$-qubit examples. For $4$-qubit examples, we have performed the simulations using supercomputer PARAM Shakti of IIT Kharagpur.   

We mention that the issue with the methods described in  \cite{krol2022,kraus2001} lies in the fact that calculating $\zeta_k,U_A,U_B$ such that $U_A\otimes U_Be^{\iota \zeta_k}\ket{\psi_k}=\ket{\phi_k}$ is a non-trivial process and we have verified the calculation for a handful of `easy' matrices. However, for generic matrices, the process is difficult. Among the synthesized $600$ $2$-qubit Haar random unitary matrices, $300$ are used for simulating with original ordering and $300$ using modified ordering of the SRBB elements. We have use Nelder-Mead  method of minimization in our algorithm. The method proposed in \cite{vidal2004,kraus2001} gives us results with errors of order $10^{-15}$ however, it is more time consuming compared to our method since, one has to be aware of the unitary matrix in order to convert its eigenvalues into magic basis states. So the problem has to be tackled individually for each unitary matrix. For our proposed method however, one need not even know about the unitary matrix and we can reach our result. 

\begin{figure}[H]
\begin{subfigure}{.5\textwidth}
  \centering
  \includegraphics[width=0.8\linewidth]{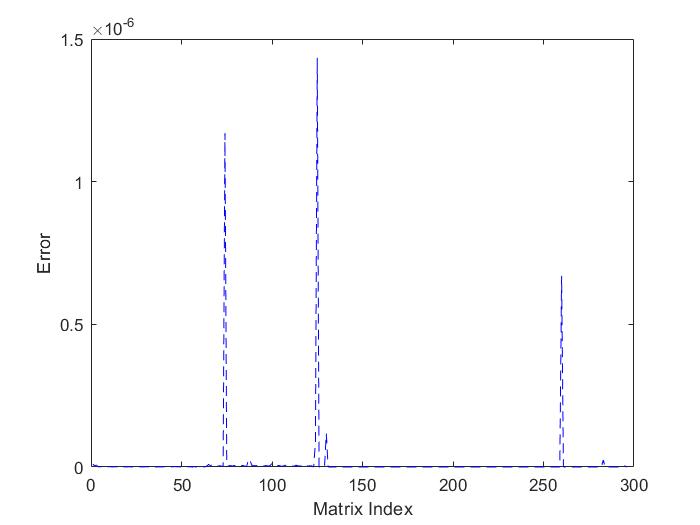}  
  \caption{Error corresponding to original ordering }
  \label{fig:sub-first}
\end{subfigure}
\begin{subfigure}{.5\textwidth}
  \centering
  \includegraphics[width=0.8\linewidth]{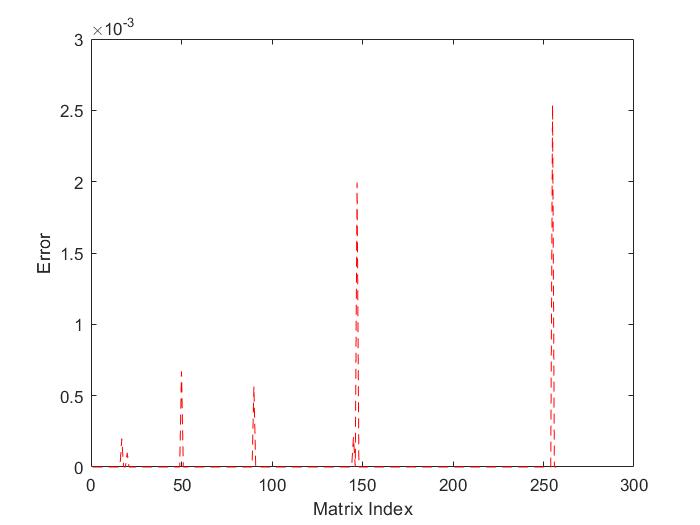}  
  \caption{Error corresponding to the modified ordering }
  \label{fig:sub-second2}
\end{subfigure}
\caption{Errors in the Frobenius norm using Algorithm \ref{algo1} (a) and Algorithm \ref{alg2} (b), considering original and modified SRB basis elements for the decomposition of $2$-qubit unitary matrices sampled from Haar distribution.}\label{fig:error2qubit}
\label{fig:randerror}
\end{figure}

\begin{figure}[H]
  \centering
  \includegraphics[width=0.6\linewidth]{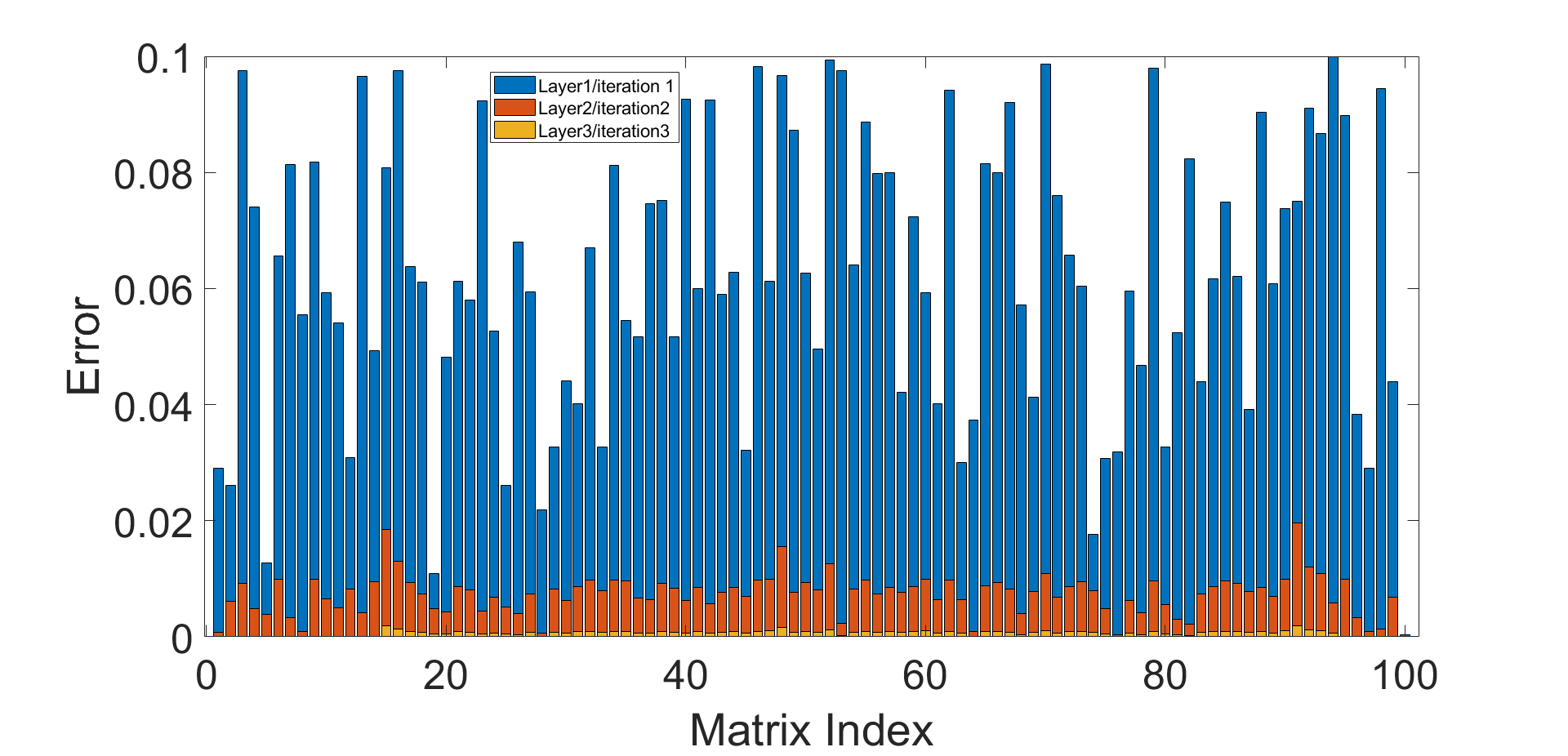}  
  \caption{The errors obtained from up to three iterations (layers) for approximating $3$-qubit Haar random unitaries. The error after $3$rd iteration lies between $10^{-4}$ to $10^{-6}.$}
  \label{fig:error3qubit}
\end{figure}

\begin{figure}[H]
\begin{subfigure}{.5\textwidth}
  \centering
  \includegraphics[width=1.0\linewidth]{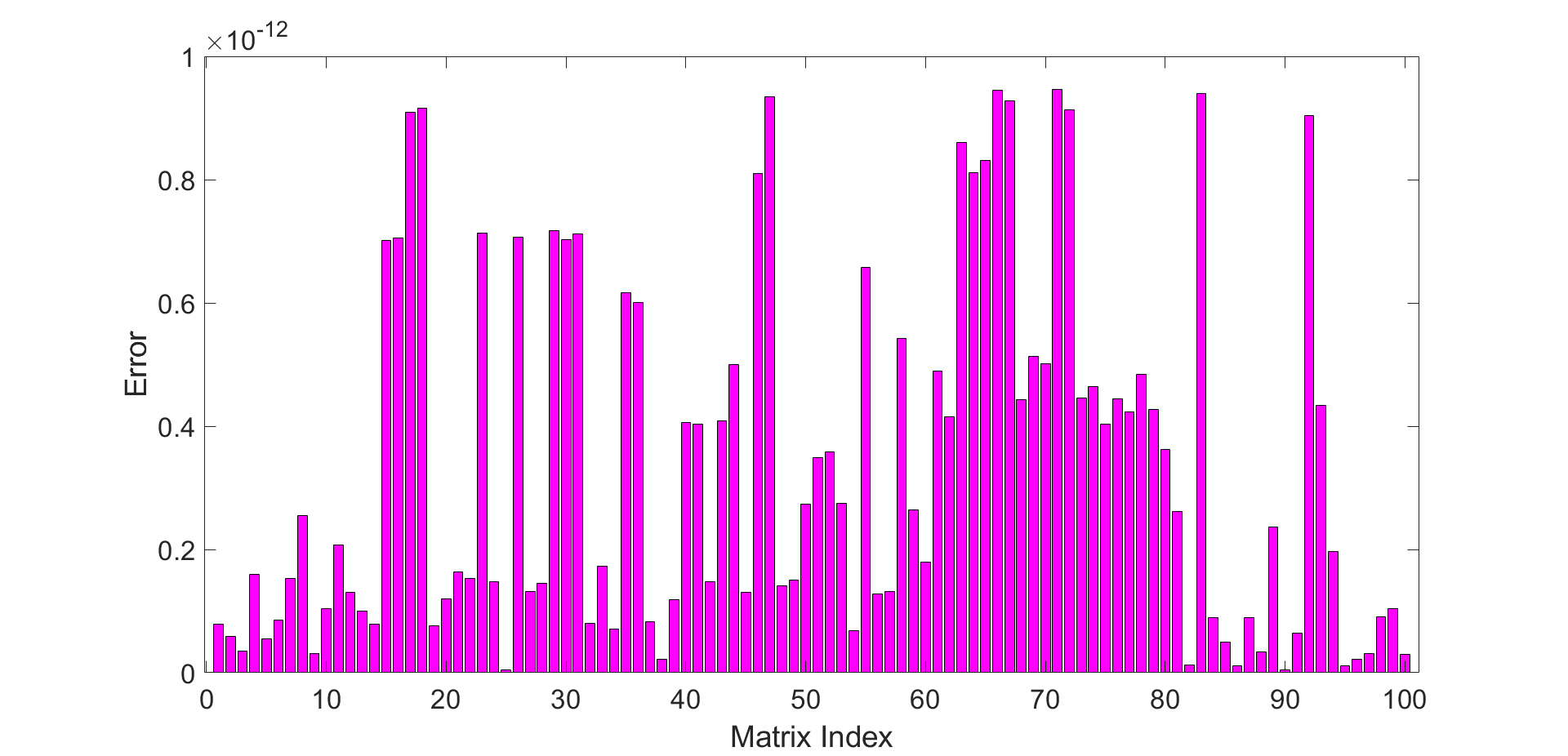}  
  \caption{Errors for $4$-sparse unitaries wrt original ordering}
  \label{fig:sub-first2}
\end{subfigure}
\begin{subfigure}{.5\textwidth}
  \centering
  \includegraphics[width=1.0\linewidth]{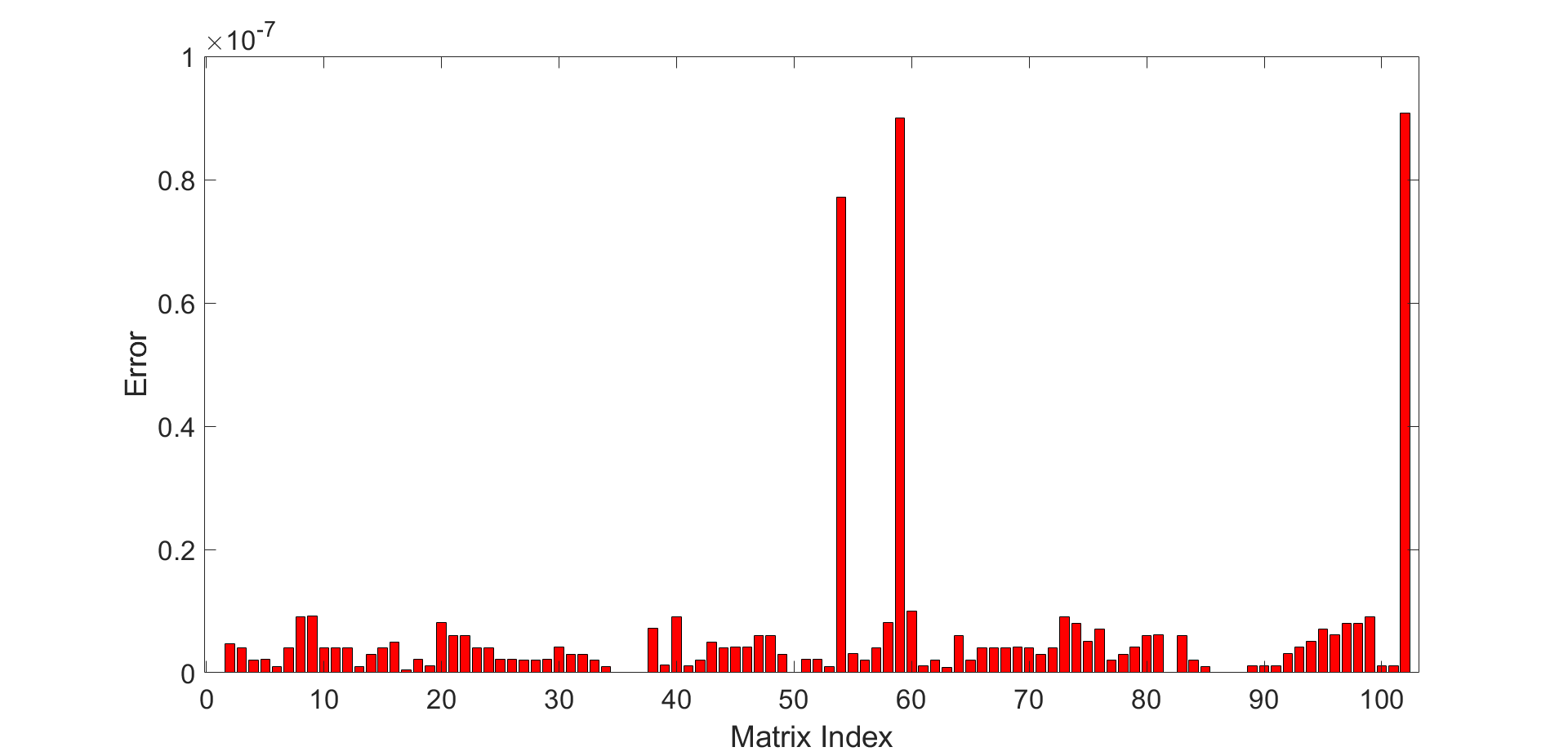}  
  \caption{Errors for $4$-sparse unitaries wrt modified ordering}
  \label{fig:sub-second3}
\end{subfigure}
\caption{Errors in Frobenius norm for approximating random $4$-sparse and $6$-sparse $3$-qubit unitaries with two ordering of the SRBB, considering only one iteration of the algorithm.}
\label{fig:error3qubit2}
\end{figure}

\begin{figure}[H]
\begin{subfigure}{.5\textwidth}
  \centering
  \includegraphics[width=1.0\linewidth]{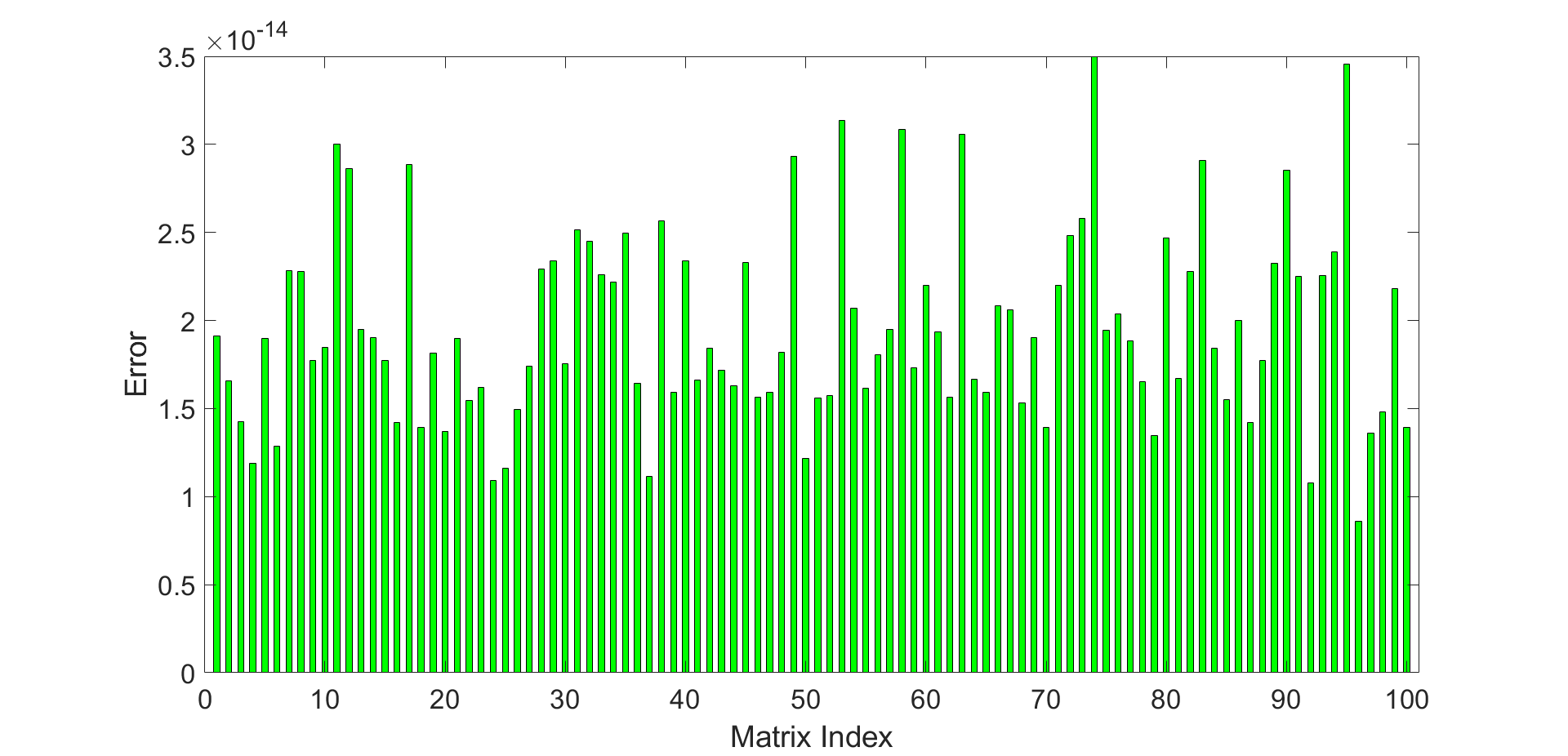}  
  \caption{Errors for unitaries of order $3$}
  \label{fig:sub-first3}
\end{subfigure}
\begin{subfigure}{.5\textwidth}
  \centering
  \includegraphics[width=1.0\linewidth]{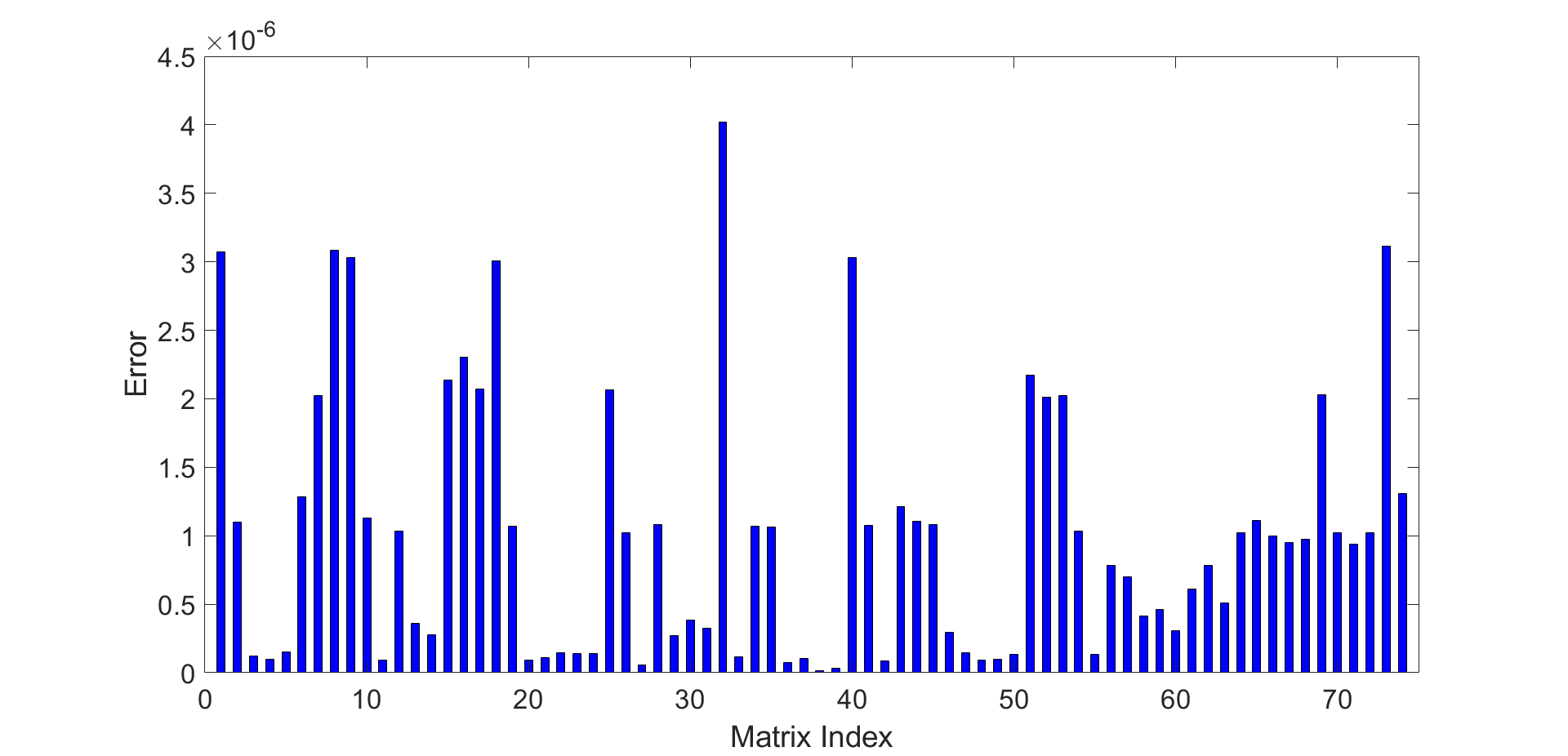}  
  \caption{Errors for unitaries of order $5$}
  \label{fig:sub-second}
\end{subfigure}
\caption{Error for approximating unitaries of order $3$ and $5$ using Algorithm \ref{algo1} up to one iteration. The unitary matrices are sampled at random from Haar distribution and Nelder-Mead is employed for optimization.}
\label{fig:error35qubit}
\end{figure}

\begin{figure}[H]
  \centering
  \includegraphics[width=0.7\linewidth]{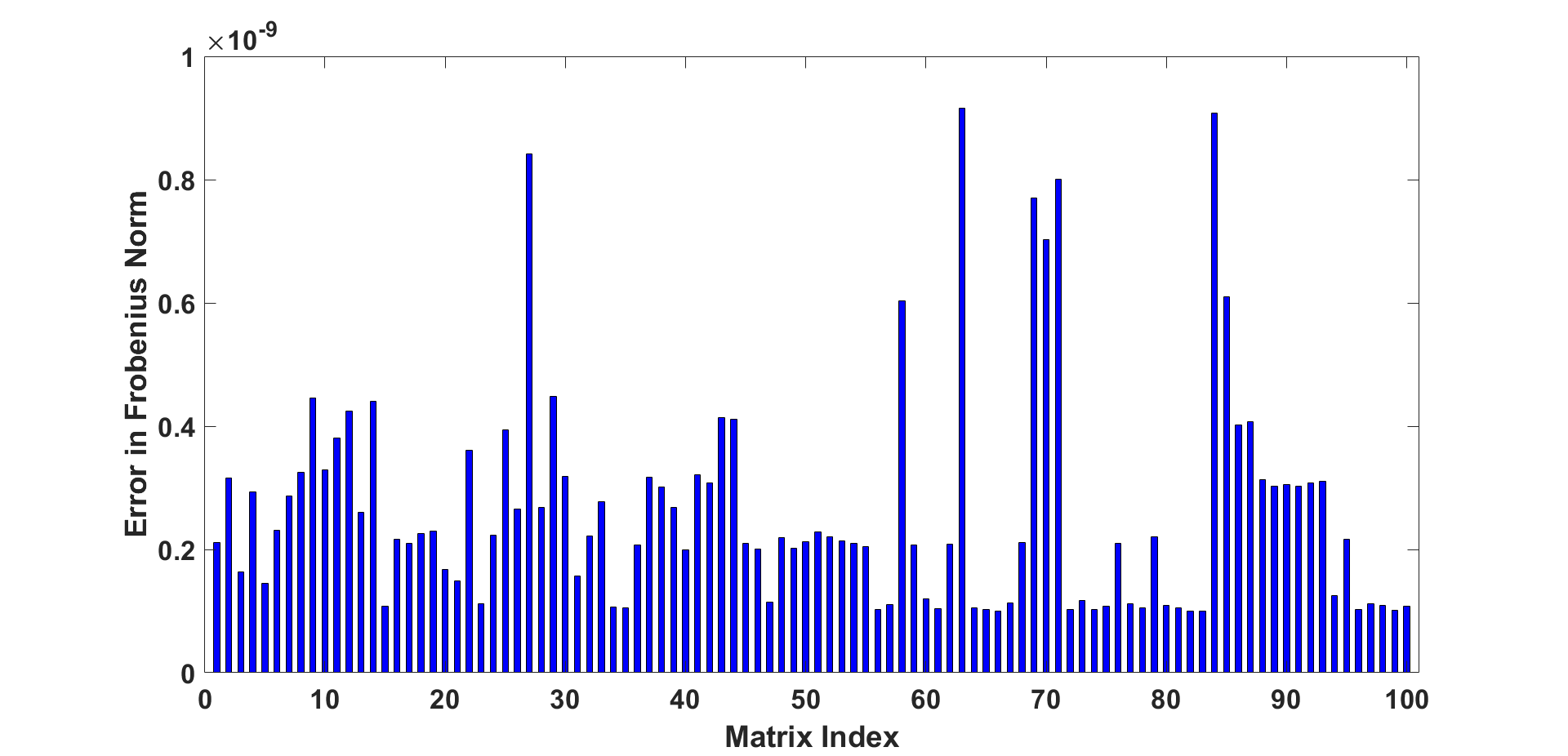}  
\caption{Errors for approximating Haar random $8$-sparse $4$-qubit block-diagonal unitaries considering only one iteration of the Algorithm \ref{alg2}.}
\label{fig:error3qubit3}
\end{figure}

\begin{figure}[H]
\begin{subfigure}{.5\textwidth}
  \centering
  \includegraphics[width=1.0\linewidth]{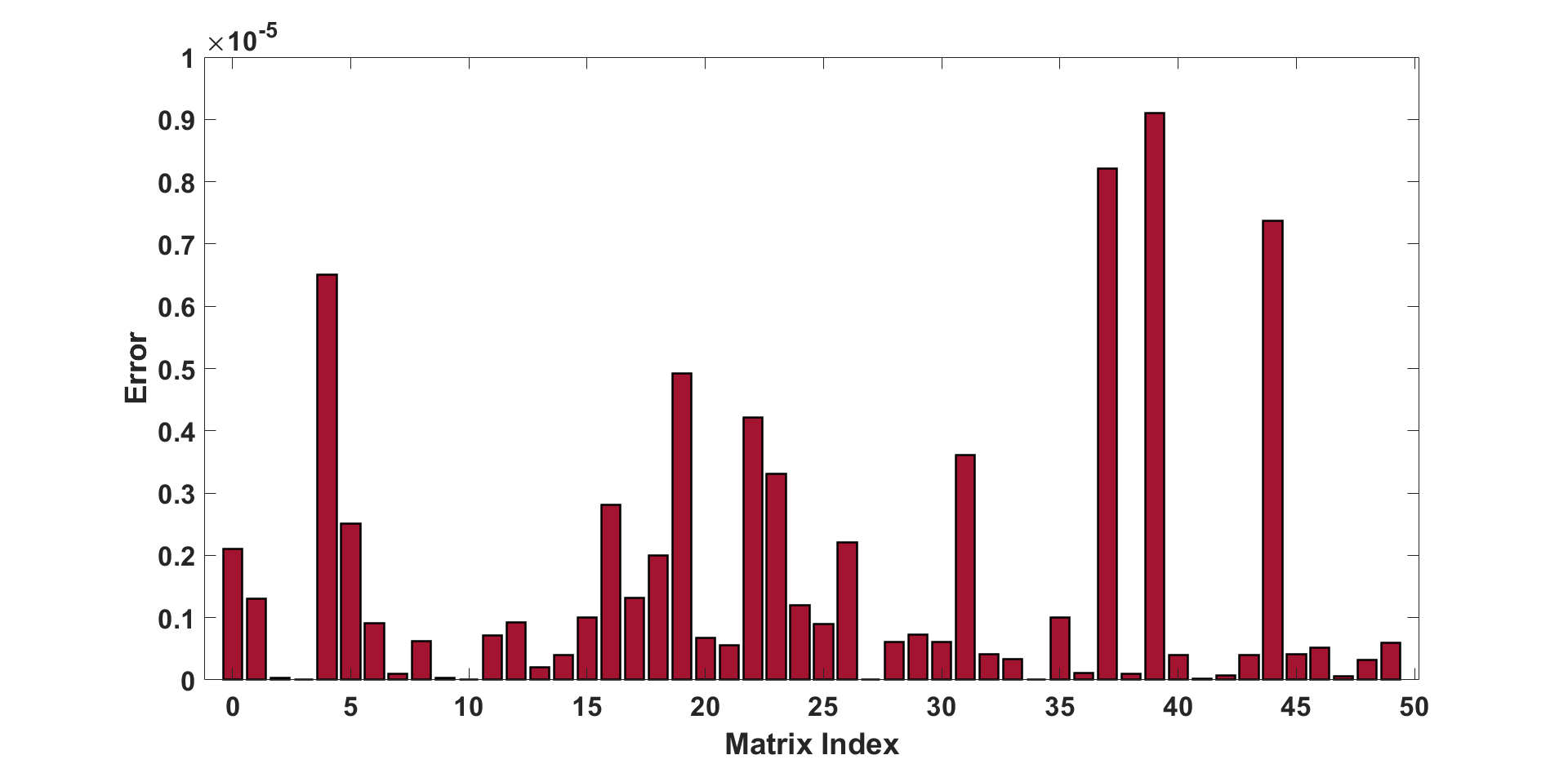}  
  \caption{Errors for approximating $5$-qubit unitaries}
  \label{fig:sub-first5}
\end{subfigure}
\begin{subfigure}{.5\textwidth}
  \centering
  \includegraphics[width=1.0\linewidth]{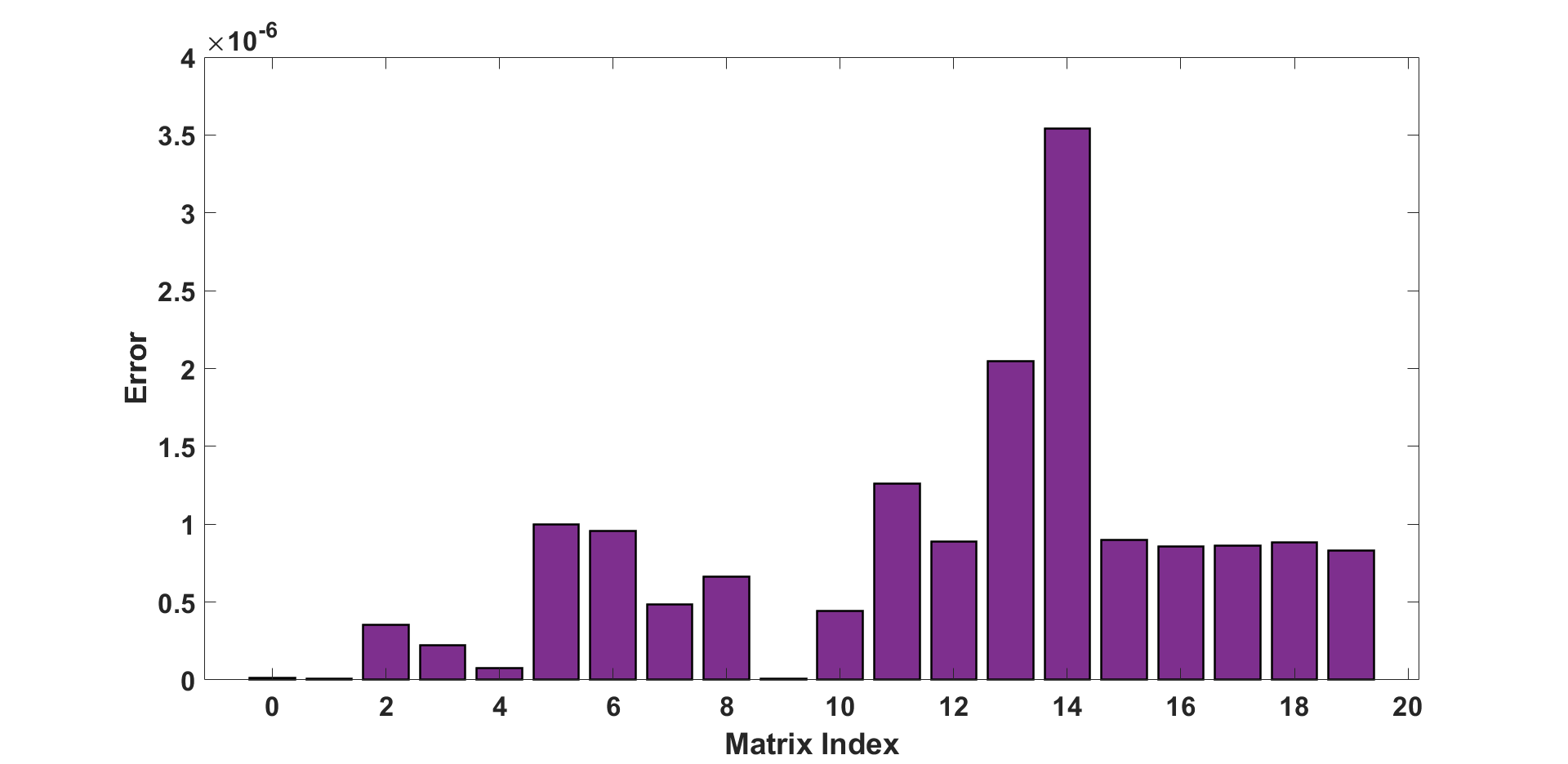}  
  \caption{Errors for approximating $6$-qubit unitaries}
  \label{fig:sub-second6}
\end{subfigure}
\caption{Error for approximating $5$ and $6$ qubit Haar random unitary matrices using Algorithm \ref{alg2} coupled with Nelder-Mead optimization method.}
\label{fig:error56qubit}
\end{figure}

We report the error and time taken for approximating certain standard $2$-qubit unitaries in Table \ref{Table:error2q}, the errors for simulating random $600$ unitary matrices are provided in \textbf{Figure} \ref{fig:randerror}, both are obtained by setting $L=1.$ We consider several standard unitaries and $100$ random unitaries sampled from Haar distribution for $3$-qubit systems. In table \ref{table:error3q} we report the error for the standard unitaries, and the errors for random unitaries are plotted in \textbf{Figure} \ref{fig:error3qubit} considering one, two and three iterations. Further, we generate a hundred $3$-qubit Haar random unitaries that are $4$-sparse and $6$-sparse. The $4$-sparse unitaries contain two blocks of order $4$ and their permutations whereas the $6$-sparse unitaries contain two blocks of order $6$ and $2$ and their permutations. The errors for these unitaries using Algorithm \ref{alg2} lies between $10^{-12}$ and $10^{-7}$ for up to one iteration/layer. The corresponding errors are depicted in \textbf{Figure} \ref{fig:error3qubit2}. Next, we consider certain standard $4$-qubit unitaries and report the error in Table \ref{table:error4q}. It is to be noted that our algorithm works better if the unitary matrix is sparse. We speculate that this is due to the performance of the optimization algorithms which need not perform well for large search spaces. We have also tested our algorithm for $5-8$ qubit systems respectively as seen in Table \ref{table:error7q}. However, we report that our program based off the proposed fails to produce results for approximating dense $8$-qubit unitaries and above while running for more than $70$ hours. Even for approximating many dense $7$-qubit unitaries, the program fails to produce results after $70$ hours. We believe this phenomenon is observed due to an exponential increase in the number of parameters as the number of qubits increase. Hence, reducing the number of parameters while increasing error at a manageable rate remains a problem for future. For sparse matrices, one can exploit the sparsity pattern of the matrix in order to get rid of redundant parameters but for dense matrices, parameter reduction still remains a primary challenge. For sparse $8$-qubit unitaries, we have obtained some results using the proposed approximation algorithm (see Table \ref{table:error7q}). However, the time taken for approximating sparse $8$-qubit matrices turned out to be about $65$ hours using sparse matrix packages (In the previous version of the work, the algorithm failed to produce any results fot $8$-qubit systems). We have performed approximation for $100$ unitaries of order three and five i.e. for unitaries which define evolution of three and five dimensional quantum states. The error are obtained after approximating the Haar random unitaries and employing Algorithm \ref{algo1} in \textbf{Figure} \ref{fig:error35qubit}. In \textbf{Figure} \ref{fig:error3qubit3}, we have approximated $100$ unitary block-diagonal $8$-sparse $4$-qubit unitaries. For this class of matrices, the error of approximation ranged from $10^{-10}$ to $10^{-7}$. In \textbf{Figure} \ref{fig:sub-first5} and \textbf{Figure} \ref{fig:sub-second6}, we have respectively chosen $50$ Haar random $5$-qubit unitaries and $20$ Haar random $6$-qubit unitaries which are to be approximated. The unitaries are $4$ and $2$ sparse respectively. The unitaries are block-diagonal in nature with non-trivial diagonal blocks. The errors for these unitaries using Algorithm \ref{alg2} lies between $10^{-10}$ and $10^{-6}$. It is also to be observed that the the program struggles to implement Nelder-Mead efficiently due to an exponential increase in the number of parameters, which, in turn, leads to an increase in error magnitude as the number of qubit increases. 
\begin{center}
\begin{table}
\footnotesize
\begin{tabular}{ |c|c|c|c| }
\hline
\textbf{Matrix} &\textbf{Time taken}  & \textbf{Error from our method}& \textbf{Error from}\cite{vidal2004}   \\
 &\textbf{in seconds} && \textbf{circuit + our method}\\
 &\textbf{in our method}&&\\
\hline \hline
$\cnot$&24 &$7.03793017 \times 10^{-14}$ & $3.9\times10^{-15}$\\
\hline
$Grover_2$&  13 & $9.87612962\times 10^{-14}$  & $1.72\times10^{-14}$\\
\hline
XX&  12& $7.33016345\times 10^{-15}$& $9.4\times10^{-13}$\\
\hline
YY&  39&$6.24698228\times 10^{-14}$ & $3.5\times10^{-14}$\\
\hline
ZZ  & 13&$6.22407276\times 10^{-14}$& $8.34\times10^{-14}$\\
\hline
SWAP  & 23&$6.15361435\times 10^{-13}$ & $3.6\times10^{-15}$\\
\hline
XZ & 28& $8.07143891\times 10^{-14}$ & $7.62\times10^{-13}$\\
\hline
ZX & 14& $3.40555621\times 10^{-14}$ &$6.91\times10^{-13}$\\
\hline
ZY& 28 & $3.36666967\times 10^{-13}$ & $5.32\times10^{-14}$\\
\hline
$\cnot_{(2,1)}$& 04& $2.12476637\times 10^{-14}$ & $1.36\times10^{-13}$\\
\hline
DCNOT & 24& $4.31202055\times 10^{-14}$ & $8.2\times10^{-14}$\\
\hline
XNOR & 15& $5.70538776\times 10^{-14}$ & $6.22\times10^{-14}$\\
\hline
iSWAP& 36 & $9.73113534\times 10^{-14}$ &$4.78\times10^{-13}$\\
\hline
fSWAP & 26& $1.64656376\times 10^{-13}$ & $5.83\times10^{-13}$\\
\hline
C-Phase& 10& $3.17597256\times 10^{-13}$ & $7.1\times10^{-14}$\\
\hline
XY & 22& $2.14722235\times 10^{-14}$ & $6.65\times10^{-13}$\\
\hline
$\sqrt{SWAP}$ & 22& $2.24302075\times 10^{-13}$ & $8.51\times10^{-13}$\\
\hline
$\sqrt{iSWAP}$ &28 & $8.22872467\times 10^{-15}$ & $6.18\times10^{-14}$\\
\hline
$QFT_2$& 42 & $5.11674305\times 10^{-13}$ & $7.83\times10^{-13}$\\
\hline
\end{tabular}\normalsize
\caption{Error and time for simulating standard $2$-qubit unitaries}
\label{Table:error2q}
\end{table}
\end{center}
The execution time for approximating the target unitaries described in Table \ref{Table:error2q} are significantly improved compared to our previous simulation that we reported in the earlier (conference) version of this paper\cite{SarmaSarkar2023}. The algorithms are implemented in Python 3.0 and the run time for approximating several unitary matrices of order $2^2$ given by $XX,YY,ZZ,ZX,\cnot_{(2,1)}$ and the phase gates are extremely fast. The justification of this phenomena lies in the fact that we have exploited the sparsity pattern of these matrices mentioned and selected a list of basis elements for the approximation as given by Table \ref{Table:error2qbasis}. We employ the Nelder-Mead method as the optimization methods to determine the values of the parameters, however, we observe that using Powell's method also produces a similar result. The choice of the initial values of the parameters is decided by a randomization techniques as follows. We generate multiple random points ($10$ to $100$) from normal distribution for the set of parameters lie in the interval $[0, 2\pi)$. Further, the algorithm is stopped immediately when the objective function goes below our specified threshold for the error bound $(\epsilon\leq 5\times 10^{-12})$ in order to account for fast run time. We speculate that the run time can be improved further in a system having a better configuration than ours. 
\vspace{-2em}
\begin{center}
\begin{table}
\begin{tabular}{ |c|c| }
\hline
 \textbf{Matrix} & \textbf{$2$-qubit Basis elements chosen from SRBB along with Identity matrix $I_4$} \\
 
\hline \hline
$\cnot$& $I_4,U_3^{(4)},U_8^{(4)},U_9^{(4)},U_{12}^{(4)},U_{15}^{(4)}$ \\
\hline
$Grover_2$&  all \\
\hline
XX& $I_4,U_3^{(4)},U_4^{(4)},U_6^{(4)},U_8^{(4)},U_9^{(4)},U_{10}^{(4)},U_{12}^{(4)},U_{13}^{(4)},U_{15}^{(4)}$ \\
\hline
YY&  $I_4,U_3^{(4)},U_4^{(4)},U_6^{(4)},U_8^{(4)},U_9^{(4)},U_{10}^{(4)},U_{12}^{(4)},U_{13}^{(4)},U_{15}^{(4)}$\\
\hline
ZZ  & $U_3^{(4)},U_8^{(4)},U_{15}^{(4)}$\\
\hline
SWAP  & $I_4,U_3^{(4)},U_4^{(4)},U_6^{(4)},U_8^{(4)},U_{15}^{(4)}$\\
\hline
XZ & $I_4,U_3^{(4)},U_4^{(4)},U_5^{(4)},U_7^{(4)},U_8^{(4)},U_9^{(4)},U_{11}^{(4)},U_{12}^{(4)},U_{14}^{(4)},U_{15}^{(4)}$\\
\hline
ZX & $I_4,U_1^{(4)},U_2^{(4)},U_3^{(4)},U_8^{(4)},U_9^{(4)},U_{12}^{(4)},U_{15}^{(4)}$\\
\hline
ZY& $I_4,U_1^{(4)},U_2^{(4)},U_3^{(4)},U_8^{(4)},U_9^{(4)},U_{12}^{(4)},U_{15}^{(4)}$\\
\hline
$\cnot_{(2,1)}$& $I_4,U_3^{(4)},U_8^{(4)},U_{11}^{(4)},U_{14}^{(4)},U_{15}^{(4)}$\\
\hline
DCNOT & $U_1^{(4)},U_2^{(4)},U_3^{(4)},U_4^{(4)},U_6^{(4)},U_8^{(4)},U_9^{(4)},U_{11}^{(4)},U_{12}^{(4)},U_{14}^{(4)},U_{15}^{(4)}$\\
\hline
XNOR & $U_1^{(4)},U_2^{(4)},U_3^{(4)},U_8^{(4)},U_{15}^{(4)}$\\
\hline
iSWAP& $I_4,U_3^{(4)},U_4^{(4)},U_6^{(4)},U_8^{(4)},U_{15}^{(4)}$\\
\hline
fSWAP & $I_4,U_3^{(4)},U_4^{(4)},U_6^{(4)},U_8^{(4)},U_{15}^{(4)}$\\
\hline
C-Phase& $I_4,U_3^{(4)},U_8^{(4)},U_{15}^{(4)}$\\
\hline
XY & $I_4,U_3^{(4)},U_4^{(4)},U_6^{(4)},U_8^{(4)},U_9^{(4)},U_{10}^{(4)},U_{12}^{(4)},U_{13}^{(4)},U_{15}^{(4)}$\\
\hline
$\sqrt{SWAP}$ & $I_4,U_3^{(4)},U_4^{(4)},U_6^{(4)},U_8^{(4)},U_{15}^{(4)}$\\
\hline
$\sqrt{iSWAP}$ &$I_4,U_3^{(4)},U_4^{(4)},U_6^{(4)},U_8^{(4)},U_{15}^{(4)}$ \\
\hline
$QFT_2$& all \\
\hline
\end{tabular}\normalsize
\caption{List of SRBB elements that are used for implementing the approximation algorithms}
\label{Table:error2qbasis}
 \end{table}\end{center}
\begin{center}
\begin{table}
\begin{tabular}{ |c|c|c|c|c|c|c|c| }
\hline
\textbf{Matrix} & \textbf{1st iteration Error}  & \textbf{QFAST}  & \textbf{UniversalQ}  & \textbf{Search Compiler} \\
&\textbf{from our method} & \textbf{+ KAK} \cite{younis2020}&\cite{younis2020}&\cite{younis2020} \\
\hline \hline
Toffoli&$4.48 \times 10^{-9}$ & $1.5\times 10^{-6}$ & $2.6\times 10^{-8}$& $2.4\times 10^{-7}$\\
\hline
Fredkin& $1.6 \times 10^{-8}$ & $2.2\times 10^{-6}$&$0$&$5.8\times 10^{-6}$\\
\hline
$Grover_3$& $4.602 \times 10^{-9}$ & $8.1\times 10^{-7}$&$0$&$5.5\times 10^{-7}$\\
\hline
Peres&$2 \times 10^{-8}$ & $6.8\times 10^{-7}$&$2.1\times 10^{-8}$&$2.3\times 10^{-7}$\\
\hline
$QFT_3$&$ 3.1 \times 10^{-9}$& $3\times 10^{-7}$&$3\times 10^{-8}$&$4.9\times 10^{-7}$\\
\hline
\end{tabular}\normalsize
\caption{Error in the Frobenius norm after simulation using one iteration/layer for $3$-qubit standard unitaries}
\label{table:error3q}
\end{table}
\end{center}    
    \begin{center}
\begin{table}
\begin{tabular}{|c|c|c|c|c|c|c|c|}
\hline
\textbf{Matrix} & \textbf{1st iteration Error}  & \textbf{QFAST} & \textbf{QFAST + UQ} & \textbf{UniversalQ} \\
& \textbf{from our method}&\textbf{+ KAK} \cite{younis2020}&\cite{younis2020}&\cite{younis2020} \\
\hline \hline
CCCX &$1.97 \times 10^{-8}$ &  $2.2\times 10^{-5}$ & $1.3\times 10^{-6}$&$2.1\times 10^{-8}$\\
\hline
$Grover_4$& $2.12 \times 10^{-9}$ &  $-$ & $-$&$-$\\
\hline
$QFT_4$&$ 9.331 \times 10^{-8}$ &  $7.9\times 10^{-7}$ & $8.5\times 10^{-7}$&$3.9\times 10^{-8}$\\
\hline
\end{tabular}\normalsize
\caption{Error in the Frobenius norm after simulation using one iteration/layer for $4$-qubit standard unitaries}
\label{table:error4q} 
\end{table}    
\end{center}


\begin{center}
\begin{table}
\begin{tabular}{|c|c|c|c|c|c|c|c|}
\hline
\textbf{Matrix} & \textbf{1st iteration Error} \\
& \textbf{from our method} \\
\hline \hline
$Grover_5$ &$ 3.82\times 10^{-7}$ \\

($5$-qubits) &  \\
\hline
$2$-sparse & $8.29 \times 10^{-8}$ \\
Generalized Toffoli &  \\
($6$-qubits) &  \\
\hline

$X^{\otimes 6}\otimes Y$ & $1.55 \times 10^{-7}$ \\
($7$-qubits) &  \\
\hline

$Z^{\otimes 7}\otimes X$ & $5.81 \times 10^{-8}$ \\
($8$-qubits) &  \\
\hline
\end{tabular}\normalsize
\caption{Error in the Frobenius norm after simulation using one iteration/layer for some known $5,6,7$ and $8$-qubit standard unitaries}
\label{table:error7q} 
\end{table}    
\end{center}

\section{Quantum circuit representation of unitary matrices of order $2^n$}\label{sec:circuit}

In the previous section, we have introduced a modified ordering while multiplying for approximation of unitary matrices for $n$-qubit systems. The modified ordering is introduced to incorporate a structure for approximating a target unitary through product of permutation matrices, $M_nZYZ$ type matrices, and block diagonal matrices when we write a given target unitary as product of exponentials of SRBB elements. Further, we provided a neural network framework for bettering the approximation, where a layer is one iteration of the Algorithm \ref{algo1}. Thus, in order to provide a quantum circuit representation of the unitary matrices, we need to provide quantum circuit representation of permutation matrices, which are product of transpositions of particular type, and $M_nZYZ$ matrices, which are block matrices with each block is a special unitary matrix of order $2,$ and block diagonal unitary matrices with blocks are of size $2.$ Below, we discuss circuit construction for each of these structured matrices.   
\subsection{Quantum circuits for product of transpositions}

Now, we construct quantum circuit for the matrix $\Pi\mathsf{T}^g_{n,x},$ $1\leq x\leq 2^{n-1}-1, g\in \{e,o\}$.  First consider an $n$-qubit quantum circuit consisting of only $(\cnot)_{(n,i)},$ $1\leq i\leq n-1$ gates as follows. {Let us choose $x\in\{0,\hdots,2^{n-1}-1\}$ with its binary representation $(x_{n-2},\hdots,x_0)$ such that $x=\sum_{j=0}^{n-2} x_j2^{j}, x_{j}\in\{0,1\} $, we define a circuit in the following way. For each $x_j,$ $0\leq j\leq n-2$ the circuit contains a $\cnot_{(n,n-j-1)}$ gate if the $x_j=1,$ where $\cnot_{(n,n-j-1)}$ denotes a $\cnot$ gate with $n$-th qubit as the control and $(n-j-1)$-th qubit as target. Since, any $\cnot$ gate represents a permutation matrix, we redefine the permutation matrices $\Pi \mathsf{T}^{g}_{n,x},g\in \{e,o\},1\leq x\leq 2^{n-1}-1$, introduced in equation (\ref{eqn:pieox}). These permutations are product of $2^{n-2}$ disjoint permutations and are heavily used in the later paper as well.}  
We denote \begin{eqnarray}\label{permusch1}
    \Pi \mathsf{T}^{e}_{n,x}&=&\prod_{j=0}^{n-2}  (\cnot_{(n,n-j-1)})^{\delta_{1,x_j} } \\
    \Pi \mathsf{T}^o_{n,x}&=&(\cnot_{(n-m-1,n)})(\Pi \mathsf{T}^e_{n,x})(\cnot_{(n-m-1,n)}) \nonumber
\end{eqnarray} {where $m$ is the greatest integer $0\leq m\leq n-2$ such that $x_{m}=1$ in the binary string of $x=(x_{n-2}\hdots x_0)$ i.e. $m=\mbox{max}\{j | \delta_{1,x_j}=1\}$ and $\delta$ denotes Kronecker delta function and $(\cnot_{(n,n-j-1)})^{0}$ is considered to be the Identity matrix. For $x=0$, we consider $\Pi \mathsf{T}^e_{n,x}$ and $\Pi \mathsf{T}^e_{n,x}$ as the Identity matrix i.e. absence of any $\cnot$ gates.}
For example, if $n=2$ and $x=1$ then the corresponding circuit is  

\begin{eqnarray}
    {{\Qcircuit @C=1em @R=.7em {&\lstick{1}&\qw&\targ &\qw&\qw&\qw\\
    &\lstick{2}&\qw&\ctrl{-1} &\qw&\qw&\qw\\}}}
\end{eqnarray}

For $n=3$, the circuits corresponding to $x=1,2,3$ are given by respectively.
\begin{eqnarray}
{\Qcircuit @C=1em @R=.7em {&\lstick{1}&\qw&\qw &\qw&\qw&\qw\\
    &\lstick{2}&\qw&\targ &\qw&\qw&\qw\\ &\lstick{3}&\qw&\ctrl{-1} &\qw&\qw&\qw\\} \hspace{0.3cm}
    \Qcircuit @C=1em @R=.7em {&\lstick{1}&\qw&\targ &\qw&\qw&\qw\\
    &\lstick{2}&\qw&\qw &\qw&\qw&\qw\\ &\lstick{3}&\qw&\ctrl{-2} &\qw&\qw&\qw\\} \hspace{0.3cm}
    \Qcircuit @C=1em @R=.7em {&\lstick{1}&\qw&\targ &\qw&\qw&\qw\\
    &\lstick{2}&\qw&\qw &\targ&\qw&\qw\\ &\lstick{3}&\qw&\ctrl{-2} &\ctrl{-1}&\qw&\qw\\}}    
\end{eqnarray}

Similarly, in $3$-qubit system, the circuits of $\Pi\mathsf{T}^o_1, \Pi\mathsf{T}^o_2$ and $\Pi\mathsf{T}^o_3$ are given by respectively 
\begin{eqnarray}
{\Qcircuit @C=1em @R=.7em {&\lstick{1}&\qw&\qw &\qw&\qw&\qw\\
    &\lstick{2}&\ctrl{1}&\targ &\ctrl{1}&\qw&\qw\\ &\lstick{3}&\targ&\ctrl{-1} &\targ&\qw&\qw\\} \hspace{0.3cm}
    \Qcircuit @C=1em @R=.7em {&\lstick{1}&\ctrl{2}&\targ &\ctrl{2}&\qw&\qw\\
    &\lstick{2}&\qw&\qw &\qw&\qw&\qw\\ &\lstick{3}&\targ&\ctrl{-2} &\targ&\qw&\qw\\} \hspace{0.3cm}
    \Qcircuit @C=1em @R=.7em {&\lstick{1}&\ctrl{2}&\targ &\qw&\ctrl{2}&\qw\\
    &\lstick{2}&\qw&\qw &\targ&\qw&\qw\\ &\lstick{3}&\targ&\ctrl{-2} &\ctrl{-1}&\targ&\qw\\}
}\end{eqnarray}

Now, since $\cnot_{(n,i)}$ is a permutation matrix, corresponding to each each $0\leq x\leq 2^{n-1}$, it is obvious that the matrix representation corresponding to each of the quantum circuits for $\Pi \mathsf{T}_{n,x}^g,g\in\{e,o\}$ discussed above is a product of permutation matrices. 

The set of binary strings $\{(x_{n-2},x_2,\hdots,x_{0}) : x_j\in\{0,1\}\}$ and the set of all subsets of $[n-1] :=\{1,\hdots,n-1\}$ have a one-one correspondence defined by $\chi: \{0,1\}^{n-1}\rightarrow 2^{[n-1]},$ which assigns $x=(x_{n-2},x_{n-3},\hdots,x_{0})$ to $\chi(x)=\Lambda_x :=\{j: x_j=1, 1\leq j\leq n-1\}\subseteq [n-1].$ Thus each position of the string represents a characteristic function for $\Lambda_x.$ Then we have the following theorem.

\begin{theorem}\label{constructeven1}
Let $\chi: \{0,1\}^{n-1}\rightarrow 2^{[n-1]}$ be the bijective function as defined above such that $\chi(x)=\Lambda_x$. For any $x\equiv (x_{n-2},\hdots,x_0)\in\{0,1\}^{n-1},$ define the functions $\alpha_{\Lambda_x}^g:  \{0,1\}^{n-1}\rightarrow \{0,\hdots,2^{n-1}-1\}$ and  $\beta_{\Lambda_x}^g:  \{0,1\}^{n-1}\rightarrow \{0,\hdots,2^{n-1}-1\},$ $g\in\{e,o\}$ as \begin{eqnarray*}
   && \alpha_{\Lambda_x}^g(m) =  \sum_{k\in \Lambda_x} m_k2^{k+1}+\sum_{j\not\in \Lambda_x } m_j2^{j+1}+2, \,\,
    \beta_{\Lambda_x}^e (m)= \sum_{k\in \Lambda_x} \overline{m}_k2^{k+1}+\sum_{j\not\in \Lambda_x} m_j2^{j+1}+2, \\
    && \beta_{\Lambda_x}^o (m)= \sum_{k\in \Lambda_x} \overline{m}_k2^{k+1}+\sum_{j\not\in \Lambda_x} m_j2^{j+1}+1,
\end{eqnarray*} with $m=(m_{n-2},\hdots,m_0)$ and $\overline{m}_k=m_k\oplus 1.$ Then $$\Pi\mathsf{T}^g_{n,x}=\prod_{m=0,\alpha^g_{\Lambda_x(m)}< \beta^g_{\Lambda_x}(m)}^{2^{n-1}-1} P_{(\alpha^g_{\Lambda_x}(m),\beta^g_{\Lambda_x}(m))}, \,\, g\in\{e,o\}.$$ Further  $\Pi\mathsf{T}^g_{n,x}\neq \Pi\mathsf{T}^g_{n,y}$ if $x\neq y$  and $(\alpha^g_{\Lambda_x}(m),\beta^g_{\Lambda_x}(m))\neq (\alpha^g_{\Lambda_y}(m),\beta^g_{\Lambda_y}(m))$ for all $0\leq m\leq 2^{n-1}-1$.
\end{theorem}

\pf See Appendix \ref{AppendixA}. \hfill $\square$

\subsection{Quantum circuit for diagonal unitaries}

The SRBB basis elements that are diagonal matrices, are given by $U_{j^2-1}^{(2^n)}, 2\leq j\leq 2^n$, which are of the form $\otimes_{j=1}^n A_j, A_j\in\{I_2, \sigma_3\}.$ Given such a basis element for some $j$, let $m$ be the greatest number such that $A_p=I_2$ for all $p>m$, and let $A_{m_1}=A_{m_2}=\hdots,A_{m_k} =\sigma_3$ for some $k$ with $m_1< m_2 < \hdots <m_k < m.$ Then a quantum circuit representation of $\exp\left(i \theta U^{(2^n)}_{j^2-1}\right)$ is given by  

\begin{equation}\label{diag}
{\Qcircuit @C=1em @R=.7em {
    &\lstick{1}& \qw &\qw &\qw &\qw &\qw &\qw &\qw &\qw\\
    &\lstick{\vdots}& \qw &\qw &\qw &\qw &\qw &\qw &\qw &\qw\\
    &\lstick{m_1}& \qw &\ctrl{6} &\qw &\qw &\qw &\qw &\ctrl{6} &\qw\\
    &\lstick{\vdots}& \qw &\qw &\qw &\qw &\qw &\qw &\qw &\qw\\
    &\lstick{m_2}& \qw &\qw &\ctrl{4} &\qw &\qw &\ctrl{4} &\qw &\qw\\
    &\lstick{\vdots}& \qw &\qw &\qw&\qw &\qw &\qw &\qw &\qw\\
    &\lstick{m_k}& \ctrl{2} &\qw &\qw &\qw &\qw &\qw &\qw &\ctrl{2}\\
    &\lstick{\vdots}& \qw &\qw &\qw&\qw &\qw &\qw &\qw &\qw\\
    &\lstick{m}& \targ &\targ &\targ &\gate{R_z((\theta)} &\qw &\targ &\targ &\targ\\
    &\lstick{\vdots}& \qw &\qw &\qw&\qw &\qw &\qw &\qw &\qw\\
    &\lstick{n}& \qw &\qw &\qw&\qw &\qw &\qw &\qw &\qw\\}}\end{equation} which represents the unitary matrix
   \small\begin{eqnarray*}
   \left( \prod_{l=1}^k (I_2^{\otimes m_l-1}\otimes (\cnot)_{(m_l,m)} \otimes I_2^{\otimes n-m})\right) \left(I_2^{\otimes m-1}\otimes R_z(\theta) \otimes I_2^{\otimes n-m}\right)
   \left(\prod_{l=1}^k (I_2^{\otimes m_l-1}\otimes (\cnot)_{(m_l,m)}\otimes I_2^{\otimes n-m}\right)
    \end{eqnarray*}\normalsize
    
    \begin{eqnarray*}
        \prod_{i=1}^k(I_2^{\otimes m_i-1}\otimes (\cnot)_{(m_i,m)}\otimes I_2^{\otimes n-m})(I_2^{\otimes m-1}\otimes R_z(\theta)\otimes I_2^{\otimes n-m})\prod_{i=1}^k(I_2^{\otimes m_i-1}\otimes (\cnot)_{(m_i,m)}\otimes I_2^{\otimes n-m}).
    \end{eqnarray*} 

\subsection{Quantum circuit for multi-controlled rotation gates}

In this section, we propose and analyze quantum circuit for $M_nZYZ$ matrices. First we have the following theorem.

\begin{theorem}\label{constructmzyz}
      A quantum circuit for a $M_nZYZ$ matrix requires at most $\left(3.2^{(n-1)}-2\right)$ $\cnot$, and $3\cdot 2^{n-1}$ rotation gates. 
    \end{theorem}

\pf From equation (\ref{mzyz2qb}), the circuit representation of a matrix in the $M_nZYZ$ form can be written as
\small\begin{eqnarray}
{ \Qcircuit @C=0.6em @R=.7em {
    &\lstick{1}&\qw& \gate{}&\qw& \gate{} &\qw& \gate{} &\qw\\
    &\lstick{\vdots}&\qw& \gate{\vdots}\qwx[-1]&\qw& \gate{\vdots}\qwx[-1]&\qw& \gate{\vdots}\qwx[-1]&\qw \\
    &\lstick{n-1}&\qw&\gate{ }\qwx[-1]&\qw&\gate{ }\qwx[-1]&\qw&\gate{ }\qwx[-1]&\qw\\
    &\lstick{n}&\qw&\gate{F_n(R_z(\alpha_1,\hdots,\alpha_{2^{n-1}}))}\qwx[-1]&\qw&\gate{F_n(R_y(\gamma_1,\hdots,\gamma_{2^{n-1}}))}\qwx[-1]&\qw&\gate{F_n(R_z(\beta_1,\hdots,\beta_{2^{n-1}}))}\qwx[-1]&\qw\\}}  
\end{eqnarray}\normalsize
    Further, from Lemma \ref{mzyz}, 

\begin{eqnarray}
    {\Qcircuit @C=0.6em @R=.7em {
    &\lstick{1}&\qw& \gate{} &\qw\\
    &\lstick{2}&\qw&\gate{ }\qwx[-1]&\qw\\
    &\lstick{\vdots}&\qw&\gate{ \vdots}\qwx[-1]&\qw\\
    &\lstick{n-1}&\qw&\gate{ }\qwx[-1]&\qw\\
    &\lstick{n}&\qw&\gate{F_n(R_a(\psi_1,\hdots,\psi_{2^{n-1}}))}\qwx[-1]&\qw\\}}\end{eqnarray}   can be decomposed as
    \begin{eqnarray}
    {\Qcircuit @C=1em @R=.7em {
    &\lstick{1}&\qw& \qw &\ctrl{4} &\qw& \qw &\ctrl{4} &\qw \\
    &\lstick{2}&\qw&\gate{ }&\qw&\qw&\gate{ }&\qw&\qw\\
    &\lstick{\vdots}&\qw&\gate{ \vdots}\qwx[-1]&\qw&\qw&\gate{\vdots }\qwx[-1]&\qw&\qw\\
    &\lstick{n-1}&\qw&\gate{ }\qwx[-1]&\qw&\qw&\gate{ }\qwx[-1]&\qw&\qw\\
    &\lstick{n}&\qw&\gate{F_{n-1}(R_a(\theta_1,\hdots,\theta_{2^{n-2}}))}\qwx[-1]&\targ&\qw&\gate{F_{n-1}(R_a(\phi_1,\hdots,\phi_{2^{n-2}}))}\qwx[-1]&\targ&\qw\\}}\end{eqnarray}  
    where $\psi_k=\begin{cases}
        \theta_i+\phi_i  \mbox{ where } 1\leq j\leq 2^{n-2}, k=j\\
        \theta_i-\phi_i \mbox{ where } 1\leq j\leq 2^{n-2}, k=2^{n-2}+j
    \end{cases}$ or
    
    \begin{eqnarray}
    {\Qcircuit @C=1em @R=.7em {
    &\lstick{1}&\ctrl{4}& \qw &\ctrl{4} &\qw& \qw &\qw &\qw \\
    &\lstick{2}&\qw&\gate{ }&\qw&\qw&\gate{ }&\qw&\qw\\
    &\lstick{\vdots}&\qw&\gate{\vdots }\qwx[-1]&\qw&\qw&\gate{ \vdots}\qwx[-1]&\qw&\qw\\
    &\lstick{n-1}&\qw&\gate{ }\qwx[-1]&\qw&\qw&\gate{ }\qwx[-1]&\qw&\qw\\
    &\lstick{n}&\targ&\gate{F_{n-1}(R_a(\theta_1,\hdots,\theta_{2^{n-2}}))}\qwx[-1]&\targ&\qw&\gate{F_{n-1}(R_a(\phi_1,\hdots,\phi_{2^{n-2}}))}\qwx[-1]&\qw&\qw\\}}\end{eqnarray}  where $\psi_k=\begin{cases}
        \theta_i+\phi_i  \mbox{ where } 1\leq j\leq 2^{n-2}, k=j\\
        -\theta_i+\phi_i \mbox{ where } 1\leq j\leq 2^{n-2}, k=2^{n-2}+j
    \end{cases}$ .

Hence, the following circuit

\begin{eqnarray}
    {\Qcircuit @C=1em @R=.7em {
    &\lstick{1}&\qw& \gate{} &\qw&\qw&\qw& \gate{} &\qw&\qw&\qw& \gate{} &\qw&\qw\\
    &\lstick{2}&\qw& \gate{}\qwx[-1]&\qw &\qw&\qw& \gate{}\qwx[-1]&\qw &\qw&\qw& \gate{}\qwx[-1]&\qw &\qw\\
    &\lstick{\vdots}&\qw&\gate{ \vdots}\qwx[-1]&\qw&\qw&\qw&\gate{ \vdots}\qwx[-1]&\qw&\qw&\qw&\gate{\vdots }\qwx[-1]&\qw&\qw\\
    &\lstick{n}&\qw&\gate{F_n(R_z)}\qwx[-1]&\qw&\qw&\qw&\gate{F_n(R_y)}\qwx[-1]&\qw&\qw&\qw&\gate{F_n(R_z)}\qwx[-1]&\qw&\qw\\}}
\end{eqnarray} can be written as 
    
\small\begin{eqnarray}\label{dec1}
{\Qcircuit @C=0.4em @R=.7em {
    &\lstick{1}&\ctrl{3}& \qw &\ctrl{3}&\qw&\qw& \ctrl{3} &\qw&\ctrl{3}&\ctrl{3}&\qw& \qw &\ctrl{3}&\qw&\qw\\
    &\lstick{2}&\qw& \gate{}&\qw &\gate{}&\gate{}&\qw& \gate{}&\qw &\qw&\qw& \gate{}&\qw &\gate{}&\qw\\
    &\lstick{\vdots}&\qw&\gate{ \vdots}\qwx[-1]&\qw&\gate{\vdots}\qwx[-1]&\gate{\vdots}\qwx[-1]&\qw&\gate{ \vdots}\qwx[-1]&\qw&\qw&\qw&\gate{ \vdots}\qwx[-1]&\qw&\gate{\vdots}\qwx[-1]&\qw\\
    &\lstick{n}&\targ&\gate{F_{n-1}(R_z)}\qwx[-1]&\targ&\gate{F_{n-1}(R_z)}\qwx[-1]&\gate{F_{n-1}(R_y)}\qwx[-1]&\targ&\gate{F_{n-1}(R_y)}\qwx[-1]&\targ&\targ&\qw&\gate{F_{n-1}(R_z)}\qwx[-1]&\targ&\gate{F_{n-1}(R_z)}\qwx[-1]&\qw\\}}\hspace{0.6cm}
\end{eqnarray}\normalsize

Further, each circuit of the form

\begin{eqnarray}
    {\Qcircuit @C=1em @R=.7em { &\lstick{1}&\qw& \gate{}&\qw\\ 
&\lstick{\vdots}&\qw&\gate{ \vdots}\qwx[-1]&\qw\\
&\lstick{n}&\qw&\gate{F_n(R_a)}\qwx[-1]&\qw\\}}
\end{eqnarray} 

at least requires $2^{n-1}$ gates \cite{krol2022}. Thus the number of $\cnot$s in the circuit given by equation (\ref{dec1}) is $6.2^{n-2}+4=3.2^{n-1}+4$. Now in section of the circuit 
\begin{eqnarray}
{\Qcircuit @C=1em @R=.7em { &\lstick{1}&\qw& \qw&\ctrl{4}&\qw&\qw\\&\lstick{2}&\qw& \gate{}&\qw&\gate{}&\qw\\ 
&\lstick{\vdots}&\qw&\gate{ \vdots}\qwx[-1]&\qw&\gate{ \vdots}\qwx[-1]&\qw\\
&\lstick{n-1}&\qw& \gate{}\qwx[-1]&\qw&\gate{}\qwx[-1]&\qw\\
&\lstick{n}&\qw&\gate{F_{n-1}(R_z)}\qwx[-1]&\targ&\gate{F_{n-1}(R_z)}\qwx[-1]&\qw\\}}    
\end{eqnarray} the left most $\cnot$ gate of 
\begin{eqnarray}
{\Qcircuit @C=1em @R=.7em { &\lstick{2}&\qw& \gate{}&\qw\\ 
&\lstick{\vdots}&\qw&\gate{ \vdots}\qwx[-1]&\qw\\
&\lstick{n-1}&\qw& \gate{}\qwx[-1]&\qw&\\
&\lstick{n}&\qw&\gate{F_{n-1}(R_z)}\qwx[-1]&\qw\\}}    
\end{eqnarray}

obtained by decomposing it into the following circuit.

\begin{eqnarray}
    {\Qcircuit @C=1em @R=.7em { &\lstick{2}&\ctrl{4}& \qw&\ctrl{4}&\qw&\qw\\ 
&\lstick{3}&\qw&\gate{ }&\qw&\gate{ }&\qw\\
&\lstick{\vdots}&\qw&\gate{ \vdots}\qwx[-1]&\qw&\gate{\vdots }\qwx[-1]&\qw\\
&\lstick{n-1}&\qw&\gate{ }\qwx[-1]&\qw&\gate{ }\qwx[-1]&\qw\\
&\lstick{n}&\targ&\gate{F_{n-2}(R_z)}\qwx[-1]&\targ&\gate{F_{n-2}(R_z)}\qwx[-1]&\qw\\}}
\end{eqnarray}  and the rightmost $\cnot$ gate of 

\begin{eqnarray}
    {\Qcircuit @C=1em @R=.7em { &\lstick{2}&\qw& \gate{}&\qw\\ 
&\lstick{3}&\qw&\gate{ }\qwx[-1]&\qw\\
&\lstick{\vdots}&\qw&\gate{ \vdots}\qwx[-1]&\qw\\
&\lstick{n-1}&\qw&\gate{ }\qwx[-1]&\qw\\
&\lstick{n}&\qw&\gate{F_{n-1}(R_z)}\qwx[-1]&\qw\\}}
\end{eqnarray} 

obtained by decomposing into the following circuit 
\begin{eqnarray}
    {\Qcircuit @C=1em @R=.7em { &\lstick{2}&\qw& \qw&\ctrl{4}&\qw&\ctrl{4}&\qw\\ 
&\lstick{3}&\qw&\gate{ }&\qw&\gate{ }&\qw&\qw\\
&\lstick{\vdots}&\qw&\gate{\vdots }\qwx[-1]&\qw&\gate{ \vdots}\qwx[-1]&\qw&\qw\\
&\lstick{n-1}&\qw&\gate{ }\qwx[-1]&\qw&\gate{ }\qwx[-1]&\qw&\qw\\
&\lstick{n}&\qw&\gate{F_{n-2}(R_z)}\qwx[-1]&\targ&\gate{F_{n-2}(R_z)}\qwx[-1]&\targ&\qw\\}}
\end{eqnarray} 
cancels each other out after further decomposition. This is because $$(\cnot)_{(2,n)}(\cnot)_{(1,n)}(\cnot)_{(2,n)}=(\cnot)_{(1,n)}.$$ The similar cancellation happens for the part of the circuit given by 

\begin{eqnarray}
    {\Qcircuit @C=1em @R=.7em { &\lstick{1}&\qw& \qw&\ctrl{4}&\qw&\qw\\&\lstick{2}&\qw& \gate{}&\qw&\gate{}&\qw\\ 
&\lstick{\vdots}&\qw&\gate{\vdots }\qwx[-1]&\qw&\gate{ \vdots}\qwx[-1]&\qw\\
&\lstick{n-1}&\qw&\gate{ }\qwx[-1]&\qw&\gate{ }\qwx[-1]&\qw\\
&\lstick{n}&\qw&\gate{F_{n-1}(R_y)}\qwx[-1]&\targ&\gate{F_{n-1}(R_y)}\qwx[-1]&\qw\\}}.
\end{eqnarray} Therefore, the total number of $\cnot$ gates that cancels out each other is $6$. Hence, there are at most $3.2^{n-1}-2$ $\cnot$ gates. \hfill{$\square$}
\subsection{Quantum circuit for unitary block diagonal matrices}   
Now, we consider circuit implementation of block diagonal unitary matrices, each block of which is a special unitary matrix of order $2.$

\begin{corollary}\label{constructblkdg}
   A quantum circuit for a block diagonal matrix $U\in \Sf\Uf(2^n)$ of the form $$\left[ 
\begin{array}{c|c|c|c|c} 
      {U}_2 &  0 &0 & 0&0\\ 
      \hline 
      0&U_4 &0&0&0 \\
      \hline
       0 & 0& 0&\ddots &0\\
        \hline
       0 & 0& 0&0& {U}_{2^{n}}
    \end{array} 
    \right],$$ where $U_{2j}\in \Uf(2), 1\leq j\leq 2^{n-1}$, requires at most $5.2^{n-1}-6$ $\cnot$ gates. 
\end{corollary}

\pf From Theorem \ref{blockdiagonal2}, any block diagonal matrix $U\in \Sf\Uf(2^n)$ consisting of $2\times 2$ blocks is of the form \begin{equation}\label{pr}\hspace{-0.85cm}\left(\prod_{t=2}^{2^n} \exp\left(i \theta_{t^2-1}U^{(2^n)}_{t^2-1}\right)\right) \left(\prod_{j=1}^{2^{n-1}}\exp\left(i \theta_{4j^2-2j}U^{(2^n)}_{4j^2-2j}\right)\right) \left(\prod_{t=2}^{2^n} \exp\left(i \theta_{t^2-1} U^{(2^n)}_{t^2-1}\right)\right)\end{equation} where $\theta_{4j^2-2j}\in \mathbb{R},1\leq j\leq 2^{n-1}, \theta_{t^2-1},\theta'_{t^2-1}\in \R$ can be obtained by employing the methods from the proofs of Theorem \ref{2levSun} and Theorem \ref{2mult}. This means that exponentials of all diagonal matrices in the basis of $su(2^n)$ needs to be multiplied on both sides. i.e. we are using the product \begin{eqnarray*}
    (\prod_{p=1}^{2^{n}-1} \exp\left(i t_{p}(\chi_{n}^{-1}(p))\right) \left(\prod_{j=1}^{2^{n-1}}\exp\left(\iota\theta_{4j^2-2j} U^{(2^n)}_{4j^2-2j}\right)\right) \left(\prod_{p=1}^{2^{n}-1} \exp\left(i t'_{p}(\chi_{n}^{-1}(p)\right)\right)
\end{eqnarray*} i.e. we are multiplying all diagonal matrices of the form $\bigotimes_{i=1}^{n} A_i$, $A_i\in \{I, \sigma_3\}$ barring the identity matrix. Now the set $\left\{\bigotimes_{i=1}^{n} A_i| A_i\in \{I_2, \sigma_3\}, 1\leq i\leq n \right\}\setminus\{I_{2^n}\}$ $=\left\{\bigotimes_{i=1}^{n-1} A_i\otimes I_2 \,|\, A_i\in \{I_2, \sigma_3\}, 1 \leq i \leq n-1 \right\}$ $\cup$  $\left\{\bigotimes_{i=1}^{n-1} A_i\otimes Z \,|\, A_i\in \{I, \sigma_3\},1\leq i\leq n-1 \right\}\setminus\{I_{2^n}\}.$

Hence using Theorem \ref{2mult} the product in equation (\ref{pr}) this product can alternatively be written as $$\left(\prod_{p=1}^{2^{n-1}-1} \exp\left(i t_{p}(\chi_{n-1}^{-1}(p)\otimes \sigma_3)\right)\right) \Tilde{U} \left(\prod_{p=1}^{2^{n-1}-1} \exp\left(i t'_{p}(\chi_{n-1}^{-1}(p)\otimes \sigma_3)\right)\right)$$
 for some $t_p,t_p'\in\R$ where $\Tilde{U}$ is a $M_nZYZ$ matrix, $\chi_{n-1}$ is discussed in Definition \ref{definition2}. This is because by Theorem \ref{2mult}  \begin{eqnarray*}
\Tilde{U}=\left(\prod_{p=0}^{2^{n-1}-1} \exp\left(i \tilde{t}_{p}(\chi_{n-1}^{-1}(p)\otimes Z\right)\right) \left(\prod_{j=1}^{2^{n-1}} \exp\left(i\theta_{4j^2-2j} U^{(2^n)}_{4j^2-2j}\right)\right) \left(\prod_{p=0}^{2^{n-1}-1} \exp\left(i \Tilde{t'}_{p}(\chi_{n-1}^{-1}(p)\otimes Z\right)\right)
 \end{eqnarray*} for some real $\Tilde{t}_p,\Tilde{t}_p'$
 
 Moreover, we have shown how to define a quantum circuit for the exponentials of matrices of the form $\bigotimes_{i=1}^{n-1}A_i\otimes I,A_{i}\in \{I_2, \sigma_3\}.$ For each $A_i=\sigma_3$, we apply $2$ $\cnot$ gates. Also the exponentials of the matrix $\sigma_3 \otimes I_2^{(\otimes n-1)}$  does not require $\cnot$ gates. Hence the number of $\cnot$ gates for a given $\bigotimes_{i=1}^{n-1}A_i\otimes I_2, A_{i}\in \{I_2, \sigma_3\}$ is $2+4+\hdots +2^{n-2}=2^{n-1}-2$. This is because the product  $\left(\prod_{p=1}^{2^{n-k}-1} \exp\left(i t_{p}(\chi_{n-k}^{-1}(p)\otimes I_2^{(\otimes k)}\right)\right)$ requires $2^{n-k}$ $\cnot$ gates from \cite{krol2022} and Theorem \ref{constructmzyz}. Therefore the total number of $\cnot$ gates for the product $\left(\prod_{p=1}^{2^{n-1}-1} \exp\left(i t_{p}(\chi_{n-1}^{-1}(p)\otimes I_2\right)\right)$ is $2^{n}-4$. Since the product is applied on both sides of a $M_nZYZ$ matrix, the total number of $\cnot$ gates becomes $2^{n+1}-4$. The rest of the proof follows from Theorem \ref{constructmzyz} since $3.2^{n-1}-2+2^{n+1}-4$ gives us the result. \hfill{$\square$}

Now, we provide a quantum circuit corresponding to the above block diagonal matrix in $\Sf\Uf(2^n)$ is given by
 \begin{eqnarray}\label{blkdiagcircuit}     
{\tiny{\centerline{\Qcircuit @C=1em @R=.7em {
    &\lstick{1}&\gate{R_z}&\gate{}&\gate{}&\gate{}& \gate{}&\gate{}&\gate{}&\gate{}&\gate{}&\gate{}&\gate{R_z}\\
    &\lstick{2}&\qw&\gate{F_2(R_z)}\qwx[-1]&\gate{}\qwx[-1]\qwx[-1]&\gate{}\qwx[-1]& \gate{}\qwx[-1]&\gate{}\qwx[-1]&\gate{}\qwx[-1]&\gate{}\qwx[-1]&\gate{}\qwx[-1]&\gate{F_2(R_z)}\qwx[-1]&\qw\\
    &\lstick{\vdots}&\qw&\qw&\gate{\vdots F_i(R_z),3\leq i\leq n-2}\qwx[-1]&\gate{\vdots}\qwx[-1]& \gate{\vdots}\qwx[-1]&\gate{\vdots}\qwx[-1]&\gate{\vdots}\qwx[-1]&\gate{\vdots}\qwx[-1]&\gate{\vdots F_i(R_z),3\leq i\leq n-2}\qwx[-1]&\qw&\qw\\
    &\lstick{n-1}&\qw&\qw&\qw&\gate{F_{n-1}(R_z)}\qwx[-1]& \gate{}\qwx[-1]&\gate{}\qwx[-1]&\gate{}\qwx[-1]&\gate{F_{n-1}(R_z)}\qwx[-1]&\qw&\qw&\qw\\
    &\lstick{n}&\qw&\qw&\qw&\qw& \gate{F_n(R_z)}\qwx[-1]&\gate{F_n(R_y}\qwx[-1]&\gate{F_n(R_z)}\qwx[-1]&\qw&\qw&\qw&\qw\\}}}}\nonumber\\.\normalsize\end{eqnarray} 
    
    The circuit represents a block diagonal matrix since it represents the product  $$\left(\prod_{k=2}^{n-1} \left(\prod_{p=1}^{2^{n-k}-1} \exp\left(i t_{p}(\chi_{n-k}^{-1}(p)\otimes I_2^{\otimes k})\right)\right)\right)V \left(\prod_{k=2}^{n-1}\left(\prod_{p=1}^{2^{n-k}-1} \exp\left(i t'_{p}(\chi_{n-k}^{-1}(p)\otimes I_2^{\otimes k})\right)\right)\right)$$ where \begin{eqnarray*}
        V=\left(\prod_{p=0}^{2^{n-1}-1} \exp\left(i t_{p}(\chi_{n-k}^{-1}(p)\otimes \sigma_3)\right)\right) \left(\prod_{j=1}^{2^{n-1}}\exp\left(i\theta_{4j^2-2j} U^{(2^n)}_{4j^2-2j}\right)\right) \left(\prod_{p=0}^{2^{n-1}-1} \exp\left(i t'_{p}(\chi_{n-k}^{-1}(p)\otimes \sigma_3)\right)\right)
    \end{eqnarray*}, which gives us the form described in Theorem \ref{blockdiagonal2}
\subsection{Scalable quantum circuits for approximating special unitary matrices}  

From Algorithm \ref{alg2}, we see that, a special unitary matrix $U\in \Sf\Uf(2^n)$ can be approximated in the circuit form with one layer in the following way. 

\begin{equation}\label{alg3}
{\Qcircuit @C=1em @R=.7em {
&\lstick{1}& \multigate{3}{\Phi(\Theta_{\phi})}&\multigate{3}{\Psi(\Theta_{\psi})}&\multigate{3}{\zeta(\Theta_{\zeta})}&\qw\\
&\lstick{\vdots}& \ghost{{\phi(\Theta)}}&\ghost{{\psi(\Theta)}}&\ghost{{\zeta(\Theta)}}&\qw\\
&\lstick{n-1}& \ghost{{\phi(\Theta)}}&\ghost{{\psi(\Theta)}}&\ghost{{\zeta(\Theta)}}&\qw\\
&\lstick{n}& \ghost{{\phi(\Theta)}}&\ghost{{\psi(\Theta)}}&\ghost{{\zeta(\Theta)}}&\qw\\}}
\end{equation}

where writing $\zeta(\Theta_{\zeta})$, $\Psi(\Theta_{\psi})$ and $\Phi(\Theta_{\psi})$ as quantum circuits respectively are given by 
\begin{eqnarray*}
 \Qcircuit @C=1em @R=.7em {
    &\gate{\prod_{l=1}^{2^n} \exp\left(i \theta_{l^2-1}B^{(2^n)}_{l^2-1}\right)}&\qw\\}\end{eqnarray*}\normalsize\footnotesize\begin{eqnarray*}
   \Qcircuit @C=1em @R=.7em {
    &\gate{\Pi\mathsf{T}^e_{n,2^{n-1}-1} M^e_{{(2^{n-1}-1)}} \Pi\mathsf{T}^e_{n,2^{n-1}-1}} &\gate{\hdots}&\gate{\Pi\mathsf{T}^e_{n,1} M^e_1 \Pi\mathsf{T}^e_{n,1}} &\gate{\prod_{j=1}^{2^{n-1}} \exp\left(i \theta_{(2j-1)^2}B^{(2^n)}_{(2j-1)^2}\right) \exp\left(i \theta_{(4j^2-2j)}B^{(2^n)}_{(4j^2-2j)}\right)}&\qw\\ } \end{eqnarray*}\normalsize\small\begin{eqnarray*}
  \Qcircuit @C=1em @R=.7em {
  &\gate{\Pi\mathsf{T}^o_{n,2^{n-1}-1} M^o_{2^{n-1}-1} \Pi\mathsf{T}^o_{n,2^{n-1}-1}}&\gate{\hdots}&\gate{\Pi\mathsf{T}^o_1 M^o_{n,1} \Pi\mathsf{T}^o_{n,1}} &\qw\\}   
\end{eqnarray*}\normalsize

Recall that $M^o_x\in \Sf\Uf(2^n)$ is a block diagonal matrix with $2\times 2$ blocks and $M^e_x, 1\leq x\leq 2^{n-1}-1$ is a $M_nZYZ$ matrix. Further, since 
 \begin{eqnarray*}
     \prod_{j=1}^{2^{n-1}} \exp\left(i \theta_{(2j-1)^2} U^{(2^n)}_{(2j-1)^2}\right) \exp\left(i \theta_{(4j^2-2j)} U^{(2^n)}_{(4j^2-2j)}\right)
 \end{eqnarray*} is  $M_nZYZ$ type matrix, a quantum circuit representation is given by 
 
\begin{eqnarray}
    {\Qcircuit @C=1em @R=.7em {
    &\lstick{1}&\qw& \gate{} &\qw&\qw&\qw& \gate{} &\qw&\qw&\qw& \gate{} &\qw&\qw\\
    &\lstick{2}&\qw& \gate{}\qwx[-1]&\qw &\qw&\qw& \gate{}\qwx[-1]&\qw &\qw&\qw& \gate{}\qwx[-1]&\qw &\qw\\
    &\lstick{\vdots}&\qw&\gate{ \vdots}\qwx[-1]&\qw&\qw&\qw&\gate{ \vdots}\qwx[-1]&\qw&\qw&\qw&\gate{\vdots }\qwx[-1]&\qw&\qw\\
    &\lstick{n}&\qw&\gate{F_n(R_z)}\qwx[-1]&\qw&\qw&\qw&\gate{F_n(R_y)}\qwx[-1]&\qw&\qw&\qw&\gate{F_n(R_z)}\qwx[-1]&\qw&\qw\\}}
\end{eqnarray}
Next,, for $1\leq x\leq 2^{n-1}-1$, $\Pi\mathsf{T}^e_{n,x} M^e_x \Pi\mathsf{T}^e_{n,x}$ can have the quantum circuit representation as
\begin{eqnarray}
{\Qcircuit @C=1em @R=.7em {
    &\lstick{1}&\multigate{3}{\Pi\mathsf{T}^e_{n,x}}& \gate{} &\qw& \gate{} &\qw& \gate{} &\multigate{3}{\Pi\mathsf{T}^e_{n,x}}&\qw\\
    &\lstick{\vdots}&\ghost{P_{(x,2^n,even)}}& \gate{}\qwx[-1]&\qw& \gate{}\qwx[-1]&\qw& \gate{}\qwx[-1]&\ghost{P_{(x,2^n,even)}}&\qw\\
    &\lstick{\vdots}&\ghost{P_{(x,2^n,even)}}&\gate{ }\qwx[-1]&\qw&\gate{ }\qwx[-1]&\qw&\gate{ }\qwx[-1]&\ghost{P_{(x,2^n,even)}}&\qw\\
    &\lstick{n}&\ghost{P_{(x,2^n,even)}}&\gate{F_n(R_z)}\qwx[-1]&\qw&\gate{F_n(R_y)}\qwx[-1]&\qw&\gate{F_n(R_z)}\qwx[-1]&\ghost{P_{(x,2^n,even)}}&\qw\\}}    
\end{eqnarray}
 Finally, for $1\leq x\leq 2^{n-1}-1$ a quantum circuit representation of $\Pi\mathsf{T}^o_{n,x} M^o_x \Pi\mathsf{T}^o_{n,x}$ is given by
    
\begin{eqnarray}
    {\Qcircuit @C=1em @R=.7em {
    &\lstick{1}&\multigate{3}{\Pi\mathsf{T}^o_{n,x}}& \multigate{3}{M^o_x}&\multigate{3}{\Pi\mathsf{T}^o_{n,x}}&\qw\\
    &\lstick{\vdots}&\ghost{P_{(x,2^n,odd)}}& \ghost{D'_x}&\ghost{P_{(x,2^n,odd)}} &\qw\\
    &\lstick{\vdots}&\ghost{P_{(x,2^n,odd)}}& \ghost{D'_x}&\ghost{P_{(x,2^n,odd)}} &\qw\\
    &\lstick{n}&\ghost{P_{(x,2^n,odd)}}& \ghost{D'_x}&\ghost{P_{(x,2^n,odd)}} &\qw\\}}
\end{eqnarray} where the circuit representation of $M^o_x$ is  of the form  mentioned in equation (\ref{blkdiagcircuit}). Finally, the circuits for $\Pi\mathsf{T}^o_{n,x}$ and $\Pi\mathsf{T}^e_{n,x}$ can be determined by Theorem \ref{constructeven1}. 

Now, we consider scaling the proposed $n$-qubit circuit into an $(n+1)$-qubit circuit for approximating special unitary matrices. Since the proposed circuit consists of mainly three types of circuits: circuits for product of transpositions, $M_nZYZ$ circuit, and circuit for block diagonal unitary matrices, it is enough to describe the techniques for extending these circuits from $n$-qubit to $(n+1)$-qubit systems as follows.  We denote $\Pi\mathsf{T}^s_{m,y}$ for $\Pi\mathsf{T}^s_y$ with $m$-qubit systems, $s\in\{e,o\}.$ $(\cnot)_{(i,j)}$ represents a $\cnot$ gate with $i$-th qubit as control qubit and $j$-th qubit is the target qubit.

    \begin{itemize}
        \item[$\blacksquare$] \textbf{Construction of scalable circuits for $\Pi\mathsf{T}^e_{n+1,x}:$} If the circuit representation of $\Pi\mathsf{T}^e_{n,x}$ for $n$-qubit system given in Theorem \ref{constructeven1} as \Qcircuit @C=1em @R=.7em {
    &\gate{\Pi\mathsf{T}^{e}_{n,x}}&\qw\\}  for some $x\in\{0,\hdots, 2^{n-1}-1\}$ then the circuit for $\Pi\mathsf{T}^{e}_{n+1,y}$, $1\leq y\leq 2^{n}-1$ is given by  
  
    \begin{equation}\label{Pevennoadd}\centerline{\Qcircuit @C=1em @R=.7em {
    &\lstick{1} & \qw & \qw& \qw\\
    &\lstick{2}&\qw &\multigate{2}{\Pi\mathsf{T}^{e}_{n,x}}&\qw\\
    &\lstick{\vdots}&\qw&\ghost{P_{(x,2^n,even)}}&\qw\\
    &\lstick{n+1}&\qw&\ghost{P_{(x,2^n,even)}}&\qw\\}}
    \end{equation}
    
    if $y=x,$ and 
        \vspace{-0.5em}
   \begin{equation}\label{Pevenadd} \centerline{\Qcircuit @C=1em @R=.7em {
    &\lstick{1} & \targ & \qw& \qw&\qw\\
    &\lstick{2}&\qw&\multigate{2}{\Pi\mathsf{T}^{e}_{n,x}}&\qw&\qw\\
    &\lstick{\vdots}&\qw&\ghost{P_{(x,2^n,even)}}&\qw&\qw\\
    &\lstick{n+1}&\ctrl{-3}&\ghost{P_{(x,2^n,even)}}&\qw&\qw\\}}\end{equation}
    
    if $y=2^{n-1}+x.$ 
    
    \item[$\blacksquare$] \textbf{Construction of scalable circuits for $\Pi\mathsf{T}^o_{n+1,x}:$} As above, the circuit for $\Pi\mathsf{T}^{o}_{n+1,y},$ $1\leq y\leq 2^{n}-1$ is given by  
        \vspace{-0.5em}
    \begin{equation}\label{Poddnoadd} \centerline{\Qcircuit @C=1em @R=.7em {
    &\lstick{1} & \qw & \qw& \qw\\
    &\lstick{2}&\qw& \multigate{2}{\Pi\mathsf{T}^{o}_{n,x}}&\qw\\
    &\lstick{\vdots}&\qw&\ghost{P_{(x,2^n,odd)}}&\qw\\
    &\lstick{n+1}&\qw&\ghost{P_{(x,2^n,odd)}}&\qw\\}} \end{equation}
    
    if $y=x,$ and
        \vspace{-0.5em}
   \begin{equation}\label{Poddadd} \centerline{\Qcircuit @C=1em @R=.7em {
    &\lstick{1} & \ctrl{3} & \targ&\qw& \qw&\ctrl{3}&\qw\\
    &\lstick{2}&\qw& \qw&\multigate{2}{\Pi\mathsf{T}^{o}_{n,x}}&\qw&\qw&\qw\\
    &\lstick{\vdots}&\qw&\qw&\ghost{P_{(x,2^n,even)}}&\qw&\qw&\qw\\
    &\lstick{n+1}&\targ&\ctrl{-3}&\ghost{P_{(x,2^n,even)}}&\qw&\targ&\qw\\}}\end{equation}
    
     if $y=2^{n-1}+x,$ $x\in\{0,\hdots, 2^{n-1}-1\}.$ 
     
      \item[$\blacksquare$] \textbf{Construction of scalable circuits for $M_nZYZ:$} This follows from equivalence of circuits given in equation (\ref{multi1}) and equation (\ref{multidec}). Indeed, $M_{n+1}ZYZ$ is of the form $$\Qcircuit @C=1em @R=.7em {
    &\lstick{1}&\qw& \gate{} &\qw&\qw&\qw& \gate{} &\qw&\qw&\qw& \gate{} &\qw&\qw\\
    &\lstick{\vdots}&\qw& \gate{}\qwx[-1]&\qw &\qw&\qw& \gate{}\qwx[-1]&\qw &\qw&\qw& \gate{}\qwx[-1]&\qw &\qw\\
    &\lstick{\vdots}&\qw&\gate{ }\qwx[-1]&\qw&\qw&\qw&\gate{ }\qwx[-1]&\qw&\qw&\qw&\gate{ }\qwx[-1]&\qw&\qw\\
    &\lstick{n+1}&\qw&\gate{F_{n+1}(R_z)}\qwx[-1]&\qw&\qw&\qw&\gate{F_{n+1}(R_y)}\qwx[-1]&\qw&\qw&\qw&\gate{F_{n+1}(R_z)}\qwx[-1]&\qw&\qw\\}$$
     and \begin{eqnarray}\label{multi12}
\centering{\Qcircuit @C=1em @R=.7em {
    &\lstick{1}&\qw& \gate{} &\qw\\
    &\lstick{2}&\qw&\gate{ }\qwx[-1]&\qw\\
    &\lstick{\vdots}&\qw&\gate{ }\qwx[-1]&\qw\\
    &\lstick{n}&\qw&\gate{ }\qwx[-1]&\qw\\
    &\lstick{n+1}&\qw&\gate{F_{n+1}(R_a(\psi_1,\hdots,\psi_{2^{n}}))}\qwx[-1]&\qw\\}}\end{eqnarray} is equivalent to
    
    \begin{eqnarray}\label{multidec2}
    \centering{\Qcircuit @C=1em @R=.7em {
    &\lstick{1}&\qw& \qw &\ctrl{4} &\qw& \qw &\ctrl{4} &\qw \\
    &\lstick{2}&\qw&\gate{ }&\qw&\qw&\gate{ }&\qw&\qw\\
    &\lstick{\vdots}&\qw&\gate{ }\qwx[-1]&\qw&\qw&\gate{ }\qwx[-1]&\qw&\qw\\
    &\lstick{n}&\qw&\gate{ }\qwx[-1]&\qw&\qw&\gate{ }\qwx[-1]&\qw&\qw\\
    &\lstick{n+1}&\qw&\gate{F_{n}(R_a(\theta_1,\hdots,\theta_{2^{n}}))}\qwx[-1]&\targ&\qw&\gate{F_{n}(R_a(\phi_1,\hdots,\phi_{2^{n}}))}\qwx[-1]&\targ&\qw\\}}\end{eqnarray}  where $$\psi_k=\begin{cases}
        \theta_j+\phi_j  \mbox{ where } 1\leq j\leq 2^{n-1},k=j\\
        \theta_j-\phi_j \mbox{ where } 1\leq j\leq 2^{n},k=j+2^{n-1}.
    \end{cases}$$

     \item[$\blacksquare$] \textbf{Construction of scalable circuits for block diagonal matrices:} This follows similarly due to the above property of $M_nZYZ,$ $1\leq n$ which define the quantum circuit for a block diagonal unitary matrix given by equation (\ref{blkdiagcircuit}). 
    \end{itemize}

\begin{theorem}\label{count}
The circuit implementation of a special unitary matrix on $n$-qubits with $L$ layers using Algorithm $2$  requires at most $L(2.4^n+(n-5)2^{n-1})$ $\cnot$ gates, $L({\frac{3}{2}\cdot} 4^n-\frac{5}{2}2^n+1)$ $R_z$ gates where $L$ is the number of iterations/layers.
\end{theorem}

\pf To prove this theorem, we need to consider the matrices, the number of rotation gates and $\cnot$ gates for circuit implementation of $\zeta(\Theta_{\zeta}),$ $\Psi(\Theta_{\psi}),$ and $\Phi(\Theta_{\phi}).$ From equations (\ref{eqn:thetazeta}), (\ref{eqn:thetapsi}) and (\ref{eqn:thetaphi}), we have 
\begin{eqnarray*}
    &&\zeta(\Theta_{\zeta}) = \prod_{j=1}^{2^n} \exp\left(i \theta_{j^2-1} U^{(2^n)}_{j^2-1}\right), \,\, \Psi(\Theta_{\psi}) = M_0^e \left(\prod_{x=1}^{2^{n-1}-1} (\Pi\mathsf{T}^e_{n,x}) M^e_x (\Pi\mathsf{T}^e_{n,x})\right), \,\, \\&&\Phi(\Theta_{\phi}) = \prod_{x=1}^{2^{n-1}-1} (\Pi\mathsf{T}^o_{n,x}) M^o_x (\Pi\mathsf{T}^o_{n,x}),
\end{eqnarray*} where $M_0^e=\left(\prod_{j=1}^{2^{n-1}} \exp\left(i \theta_{(2j-1)^2}U^{(2^n)}_{(2j-1)^2}\right) \exp\left(i \theta_{(4j^2-2j)} U^{(2^n)}_{(4j^2-2j)}\right)\right)$\\

Using Lemma \ref{lemmamult}, Theorem \ref{constructmzyz} and from \cite{mottonen2004,krol2022}, a $M_nZYZ$ matrix takes $3.2^{n-1}-2$ $\cnot$ gates and $3.2^{n-1}$ $R_z,R_y$ gates. Now $M^e_x, M^e_0$ are $M_nZYZ$ matrices and $M^o_x\in \Sf\Uf(2^n)$ is a block diagonal matrix for $1\leq x\leq 2^{{n-1}-1}.$ Then $$\zeta(\Theta_{\zeta})=\left(\prod_{p=1}^{2^{n-1}-1} \exp\left(i \theta_{p}(\chi_{n-1}^{-1}(p)\otimes I_2)\right)\right) \left(\prod_{q=0}^{2^{n-1}-1} \exp\left(i \theta_{q}(\chi_{n-k}^{-1}(p)\otimes \sigma_3\right)\right)$$  where $\chi$ is described in Definition \ref{definition2}. 

Further the term $M^e_0$ in $\Psi(\Theta_{\psi})$ is a $M_nZYZ$ matrix and from Theorem \ref{2mult} can be written as $$\left(\prod_{p=0}^{2^{n-1}-1} \exp\left(i \theta_{p}(\chi_{n-1}^{-1}(p)\otimes \sigma_3)\right)\right) \left(\prod_{j=1}^{2^{n-1}} \exp\left(i\theta_{4j^2-2j} U^{(2^n)}_{4j^2-2j}\right)\right) \left(\prod_{p=1}^{2^{n-1}} \exp\left(i \theta'_{p}(\chi_{n-1}^{-1}(p)\otimes \sigma_3\right)\right)$$ where all necessary terms have been defined in Theorem \ref{2mult}. Hence, the term $\left(\prod_{p=0}^{2^{n-1}-1} \exp\left(i \theta_{p}(\chi_{n-1}^{-1}(p)\otimes \sigma_3)\right)\right)$ from $M^e_0$ and the term $\left(\prod_{q=0}^{2^{n-1}-1} \exp\left(i \theta_{q}(\chi_{n-k}^{-1}(p)\otimes \sigma_3)\right)\right)$ from $\zeta(\Theta_{\zeta})$ are multiplied to form the product $\left(\prod_{p=1}^{2^{n-1}-1} \exp\left(i \theta_{p}(\chi_{n-1}^{-1}(p)\otimes I_2)\right)\right)\Tilde{M}^e_0$ where $\Tilde{M}^e_0$ is a $M_nZYZ$ matrix such that $\Tilde{M}^e_0$ is equal to $\left(\prod_{q=0}^{2^{n-1}-1} \exp\left(i \theta_{q}(\chi_{n-k}^{-1}(p)\otimes \sigma_3)\right)\right) M^e_0$. 

Next, $\Phi(\Theta_{\phi})=\prod_{x=1}^{2^{n-1}-1} \Pi\mathsf{T}^o_{n,x} M^{o}_x  \Pi\mathsf{T}^o_{n,x} $ where $M^{e}_x$ is a block diagonal matrix, which requires $(5.2^{n-1}-6)$ $\cnot$ gates and $2(2^n-1)$ $R_z$ gates and $2^{n-1}$number of $ R_y$ gates from Corollary \ref{constructblkdg}.
   
Finally, from the construction of $\Pi\mathsf{T}^e_{n,x},$ $1\leq x\leq 2^{n-1}-1$ from Theorem \ref{constructeven1}, it can be seen that for different values of $x$, we get a quantum circuit which consists of $k$ (depending on $x$) $(\cnot)$-gates having control at the $n$-th qubit and target is at $i$-th qubit for $1\leq i\leq n-1$. Thus the number of $(\cnot)_{(n,i)}$ gates present in the construction of $\Pi\mathsf{T}^e_{n,x}$ is at most $1$ for a fixed $i$. Hence, one can either choose one target qubit from $\{1,2\hdots,n-1\}$ and in this case total number of $\cnot$ gates will be $\binom{n-1}{1}$. One can also choose $2$ target qubits from $\{1,2\hdots,n-1\}$,in this case total number of $\cnot$ gates will be $2\binom{n-1}{2}$. Continuing in this way, the number of $\cnot$ gates that is required for the permutation matrices is $\sum_{l=1}^{n-1}l\binom{n-1}{1} $ for all the permutation matrices in the set $\mathcal{P}_{2^n,\mbox{even}}$. On the other hand, $\Pi\mathsf{T}^o_{n,x}$ for each $1\leq x\leq 2^{n-1}-1$ requires two more $(\cnot)$ gates than $\Pi\mathsf{T}^e_{n,x}$ from our construction. Hence the total $\sum_{l=1}^{n-1}(l+2) \binom{n-1}{1} $ number of $\cnot$ gates are required for all the permutation matrices in the set $\mathcal{P}_{2^n,\mbox{odd}}$. Now $\sum_{l=1}^{n-1}i\binom{n-1}{l}=\sum_{l=1}^{n-1}(n-1)\binom{n-2}{l-1}=2^{n-2}(n-1)$, which gives us the total $\cnot$ gates for permutation matrices required to construct elements from $\mathcal{P}_{2^n,\mbox{odd}}$ and $\mathcal{P}_{2^n,\mbox{even}}$ to be $2(2^{n-1}-1)+(n-1)2^{n-1}$. Since permutation matrices are multiplied on both sides, the number of $\cnot$ gates becomes $4(2^{n-1}-1)+(n-1)2^n$. Including the $\cnot$ gates used for the construction of unitary diagonal matrices in the circuit in \ref{constructblkdg}, total number of $\cnot$ gates for constructing product of all unitary diagonal matrices is $2^n-2$. 

However, looking at $\zeta(\Theta_\zeta)\Psi(\Theta_\psi)\Phi(\Theta_\phi)$, we see that the diagonal matrices of the form $\bigotimes_{l=1}^{n-1}A_l\otimes \sigma_3$ gets multiplied with the first $M_nZYZ$ matrix in $\Psi(\Theta_{\psi})$ where $A_l\in \{I_2,\sigma_3\}$. Hence only diagonal matrices of the form $\otimes_{l=1}^{n-1}A_l\otimes I_2$ remain from $\zeta(\Theta_{\zeta})$. Consequently, the total number of $\cnot$ gates for constructing the product of diagonal unitary matrices i.e. $\zeta(\Theta_{\zeta})$ is $2^{n-1}-2$. The same result holds true for number of $R_z$ gates. Hence the desired result follows. \hfill{$\square$}

In order to construct a $(n+1)$-qubit circuit from an $n$-qubit circuit, we add one more qubit at the top of the current circuit. i.e. from

\begin{eqnarray*}
    \Qcircuit @C=1em @R=.7em {\lstick{1}&\multigate{4}{U}\\\lstick{2}&\ghost{U}\\\lstick{\vdots}&\ghost{U}\\\lstick{n-1}&\ghost{U}\\\lstick{n}&\ghost{U}}
\end{eqnarray*} 

to 
\begin{eqnarray*}
\Qcircuit @C=1em @R=.7em {\lstick{1}&\qw\\\lstick{2}&\multigate{4}{U}\\\lstick{3}&\ghost{U}\\\lstick{\vdots}&\ghost{U}\\\lstick{n}&\ghost{U}\\\lstick{n+1}&\ghost{U}}    
\end{eqnarray*}

Now, we present the following algorithms based on the above discussion  that will help us to create an algorithm for constructing scalable quantum circuits. 

\begin{algorithm}[H]
\caption{Creating circuit for $\Pi\mathsf{T}^e_{n+1,y}, 0\leq y\leq 2^n-1$ from circuit $\Pi\mathsf{T}^e_{n,x},0\leq x\leq 2^{n-1}-1$}\label{Createeven}
\textbf{Provided:} $\cnot$ gates, circuits $\Pi\mathsf{T}^e_{n,x},0\leq x\leq 2^{n-1}-1$ .\\
\textbf{Input:} $y\in \{0,\hdots,2^n-1\}$\\
\textbf{Output:} $\eta(y,2^{n+1},even)$ gives a circuit of $I_2\otimes \Pi\mathsf{T}^e_{n+1,y}$
\begin{algorithmic}
\For {$y=0:2^{n}-1;y++$}
\If {$y<2^{n-1}$}
\State $x=y$
\State $\eta(y,2^{n+1},even)$ $\rightarrow$  Add one qubit layer at the top.   See equation (\ref{Pevennoadd}) 
\Else
\State $x=y-2^{n-1}$
\State $\eta(y,2^{n+1},even)\rightarrow$ Add one qubit layer at the top and add a $(\cnot)_{(n+1,1)}$ to left of $\Pi\mathsf{T}^e_{n,x}$. See equation (\ref{Pevenadd}). 
\EndIf
\EndFor
\end{algorithmic}
\end{algorithm}    
\vspace{-.75cm}
    \begin{algorithm}[H]
\caption{Creating circuit for $\Pi\mathsf{T}^o_{n+1,y},0\leq y\leq 2^n-1$ from circuit $\Pi\mathsf{T}^o_{n,x},0\leq x\leq 2^{n-1}-1$}\label{Createodd}
\textbf{Provided:} $\cnot$ gates, circuits $\Pi\mathsf{T}^e_{n,x},\Pi\mathsf{T}^o_{n,x},0\leq x\leq 2^{n-1}-1$ .\\
\textbf{Input:} $y\in \{0,\hdots,2^n-1\}$\\
\textbf{Output:}$\eta(y,2^{n+1},odd)$ gives a circuit of $\Pi\mathsf{T}^o_{n+1,x}$
\begin{algorithmic}
\For{$y=0:2^{n}-1;y++$}
\If{$y<2^{n-1}$}
\State $x=y$
\State $\eta(y,2^{n+1},odd)\rightarrow$  Add one qubit layer at the top. See equation (\ref{Poddnoadd}) 
\Else
\State $x=y-2^{n-1}$
\State $\eta(y,2^{n+1},odd)\rightarrow$ Add one qubit layer at the top and add a $(\cnot)_{(n+1,1)}$ gate,$(\cnot)_{(1,n+1)}$ gate to the left of $I_2\otimes \Pi\mathsf{T}^e_{n,x}$. Add another $(\cnot)_{(n+1,1)}$ gate to the right of $\Pi\mathsf{T}^e_{n,x}$. See equation (\ref{Poddadd}). 
\State End If
\EndIf
\State End For
\EndFor
\State End
\end{algorithmic}
\end{algorithm}
\vspace{-.75cm}
    \begin{algorithm}[H]
\caption{Creating circuit for $(n+1)$-qubit rotation gates $F_{(n+1)}(R_z)$ from multi-qubit rotation gates $F_{(n)}(R_z)$}\label{multialgoz}
\textbf{Provided:} $\cnot$ gates, circuits $F_n(R_z)$ .\\
\textbf{Input:}$a_1,a_2,a_4\hdots,a_{2^{n-1}}$ for $Fn(R_z):=F_n(R_z(a_1,a_2\hdots,a_{2^{n-1}}))$ and $b_1,b_2,b_3\hdots,b_{2^{n-1}}$ for $Fn(R_z):=F_n(R_z(b_1,b_2\hdots,b_{2^{n-1}}))$ \\
\textbf{Output:}$\xi(F_n(R_z(a_1,\hdots,a_{2^{n-1}})),F_n(R_z(b_1,\hdots,b_{2^{n-1}}))):=\xi(F_n(R_z),F_n(R_z))$ gives a circuit of $F_{n+1}(R_z)$
\begin{algorithmic}
\State Add one layer of qubit at the top. Add a $(\cnot)_{(1,n+1)}$ to the left of $I_2\otimes F_n(R_z)$. Then add another $(\cnot)_{(1,n+1)}$ and a $I_2\otimes F_n(R_z)$. See equation (\ref{multi12}) and equation (\ref{multidec2}). 
\State End
\end{algorithmic}
\end{algorithm}\vspace{-.75cm}
    \begin{algorithm}[H]
\caption{Creating circuit for $(n+1)$-qubit rotation gates $F_{(n+1)}(R_y)$ from multi-qubit rotation gates $F_{(n)}(R_y)$}\label{multialgoy}
\textbf{Provided:} $\cnot$ gates, circuits $F_n(R_y)$ .\\
\textbf{Input:}$a_1,a_2,a_4\hdots,a_{2^{n-1}}$ for $Fn(R_y):=F_n(R_y(a_1,a_2\hdots,a_{2^{n-1}}))$ and $b_1,b_2,b_3\hdots,b_{2^{n-1}}$ for $Fn(R_y):=F_n(R_y(b_1,b_2\hdots,b_{2^{n-1}}))$ \\
\textbf{Output:}$\xi(F_n(R_y(a_1,\hdots,a_{2^{n-1}})),F_n(R_y(b_1,\hdots,b_{2^{n-1}}))):=\xi(F_n(R_y),F_n(R_y))$ gives a circuit of $F_{n+1}(R_z)$
\begin{algorithmic}
\State Add one layer of qubit at the top. Add a $(\cnot)_{(1,n+1)}$ to the left of $I_2\otimes F_n(R_y)$. Then add another $(\cnot)_{(1,n+1)}$ and a $I_2\otimes F_n(R_y)$. See equation (\ref{multi1}) and equation (\ref{multidec}). 
\State End
\end{algorithmic}
\end{algorithm}

Now, we provide Algorithm \ref{multialgoy2} by combining all the Algorithms \ref{Createeven}-\ref{multialgoz} for the generation of $(n+1)$-qubit circuit from $n$-qubit circuit.

\begin{algorithm}[H]
\caption{Creating a $(n+1)$-qubit circuit to approximate any $U\in \Sf\Uf(2^{n+1})$ from a $n$-qubit circuit that approximates any $\hat{U}\in SU(2^{n})$}\label{multialgoy2}
\textbf{Provided:} $\cnot$ gates and 1 qubit rotation gates.\\
\textbf{Input:} $n$-qubit circuit that approximates any $\hat{U}\in \Sf\Uf(2^{n})$ and of the form mentioned in equation (\ref{alg3}) i.e. $\zeta(\Theta_\zeta)\Psi(\Theta_\psi)\Phi(\Theta_\phi)$ where all the terms have been defined in equation (\ref{alg3})\\
\textbf{Output:}$(n+1)$-qubit circuit that approximates any ${U}\in \Sf\Uf(2^{n+1})$ \\
\small\begin{algorithmic}
\Procedure{}{}       \Comment{}
\State Add a qubit layer at the top/beginning of the circuit.
\State Create product of all $2^{n+1}$ special unitary diagonal matrices from product of all $2^{n}$ special unitary diagonal matrices using $\xi(F_i(R_z),F_i(R_z)),1\leq i\leq n$ in Algorithm \ref{multialgoz}.  
 \\

\For{$y=1:2^{n}-1;y++$}
\State Use Algorithm \ref{Createeven} create $\Pi\mathsf{T}^e_{n+1,y}$ using the function $\eta(y,2^{n+1},even)$
\State \State Use Algorithm \ref{Createodd} create $\Pi\mathsf{T}^o_{n+1,y}$ using the function $\eta(y,2^{n+1},odd)$
\State Add $\cnot$ gates to convert $\Pi\mathsf{T}^o_{n,y}\rightarrow \Pi\mathsf{T}^o_{n+1,y}$\\ 
\State\textbf{End}
\EndFor
\State $\zeta(\Theta)\rightarrow \Pi_{i=1}^{(2^{n+1}-1)}\exp{(\iota \theta_{a}\chi^{-1}_{n+1}(a))}$, (see definition of $\chi$ at equation (\ref{definition2}))
\State Create a $(n+1)$-qubit MZYZ matrix $M^e_{0}$ from a $n$ qubit $MZYZ$ matrix using $\xi(F_n(R_z),F_n(R_z)),\xi(F_n(R_y),F_n(R_y))$ in Algorithm \ref{multialgoz} and Algorithm \ref{multialgoy}\\
\For{$y=1:2^{n}-1;i++$}
\State Create a $(n+1)$-qubit MZYZ matrix $M^e_{y}$ from a $n$ qubit $MZYZ$ matrix using $\xi(F_n(R_z),F_n(R_z)),\xi(F_n(R_y),F_n(R_y))$ in Algorithm \ref{multialgoz} and Algorithm \ref{multialgoy}\\
\State  Create a $(n+1)$-qubit block diagonal special unitary matrix $M^o_{y}$ from a $n$ qubit block diagonal special unitary matrix using $\xi(F_i(R_z),F_i(R_z)),1\leq i\leq n$ in Algorithm \ref{multialgoz}. and  $\xi(F_n(R_y),F_n(R_y))$ in Algorithm \ref{multialgoy}
\State\textbf{End}
\EndFor
\For{$y=1:2^n-1:y++$}
\State $\Psi(\Theta_\psi)\rightarrow M^e_0\Pi\mathsf{T}^e_{n+1,y} M^e_y \Pi\mathsf{T}^e_{n+1,y}$\\
\State $\Psi(\Theta_\psi)\rightarrow \Psi(\Theta_\psi)$\\
\State $\Phi(\Theta_\phi)\rightarrow \Pi\mathsf{T}^o_{n+1,x} M^o_x \Pi\mathsf{T}^o_{n+1,x}$\\
\State $\Phi(\Theta_\phi)\rightarrow \Phi(\Theta_\phi)$
\State\textbf{End}
\EndFor
\State  $\zeta(\Theta_\zeta)\psi(\Theta_\psi)\Phi(\Theta_\phi)$
\State \textbf{End Procedure}
\EndProcedure
\end{algorithmic}\normalsize
\end{algorithm}

\begin{figure}
 \begin{center}
  \includegraphics[height=6.5 cm,width=12.6 cm]{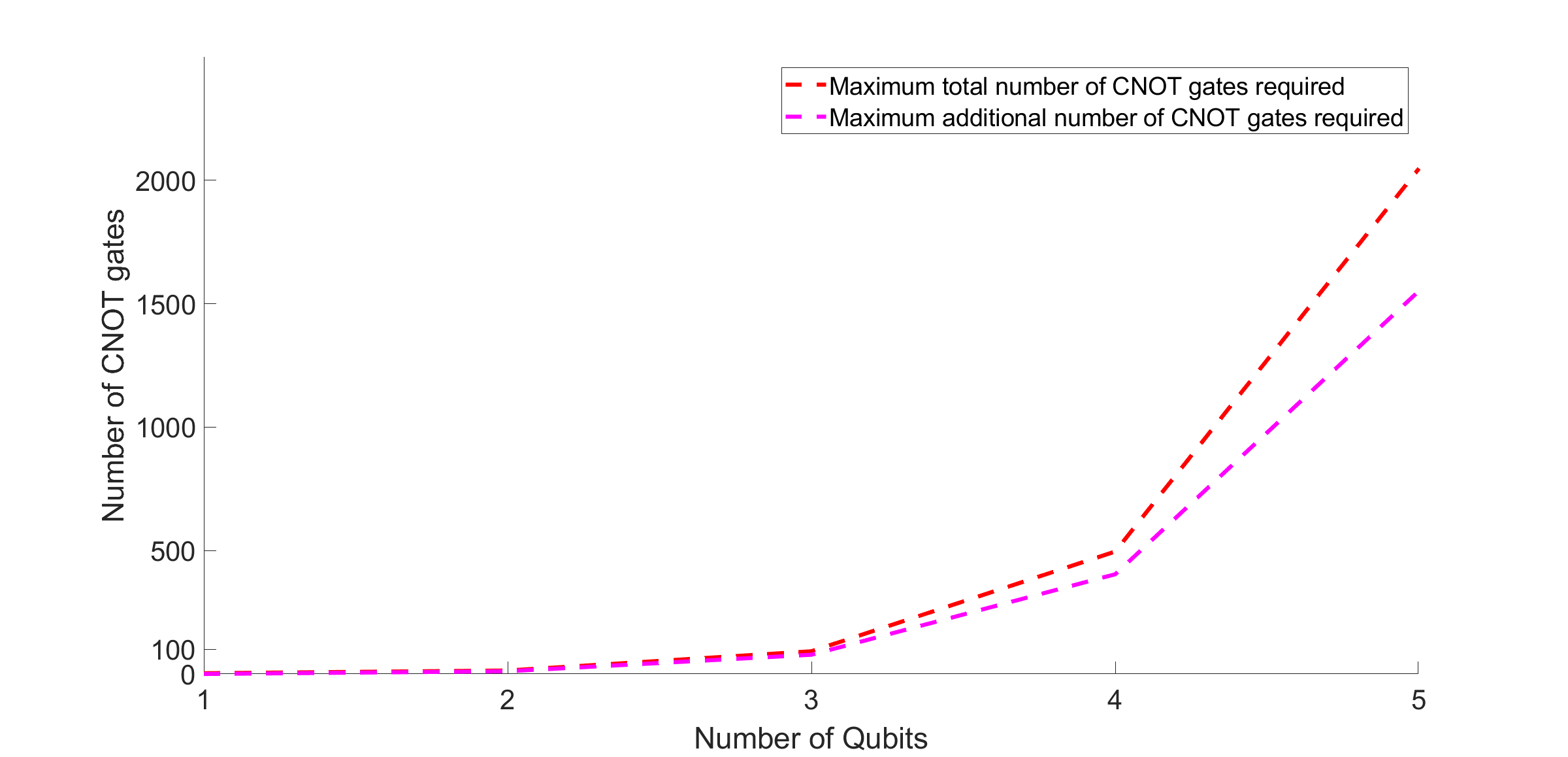}    
 \end{center}
  \caption{Red colored curve denotes the total number of $\cnot$ gates required for one layer of multiplication of exponentials of basis matrices in $n$-qubit system and magenta colored curve denotes the additional number of $\cnot$ gates required with the increase of number of qubits}
  \label{fig:cnotcount}
\end{figure}

In \textbf{Figure} \ref{fig:cnotcount}, we plot the growth of $\cnot$ gate count as the number of qubits increases while approximating special unitary matrices through Algorithm 2.

\subsection{Quantum circuit for two-qubit unitaries}\label{Sec:2qqc}

Now, we provide a parametric quantum circuit for approximating $2$-qubit special unitaries following Algorithm 2. The circuit given in Equation \ref{circuit:2q}  consists of $14$ $\cnot$ gates and $16$ $1$-qubit gates. Our circuit does not give minimum number of $\cnot$ gates however, our results coincide with number of $\cnot$ gates found in CS Decomposition \cite{krol2022}. 


\begin{tiny}
\begin{equation}\label{circuit:2q} \centerline{\Qcircuit @C=.5em @R=.5em {
&\lstick{1}&\ctrl{1}&\targ&\qw&\ctrl{1}&\qw&\qw&\ctrl{1}&\qw&\qw&\ctrl{1}&\qw&\targ&\qw&\ctrl{1}&\qw&\qw&\ctrl{1}&\qw&\qw&\ctrl{1}&\qw&\targ&\qw&\ctrl{1}&\qw&\qw&\ctrl{1}&\gate{R_z}&\qw&\ctrl{1}&\qw&\ctrl{1}\\
&\lstick{2}&\targ&\ctrl{-1}&\gate{R_z}&\targ&\gate{R_z}&\gate{R_y}&\targ&\gate{R_y}&\gate{R_z}&\targ&\gate{R_z}&\ctrl{-1}&\gate{R_z}&\targ&\gate{R_z}&\gate{R_y}&\targ&\gate{R_y}&\gate{R_z}&\targ&\gate{R_z}&\ctrl{-1}&\gate{R_z}&\targ&\gate{R_z}&\gate{R_y}&\targ&\gate{R_y}&\gate{R_z}&\targ&\gate{R_z}&\targ\\}}
\end{equation}\end{tiny}

The circuit given by equation (\ref{circuit:2q}) with layer $1$ represents the any unitary matrix represented as a product of exponentials of SRBB elements in the following order according to Algorithm 2. 
\begin{eqnarray*}
\zeta(\theta_3,\theta_8,\theta_{15}) &=& \exp\left(i \theta_3 U^{(4)}_{3}\right) \exp\left(i\theta_8 U^{(4)}_{8}\right) \exp\left(i \theta_{15} U^{(4)}_{15}\right) \\
\Psi (\theta_1,\theta_2,\theta_9,\theta_{12},\theta_{10},\theta_{13},\theta_4,\theta_6) &=& \exp\left(i \theta_{1} U^{(4)}_{1}\right)  \exp\left(i \theta_{2} U^{(4)}_{2}\right) \exp\left(i \theta_{9} U^{(4)}_{9}\right) \exp\left(i \theta_{12} U^{(4)}_{12}\right) \\
&&  \exp\left(i \theta_{10}U^{(4)}_{10}\right) \exp\left(i \theta_{13} U^{(4)}_{13}\right)\exp\left(i \theta_{4} U^{(4)}_{4}\right) \exp\left(i \theta_{6}U^{(4)}_{6}\right)   \\ 
\Phi (\theta_5,\theta_7,\theta_{11},\theta_{14}) &=&  \exp\left(i \theta_{5} U^{(4)}_{5}\right) \exp\left(i \theta_{7} U^{(4)}_{7}\right) \exp\left(i \theta_{11} U^{(4)}_{11}\right) \exp\left(i \theta_{14} U^{(4)}_{14}\right),
\end{eqnarray*} where $U_j^{(4)}$, $1\leq j\leq 15$ are the SRBB elements of $\C^{2^2\times 2^2}.$


The parametric quantum circuit representation of the circuit equation (\ref{circuit:2q}) is as follows: 

The quantum circuit for $\zeta(\Theta_\zeta)$ is given by equation (\ref{circuit:2qzita})

\begin{equation}\label{circuit:2qzita}\centerline{\Qcircuit @C=.5em @R=.5em {
&\lstick{1}&\qw&\ctrl{1}&\qw&\ctrl{1}&\gate{R_z(\theta_8)}\\
&\lstick{2}&\gate{R_z(\theta_{3})}&\targ&\gate{R_z(\theta_{15})}&\targ&\qw}}\end{equation}

Hence, the circuit for $\Psi(\Theta_\psi)$ in equation (\ref{circuit:2qpsi}) is
{\tiny{
\begin{equation}\label{circuit:2qpsi} \centerline{\Qcircuit @C=.5em @R=.5em {
&\lstick{1}&\targ&\qw&\ctrl{1}&\qw&\qw&\ctrl{1}& \qw&\qw&\ctrl{1}&\qw&\ctrl{1}&\targ& \qw&\ctrl{1}&\qw&\qw&\ctrl{1}& \qw&\qw&\ctrl{1}&\qw&\ctrl{1}&\qw\\
&\lstick{2}&\ctrl{-1}&\gate{R_z(a')}&\targ&\gate{R_z(b')}&\gate{R_y(\frac{{m}_1-{m}_9}{2})}&\targ&\gate{R_y(\frac{{m}_1+{m}_9}{2})}&\gate{R_z(a)}&\targ&\gate{R_z(b)}&\targ&\ctrl{-1}&\gate{R_z(\alpha')}&\targ&\gate{R_z(\beta')}&\gate{R_y(\frac{{\mu}_1-{\mu}_9}{2})}&\targ&\gate{R_y(\frac{{\mu}_1+{\mu}_9}{2})}&\gate{R_z(\alpha)}&\targ&\gate{R_z(\beta)}&\targ&\qw\\}}\end{equation}}}

The circuit of $\exp{(\iota  \theta_{1}B^{(4)}_{1})}\exp{(\iota  \theta_{2}B^{(4)}_{2})}\exp{(\iota  \theta_{9}B^{(4)}_{9})}\exp{(\iota  \theta_{12}B^{(4)}_{12})}$ is
\begin{equation}\label{circuit:2qpsi1}
\centerline{\Qcircuit @C=.5em @R=.5em {
&\lstick{1}& \qw&\ctrl{1}&\qw&\qw&\ctrl{1}& \qw&\qw&\ctrl{1}&\qw&\ctrl{1}&\qw\\
&\lstick{2}&\gate{R_z(\alpha')}&\targ&\gate{R_z(\beta')}&\gate{R_y(\frac{{\mu}_1-{\mu}_9}{2})}&\targ&\gate{R_y(\frac{{\mu}_1+{\mu}_9}{2})}&\gate{R_z(\alpha)}&\targ&\gate{R_z(\beta)}&\targ&\qw\\}}\end{equation} where

\begin{eqnarray*}
&& \alpha'=\frac{2\theta_9+2\theta_{12}-\kappa_2-\kappa_1-\gamma_2-\gamma_1}{4}, \,\, \beta'=\frac{2\theta_9+2\theta_{12}+\kappa_2+\kappa_1-\gamma_2-\gamma_1}{4} \\
&& \alpha=\frac{2\theta_1+2\theta_2+\kappa_2-\kappa_1+\gamma_2-\gamma_1}{4}, \,\, \beta=\frac{-2\theta_1-2\theta_2-\kappa_2+\kappa_1+\gamma_2-\gamma_1}{4}
\end{eqnarray*} with 
\begin{eqnarray*}
&& {\mu}_1=\arccos{\sqrt{(\cos{\theta_1}\cos{\theta_2})^2+(\sin{\theta_1}\sin{\theta_2})^2}}, \,\, \\&&{\mu}_9=\arccos{\sqrt{(\cos{\theta_9}\cos{\theta_{12}})^2+(\sin{\theta_9}\sin{\theta_{12}})^2}} \\
&& \gamma_1=\arccos{\frac{\cos{\theta_1}\cos{\theta_2}}{\cos{{\mu}_1}}}, \,\, \gamma_2=\arccos{\frac{\cos{\theta_1}\sin{\theta_2}}{\sin{{\mu}_1}}} \\
&& \kappa_1=\arccos{\frac{\cos{\theta_9}\cos{\theta_{12}}}{\cos{{\mu}_9}}}, \,\, \kappa_2=\arccos{\frac{\cos{\theta_9}\sin{\theta_{12}}}{\sin{{\mu}_9}}}.
\end{eqnarray*}

The circuit of $\exp{(\iota  \theta_{4}B^{(4)}_{4})}\exp{(\iota  \theta_{6}B^{(4)}_{6})}\exp{(\iota  \theta_{10}B^{(4)}_{10})}\exp{(\iota  \theta_{13}B^{(4)}_{13})}$ is in equation (\ref{circuit:2qpsi2}).

\begin{equation}\label{circuit:2qpsi2}\centerline{\Qcircuit @C=.5em @R=.5em {
&\lstick{1}&\targ&\qw&\ctrl{1}&\qw&\qw&\ctrl{1}& \qw&\qw&\ctrl{1}&\qw&\ctrl{1}&\targ\\
&\lstick{2}&\ctrl{-1}&\gate{R_z(a')}&\targ&\gate{R_z(b')}&\gate{R_y(\frac{{m}_4-{m}_{10}}{2})}&\targ&\gate{R_y(\frac{{m}_4+{m}_{10}}{2})}&\gate{R_z(a)}&\targ&\gate{R_z(b)}&\targ&\ctrl{-1}\\}}\end{equation} where 

\begin{eqnarray*}
&& a=\frac{2\theta_4+2\theta_6+g_1-g_2+p_2-p_1}{4}, \,\, b=\frac{2\theta_4+2\theta_6-g_1+g_2+p_2-p_1}{4} \\
&& a'=\frac{2\theta_{10}+2\theta_{13}+g_1+g_2-p_2-p_1}{4}, \,\, b'=\frac{-2\theta_{10}-2\theta_{13}-g_1-g_2-p_2-p_1}{4}
\end{eqnarray*} with 
\begin{eqnarray*}
&& m_4=\arccos{\sqrt{(\cos{\theta}_4\cos{\theta}_6)^2+(\sin{\theta}_4\sin{\theta}_6)^2}}, \,\,\\&& m_{10}=\arccos{\sqrt{(\cos{\theta}_{10}\cos{\theta}_{13})^2+(\sin{\theta}_{10}\sin{\theta}_{13})^2}}, \\
&& g_1=\arccos{\frac{\cos{\theta}_4\cos{\theta}_6}{\cos{m_4}}}, \,\, g_2=\arccos{\frac{\cos{\theta}_4\sin{\theta}_6}{\sin{m_4}}} \\
&& p_1=\arccos{\frac{\cos{\theta}_{10}\cos{\theta}_{13}}{\cos{m_{10}}}}, \,\, p_2=\arccos{\frac{\cos{\theta}_{10}\sin{\theta}_{10}}{\sin{m_{10}}}}.
\end{eqnarray*}

The quantum circuit for $\Phi(\Theta_{\phi})$ is given by equation (\ref{circuit:2qphi}).

\begin{equation}\label{circuit:2qphi}
\centerline{\Qcircuit @C=.5em @R=.5em {
&\lstick{1}&\targ&\ctrl{1}&\targ &\qw&\ctrl{1}&\qw&\qw&\ctrl{1}& \qw&\qw&\ctrl{1}&\qw&\ctrl{1}&\qw&\targ&\ctrl{1}&\targ \\
&\lstick{2}&\ctrl{-1}&\targ&\ctrl{-1}&\gate{R_z(\Tilde{\alpha'})}&\targ&\gate{R_z(\Tilde{\beta'})}&\gate{R_y(\frac{{u}_5-{u}_{11}}{2})}&\targ&\gate{R_y(\frac{{u}_5+{u}_{11}}{2})}&\gate{R_z(\Tilde{\alpha})}&\targ&\gate{R_z(\Tilde{\beta})}&\targ&\qw&\ctrl{-1}&\targ&\ctrl{-1}\\}}\end{equation}
where 
\begin{eqnarray*}
&& \Tilde{\alpha'}=\frac{2\theta_{11}+2\theta_{14}-\Tilde{\kappa}_2-\Tilde{\kappa}_1-\Tilde{\gamma}_2-\Tilde{\gamma}_1}{4}, \,\, \Tilde{\beta'}=\frac{2\theta_{11}+2\theta_{14}+\Tilde{\kappa}_2+\Tilde{\kappa}_1-\Tilde{\gamma}_2-\Tilde{\gamma}_1}{4}, \\
&& \Tilde{\alpha}=\frac{2\theta_5+2\theta_7+\Tilde{\kappa}_2-\Tilde{\kappa}_1+\Tilde{\gamma}_2-\Tilde{\gamma}_1}{4}, \,\, \Tilde{\beta}=\frac{-2\theta_5-2\theta_7-\Tilde{\kappa}_2+\Tilde{\kappa}_1+\Tilde{\gamma}_2-\Tilde{\gamma}_1}{4}
\end{eqnarray*}
with 
\begin{eqnarray*}
&& {u}_5=\arccos{\sqrt{(\cos{\theta_5}\cos{\theta_7})^2+(\sin{\theta_5}\sin{\theta_7})^2}}, \,\,\\&& {u}_{11}=\arccos{\sqrt{(\cos{\theta_{11}}\cos{\theta_{14}})^2+(\sin{\theta_{11}}\sin{\theta_{14}})^2}} \\
&& \Tilde{\gamma}_1=\arccos{\frac{\cos{\theta_5}\cos{\theta_7}}{\cos{{u}_5}}}, \,\, \Tilde{\gamma}_2=\arccos{\frac{\cos{\theta_5}\sin{\theta_7}}{\sin{{u}_5}}}, \\
&& \Tilde{\kappa}_1=\arccos{\frac{\cos{\theta_{11}}\cos{\theta_{14}}}{\cos{{u}_{11}}}}, \,\, \Tilde{\kappa}_2=\arccos{\frac{\cos{\theta_{11}}\sin{\theta_{14}}}{\sin{{u}_{11}}}}.
\end{eqnarray*}
\section{Conclusion} \label{sec:Conclusion2}
In this paper, we have introduced a recursive method for generation of a basis for the algebra of complex matrices of order $d\geq 2$ with basis elements as Hermitian, unitary and $1$-sparse matrices.  This basis is used to develop parametric representation of unitary matrices employing a Lie group theoretic approach. Further, optimized-based algorithms are proposed to approximate any target unitary matrix by determining optimal values of the parameters. Then the above results are applied to determine parametric representation of unitary matrices of order $d=2^n,$ which represent unitary evolution of $n$-qubit systems, by defining a new basis, which we call Standard Recursive Block Basis for the algebra of complex matrices of order $2^n$ obtained by changing certain elements of the above basis. Consequently, a scalable quantum circuit  model is implemented using the approximation algorithm in a quantum neural network framework for unitary evolution of $n$-qubit systems. The performance of the approximation algorithms is investigated through several examples for standard and random $2$-qubit, $3$-qubit and $4$-qubit unitaries. It is observed that the error of approximation reduces with the increase of iteration or layer of the approximation algorithm. In future, we plan to explore finding a connection between the optimal number of layers for the approximation algorithm with the error of accuracy of the algorithm for a given target unitary matrix. Besides, the performance of the proposed approximation algorithm can be investigated by implementing the proposed scalable quantum circuits in available NISQ computers with large number of qubits. Finding the efficiency of the parameterized quantum circuit with the available restricted set of quantum gates with specific quantum hardware architecture is another problem that should be explored in the future.

\section*{Acknowledgement} RSS acknowledges support through the Prime Minister's Research Fellowship (PMRF), Government of India when this work was carried out. The software implementation of the Algorithms \ref{algo1} and \ref{alg2} have been mainly carried out on the supercomputer PARAM Shakti of IIT Kharagpur, established under National Supercomputing Mission (NSM), Government of India and supported by Centre for Development of Advanced Computing (CDAC), Pune \cite{Paramshakti}.

\bibliographystyle{plain}
\bibliography{Bibliography}
\appendix
\section{Proof of Theorem \ref{constructeven1}}\label{AppendixA}

\pf Let us consider the case where $g=e$. Then, consider $n$-qubit quantum circuit given by equation (\ref{first1}).  
     \small\begin{eqnarray}\label{first1}
         {\Qcircuit @C=1em @R=.7em {
    &\lstick{1}&\qw&\qw&\qw &\qw&\qw&\qw\\
    &\lstick{\vdots}&\qw&\qw&\qw &\qw&\qw&\qw\\
    &\lstick{p_1}&\targ&\qw&\qw &\qw&\qw&\qw\\
    &\lstick{\vdots}&\qw&\qw&\qw &\qw&\qw&\qw\\
    &\lstick{p_2}&\qw&\targ&\qw &\qw&\qw&\qw\\
    &\lstick{\vdots}&\qw&\qw&\qw &\qw&\qw&\qw\\
    &\lstick{p_3}&\qw&\qw&\targ &\qw&\qw&\qw\\
    &\lstick{\vdots}&\qw&\qw&\qw &\qw&\qw&\qw\\
    &\lstick{\vdots}&\qw&\qw&\qw &\qw&\qw&\qw\\
    &\lstick{p_k}&\qw&\qw&\qw &\targ&\qw&\qw\\
    &\lstick{\vdots}&\qw&\qw&\qw &\qw&\qw&\qw\\
    &\lstick{n}&\ctrl{-9}&\ctrl{-7}&\ctrl{-5} &\ctrl{-2}&\qw&\qw\\}}
     \end{eqnarray}\normalsize We see that the circuit in equation (\ref{first1}) can be written as \begin{eqnarray*}
         \prod_{j=1}^k(\cnot)_{(n,p_{k-j+1})}= \prod_{j=1}^k(\cnot)_{(n,n-(n-p_{k-j+1}-1)-1)}
     \end{eqnarray*} where $p_1<m_2\hdots<p_k,$ $1\leq k\leq n-2$. Hence, from the definitions in equation (\ref{permusch1}), the circuit in equation (\ref{first1}) is denoted as $\Pi\mathsf{T}_{n,x}^e$ where $x=(x_{n-2},\hdots,x_{0}):=\sum_{j=1}^k2^{n-p_{k-j+1}-1}=\sum_{j=1}^k2^{n-p_{j}-1}$ and $\Lambda_x=\{n-p_j-1|j\in \{1,\hdots,k\}\}$

Now consider the canonical basis of $\C^{2^n}$ denoted as $B=\{\ket{v_1,v_2,\hdots,v_{n}}|v_l\in \{0,1\}, l\in\{1,\hdots,n\}\}.$ We also define an indexing on B via the bijective map $\mathbb{O}:B\rightarrow \mathbb{N}$ which is based on the basis elements in the following way by considering $\mathbb{O}(\ket{v_1,v_2,\hdots,v_{n}})=\sum_{j=1} 2^{n-j}v_j +1 $. That is, the map $\mathbb{O}$ produces an indexing on the basis elements (The extra $1$ in the map is added to preserve the range of the map). Then the output of the circuit \ref{first1} corresponding to the basis elements of $\C^{2^n}$ as inputs are given by \begin{eqnarray*}
    \ket{v_1,v_2,\hdots,v_{n-1},1}\rightarrow \ket{v_1,v_2,\hdots,\overline{v_{p_1}},\hdots,\overline{v_{p_2}},\hdots,\overline{v_{p_k}},\hdots,v_{n-1},1}
\end{eqnarray*} and \begin{eqnarray*}
    \ket{v_1,v_2,\hdots,v_{n-1},0}\rightarrow \ket{v_1,v_2,\hdots,v_{n-1},0}
\end{eqnarray*} where $\overline{v}=1\oplus v,$ $\oplus$ denotes the modulo $2$ addition.

 Note that the basis elements of the form $\ket{v_1,v_2,\hdots,v_{n-1},0}$ remain invariant under our linear map obtained from circuit in equation (\ref{first1}). And clearly from our ordering we see that the element $\sum_{j=1}^k2^{n-p_j}v_{p_j}+\sum_{l=1,l\neq\{p_1,\hdots,p_k\}}^{n-1}2^{n-l}v_l+2$ is mapped to the element $\sum_{j=1}^k2^{n-p_j}\overline{v_{p_j}}+\sum_{l=1,l\neq\{p_1,\hdots,p_k\}}^{n-1}2^{n-l}v_l+2$ and vice-versa for every $v_1,v_2\hdots,v_n\in\{0,1\}$. Hence, rewriting we see that the element $\sum_{j=1}^k2^{n-p_j-1+1}{v_{n-(n-p_j-1)-1}}+\sum_{l=0,l\not\in\Lambda_x}^{n-2}2^{n-l-1+1}v_{n-(n-l-1)-1}+2$ is mapped to the element indexed
$\sum_{j=1}^k2^{n-p_j-1+1}\overline{v_{n-(n-p_j-1)-1}}+\sum_{l=0,l\not\in\Lambda_x}^{n-2}2^{n-l-1+1}v_{n-(n-l-1)-1}+2$ and vice-versa. Now for each $0\leq m\leq 2^{n}-1$, consider $m=(m_{n-2},\hdots,m_0):=(v_1,v_2,\hdots,v_{n-1})$. Then, we get that the element $\sum_{k\in \Lambda_x} 2^{k+1}m_k+\sum_{j\not\in \Lambda_x } m_j2^{j+1}+2$ is mapped to the element $\sum_{\in \Lambda_x} 2^{k+1}\overline{m_k}+\sum_{j\not\in \Lambda_x } m_j2^{j+1}+2$ and vice-versa.

Let $T:\mathbb{C}^{2h}\rightarrow \mathbb{C}^{2h}$ be a bijective linear transformation on a complex vector space of even dimension $2h$ (even integer) such that $T^2=I$. Let $B=\{v_1,v_2,\hdots, v_{2h}\}$ be the standard basis of $\mathbb{C}^{2h}$ i.e. $v_j$ is a $2h-$ tupule vector with $1$ at the $j-$th poisition and rest is $0$. Then, considering $B$ as the basis for both the domain and range spaces of $T$, we introduce the mapping $T(v_{k_r})=v_{j_r},T(v_{j_r})=v_{k_r} ,T(v_l)=v_l, l\in \{1,2,\hdots,2h\}\setminus\{k_r,j_r,r\in\{1,\hdots,R\}\}$ for some $R$ such that $k_r<j_r\forall r$ and $(k_{r},j_{r})=(k_{r'},j_{r'})\implies r=r'$ i.e. $k_r,j_r$ are distinct. Then the  matrix of such a linear map gives us the product of disjoint transpositions $\prod_{r=1}^RP_{(k_{r},j_{r})}$. Our circuit is a unitary matrix and hence its map is linear. Also it is obvious that putting two identical circuits of the form in equation (\ref{first1}) gives us the identity map because the mapped elements are reverted back to itself. We define $\alpha_{\Lambda_x}(m)=\sum_{k\in \Lambda_x} 2^{k+1}m_k+\sum_{j\not\in \Lambda_x } m_j2^{j+1}+2$ and $\beta_{\Lambda_x}(m)=\sum_{k\in \Lambda_x} 2^{k+1}\overline{m_k}+\sum_{j\not\in \Lambda_x } m_j2^{j+1}+2$. Note that $m$ can take $2^{n-1}$ values and for each unique $m$ we get unique $\alpha_{\Lambda_x}(m)$ and $\beta_{\Lambda_x}(m)$. Also it is easy to see that $\alpha_{\Lambda_x}(m)\neq \beta_{\Lambda_x}(m)\forall m\in \{0,\hdots,2^{n-1}-1\}$. Since, the transposition $P_{(\alpha,\beta)}=P_{(\beta,\alpha)}$, we consider the cases where $\alpha_{\Lambda_x}(m)<\beta_{\Lambda_x}(m)$ only. From simple combinatorics, this will happens for half of $m$'s. Thus our circuit in equation (\ref{first1}) is a product of disjoint $2^{n-2}$ transpositions of the form provided in the statement of Theorem \ref{constructeven1}. Further, it is of note that the condition $\alpha^e_{\Lambda_x}<\beta^e_{\Lambda_x}$ is considered to stop the over-count since any 2-cycle permutation is also symmetric and $P_{(\alpha,\beta)}=P_{(\beta,\alpha)}$. 

Now let us take another circuit for $\Pi\mathsf{T}_{n,y}^e, y=\sum_{j=1}^{k'} 2^{n-q_j-1}\neq x$ given by equation (\ref{second1}).
\small\begin{eqnarray}\label{second1}
    {\Qcircuit @C=1em @R=.7em {
    &\lstick{1}&\qw&\qw&\qw &\qw&\qw&\qw\\
     &\lstick{\vdots}&\qw&\qw&\qw &\qw&\qw&\qw\\
    &\lstick{q_1}&\targ&\qw&\qw &\qw&\qw&\qw\\
    &\lstick{\vdots}&\qw&\qw&\qw &\qw&\qw&\qw\\
    &\lstick{q_2}&\qw&\targ&\qw &\qw&\qw&\qw\\
    &\lstick{\vdots}&\qw&\qw&\qw &\qw&\qw&\qw\\
    &\lstick{q_3}&\qw&\qw&\targ &\qw&\qw&\qw\\
    &\lstick{\vdots}&\qw&\qw&\qw &\qw&\qw&\qw\\
    &\lstick{q_{k'}}&\qw&\qw&\qw &\targ&\qw&\qw\\
    &\lstick{\vdots}&\qw&\qw&\qw &\qw&\qw&\qw\\
    &\lstick{n}&\ctrl{-8}&\ctrl{-6}&\ctrl{-4} &\ctrl{-2}&\qw&\qw\\}}
\end{eqnarray}\normalsize
In the circuits in equations (\ref{first1}) and (\ref{second1}), not all $p_j$'s and $q_j's$ are distinct. However the condition $y\neq x$ implies that the set $\Lambda_x\subset [n-1]$ and $\Lambda_y\subset [n-1]$ have at least one element that is not contained in other i.e. $\exists$ at least one $p_l\in \Lambda_{x}$ such that $p_l\not \in \Lambda_y$ i.e. $p_l\neq q_j\forall q_j\in \Lambda_y$. Let for some $0\leq m\leq 2^{n-1}-1$, $(\alpha^e_{\Lambda_x}(m),\beta^e_{\Lambda_x}(m))= (\alpha^e_{\Lambda_y}(m),\beta^e_{\Lambda_y}(m))$ i.e. $\Pi\mathsf{T}_{n,x}^e$ and $\Pi\mathsf{T}_{n,y}^e$ share some transposition. Then there exists some basis element $\ket{v_1,v_2,\hdots,v_{n-1},1}$ of  $\mathbb{C}^{2^n}$ whose image is mapped to the same element under circuits in equations (\ref{first1}) and (\ref{second1}). Under the mapping from circuit in equation (\ref{first1}), $v_{p_l}\rightarrow \overline{v_{p_l}}$ but when passed through the circuit in equation (\ref{second1}), $v_{p_l}\rightarrow v_{p_l}$. Hence $\exists$ at least one $p_l$ such that $v_{p_l}=\overline{v_{p_l}}$ which is a contradiction. Hence for $x\neq y$, the permutation matrices $\Pi\mathsf{T}_{n,x}^e$ and $\Pi\mathsf{T}_{n,y}^e$ do not share any transpositions.

The proof is similar for $\Pi\mathsf{T_{n,x}^o}$ i.e. for the case $g=o$. In such a case, we take the following circuit for $\Pi\mathsf{T}_{n,x}^o$ where $x=(x_{n-2},\hdots,x_{0}):=\sum_{j=1}^k2^{n-p_{k-j+1}-1}=\sum_{j=1}^k2^{n-p_{j}-1}$ and $\Lambda_x=\{n-p_j-1|j\in \{1,\hdots,k\}\}$ such that $n-p_1>\hdots>n-p_k$. \small\begin{eqnarray}\label{third1}
    {\Qcircuit @C=1em @R=.7em {
    &\lstick{1}&\qw&\qw&\qw&\qw &\qw&\qw&\qw&\qw\\
     &\lstick{\vdots}&\qw&\qw&\qw&\qw &\qw&\qw&\qw&\qw\\
    &\lstick{p_1}&\ctrl{8}&\targ&\qw&\qw &\qw&\qw&\ctrl{8}&\qw\\
    &\lstick{\vdots}&\qw&\qw&\qw&\qw &\qw&\qw&\qw&\qw\\
    &\lstick{p_2}&\qw&\qw&\targ&\qw &\qw&\qw&\qw&\qw\\
    &\lstick{\vdots}&\qw&\qw&\qw&\qw &\qw&\qw&\qw&\qw\\
    &\lstick{p_3}&\qw&\qw&\qw&\targ &\qw&\qw&\qw&\qw\\
    &\lstick{\vdots}&\qw&\qw&\qw&\qw &\qw&\qw&\qw&\qw\\
    &\lstick{p_k}&\qw&\qw&\qw&\qw &\targ&\qw&\qw&\qw\\
    &\lstick{\vdots}&\qw&\qw&\qw&\qw &\qw&\qw&\qw&\qw\\
    &\lstick{n}&\targ&\ctrl{-8}&\ctrl{-6}&\ctrl{-4} &\ctrl{-2}&\qw&\targ&\qw\\}}
\end{eqnarray}\normalsize
In such cases, the elements of $B$ are mapped in the following way.
\begin{eqnarray*}
    \ket{v_1,v_2,\hdots,v_{p_1},\hdots,v_{n-1},v_n}\rightarrow \ket{v_1,v_2,\hdots,\overline{v_{p_1}},\hdots,\overline{v_{p_2}},\hdots,\overline{v_{p_k}},\hdots,v_{n-1},\overline{v_n}}
\end{eqnarray*} if $v_n\oplus v_{p_1}=1$ and \begin{eqnarray*}
    \ket{v_1,v_2,\hdots,v_{p_1},\hdots,v_{n-1},v_n}\rightarrow \ket{v_1,v_2,\hdots,v_{p_1},\hdots,v_{n-1},v_n}
\end{eqnarray*} if $v_n\oplus v_{p_1}=0$. The rest of the proof follows similar to the $g=e$ case. This concludes the proof. $\hfill\square$
\end{document}